\documentstyle[amssymb,12pt,manuscript,aps,epsf]{revtex}

\begin{document}
\author{Amand Faessler$^{a)}$, C. Fuchs$^{a)}$, and M. I. Krivoruchenko$^{a,b)}$}
\address{$^{a)}${\small Institut f\"{u}r Theoretische Physik, Universit\"{a}t}\\
T\"{u}bingen, Auf der Morgenstelle 14\\
{\small D-72076 T\"{u}bingen, Germany}\\
$^{b)}${\small Institute for Theoretical and Experimental Physics, B.}\\
Cheremuskinskaya 25\\
{\small 117259 Moscow, Russia}}
\title{Dilepton Spectra from Decays of Light Unflavored Mesons}
\maketitle

\begin{abstract}
The invariant mass spectrum of the $e^{+}e^{-}$ and $\mu ^{+}\mu ^{-}$ pairs
from decays of light unflavored mesons with masses below the $\phi (1020) $%
-meson mass to final states containing along with a dilepton pair one
photon, one meson, and two mesons are calculated within the framework of the
effective meson theory. The results can be used for simulations of the
dilepton spectra in heavy-ion collisions and for experimental searches of
dilepton meson decays.
\end{abstract}

\newpage 

\section{Introduction}

\setcounter{equation}{0}

During the last decade, the problem of the description of hadrons in dense
and hot nuclear matter received great attention. It is well known that
particles change their properties when they are placed into a medium.
Already in the seventies the reduction of the nucleon masses in nuclei was
implemented into the Walecka model in the framework of effective hadron
field theory \cite{Wal,Chin}. Later on this effect was put on firmer grounds
on the basis of a partial restoration of chiral symmetry and finite-density
QCD sum rules \cite{QCD}. The change of the meson properties is also
discussed in the quantum hadrodynamics \cite{Chin} and finite-density QCD
sum rules \cite{MES}. The investigations in the Nambu-Jona-Lasinio model
provide an evidence for reduction of nucleon and meson masses at finite
density and temperature as well \cite{Fae}.

On the other hand, many-body correlations lead to a dressing of the
particles inside the medium and strongly modify spectral properties of the
mesons \cite{herrmann93,rapp97,peters98}. Thus one expects a significant
reduction of the corresponding life times which, loosely speaking, result in
melting the mesons in the nuclear environment.

The melting of the higher nucleon resonances is established experimentally
from the measurement of the total photoabsorption cross section on heavy
nuclei \cite{Bia}. There is a clear signal for a change of the shape of the $%
\Delta $-resonance in nuclei, related to the Fermi motion and, as noticed in
Ref. \cite{Kon}, to the collision broadening effect discussed first by
Weisskopf \cite{Wei} in connection with a broadening of the atomic spectral
lines in gases. The higher nucleon resonances are not seen in the cross
section due to a strong collision broadening effect.

The purpose of the current investigations is to determine mass shifts and
the broadening of hadronic resonances in nuclear matter. The formulation of
this problem can be traced back to the atomic spectroscopy where the shifts
of atomic energy levels and the broadening of atomic spectral lines in dense
and hot gases is a relatively well studied subject (see {\it e.g.} Ref. \cite
{SOB}).

The search for signatures of modified hadron properties such as reduction
masses according to the Brown-Rho scaling \cite{BR95} are presently pursued
with high experimental efforts. However, most hadronic probes loose
important information on the early and violent phase of the reaction by
strong final state interactions. One of the most promising probes to study
in-medium properties of vector mesons are dilepton pairs $\ell ^{+}\ell ^{-}$
($\ell =e,$ $\mu $) which are produced from decays of $\rho ^0$-, $\omega $%
-, and $\phi $-mesons in heavy-ion collisions. The leptonic probes have the
advantage that they are the nearly undistorted messengers from the
conditions at their creation, unlike {\it e.g.} pions.

The dilepton spectra measured by the CERES and HELIOS-3 Collaborations at
CERN SPS have attracted special interest \cite{ceres,helios}. Compared to
theoretical predictions, both experiments found a significant enhancement of
the low-energy dilepton yield below the $\rho $ and $\omega $ peaks. One way
to explain this low-energy dilepton excess is to assume the scenario of a
significant reduction of the $\rho $-meson mass in a dense medium \cite
{koch92,li95,Cassing95}. On the other hand, a more sophisticated treatment
of the $\rho $-meson spectral function which includes the broadening in
dense matter \cite{rapp97,peters98} seems also to be sufficient to account
for these data \cite{Cassing95,Cassing98}.

An excess of low-energy dileptons occurs already at moderate bombarding
energies. However, the spectra obtained by the DLS Collaboration at the
BEVALAC \cite{DLS} for the incident energies around $1$ $A\cdot $GeV cannot
be reproduced by present transport calculations \cite{ernst}. Even using the
reduction of the $\rho $-meson mass, there remains a discrepancy for the
dilepton yield by a factor of $2$ to $3$ \cite{ernst,BK}. Also the medium
dependence of the spectral function, even in combination with a dropping $%
\rho $ mass, does not provide an explanation for this so called 'DLS puzzle' 
\cite{brat98,BC}. It is interesting to note that this fact is independent on
the system size and occurs in light ($d+Ca$) as well as in heavier systems ($%
Ca+Ca$). The dilepton yield in elementary $p+p$ collisions measured by the
DLS Collaboration \cite{Wil} is also underestimated by standard theoretical
descriptions \cite{ernst}. In Ref. \cite{pppd} it was claimed that the
discrepancy between the DLS data and theoretical simulations disappears if
additional background contributions from dilepton decays of higher nucleon
resonances (mainly $N^{*}(1520)$) are taken into account. This demonstrates
that a most precise knowledge of the background is indispensable for the
interpretation of the present and future dilepton data. The HADES experiment
at GSI, Germany, will focus on these topics to a large extent \cite{hades}.

While the dilepton spectra stemming from the mesonic decays are not
distorted by final state interactions, the problem of extracting information
on the in-medium properties of the vector mesons is a specific theoretical
task, due to a large amount of decays contributing to the background. A
precise and rather complete knowledge of the relative weights for existing
decay channels is therefore indispensable in order to draw reliable
conclusions from dilepton spectra.

The study of the dilepton decays is useful also for the search on the
dilepton decay modes of the light unflavored mesons. These decays give a
deeper insight into meson structure, allowing to measure transition form
factors at the time-like (resonance) region. In four-body decays the decay
probabilities are determined by the half-off-shell meson transition form
factors which cannot be measured in other reactions. There are plans to
study the $\eta ^{\prime }$ transition form factors in the space-like region
in reactions of the photo- and electroproduction at CEBAF energies \cite{Dav}%
. Recently, the DLS Collaboration published data on the dilepton production
in the elementary $pp$ and $pd$ collisions \cite{Wil}. These data are
analyzed in Refs. \cite{ernst,pppd}. The plans from the HADES Collaboration
include measurements of the dilepton spectra from proton-proton and
pion-proton collisions \cite{hades}. These experiments are stimulated by the
already mentioned discrepancy between the number of the observed dilepton
events and results of the transport simulations, that indicate a limited
understanding of the mechanism for the dilepton emission in heavy-ion
collisions. In such experiments, there is a possibility for exclusive
measurements of the different mesonic channels.

In the present work, we perform a detailed study of possible mesonic $\ell
^{+}\ell ^{-}$ decays which appear in the SIS energy range, {\it i.e.} at
lab. energies below $2$ GeV/nucleon. At $2$ GeV/nucleon, the $\rho $-meson
is produced slightly above the threshold. Due to statistical fluctuations,
the production of heavier mesons is, however, also possible. We consider
decays of unflavored light mesons with masses below the $\phi (1020)$-meson
mass within the framework of the effective meson theory. The vertex
couplings are determined from the measured strong and radiative decay widths
and, when the experimental data are not available, from $SU(3)$ symmetry.
The transition form factors entering the decay rates are calculated using
the Vector Meson Dominance (VMD) model. In this way, we achieve generally
good agreement with the experimental branching ratios for radiative meson
decays. For the dilepton decay modes the branching ratios are, however, only
known in a few cases.

We consider the vector mesons $\rho $, $\omega $, and $\phi (1020)$ ($=V$),
the pseudo-scalar mesons $\pi $, $\eta $, $\eta ^{\prime }$ ($=P$), and the
scalar mesons $f_0(980)$ and $a_0(980)$ ($=S$). The various decay modes can
systematically be classified as follows: (i) There are, first of all, the
direct decays modes $V\rightarrow \ell ^{+}\ell ^{-}$, which contain the
information on the in-medium vector meson masses. In the next Sect., some
useful relations which simplify calculations of the decay rates to final
states with a dilepton pair are derived. In Sect.3 we make a few remarks on
the direct decay modes. There exits then a large number of processes which
mask the vector meson peaks and which should be treated as a background.
(ii)\ These are Dalitz decays of pseudoscalar mesons $P\rightarrow \gamma
\ell ^{+}\ell ^{-}$ and scalar mesons $S\rightarrow \gamma \ell ^{+}\ell
^{-} $. These decays are discussed in Sect.4. (iii) One has also Dalitz
decays with one meson in the final states $V\rightarrow P\ell ^{+}\ell ^{-}$
and $P\rightarrow V\ell ^{+}\ell ^{-}$, which we discuss in Sect. 5. The
radiative decays of these mesons are well studied both from the theoretical
and experimental points of view. The uncertainties in the estimates for the
dilepton decays are connected with the lack of experimental information on
the transition form factors. Constructing the VMD model transition form
factors, we take special care of the quark counting rules. (iv) Finally,
there exist decays to four-body final states $V\rightarrow PP\ell ^{+}\ell
^{-}$, $P\rightarrow PP\ell ^{+}\ell ^{-}$, and $S\rightarrow PP\ell
^{+}\ell ^{-}$. Sect. 6 is devoted to these decays. We calculate almost all
four-body dilepton modes of the unflavored mesons with masses below the $%
\phi (1020)$. The numerical results for radiative widths of the three-body
decays ($V\rightarrow PP\gamma $, {\it etc.}), which provide a useful test
for the model considered, for the dilepton widths ($V\rightarrow P\ell
^{+}\ell ^{-},$ $V\rightarrow PP\ell ^{+}\ell ^{-}$, {\it etc.}), and for
the dilepton spectra from the unflavored meson decays are presented in Sect.
7.


\section{Relation between the decays ${\sf M}\rightarrow {\sf M}^{\prime
}\gamma ^{*}$ and ${\sf M}\rightarrow {\sf M}^{\prime }\ell ^{+}\ell ^{-}$}

\setcounter{equation}{0}

As mentioned above, we consider decays ${\sf M}\rightarrow {\sf M}^{\prime
}\ell ^{+}\ell ^{-}$ where ${\sf M}$ is a meson, ${\sf M}^{\prime }$ is a
photon, a meson, or two mesons, and $\ell ^{+}\ell ^{-}$ is an
electron-positron or muon-antimuon pair. The results of this Sect. are
valid, however, for arbitrary states ${\sf M}^{\prime }$. The decay ${\sf M}%
\rightarrow {\sf M}^{\prime }\ell ^{+}\ell ^{-}$ proceeds through two steps: 
${\sf M}\rightarrow {\sf M}^{\prime }\gamma ^{*}$ and $\gamma
^{*}\rightarrow \ell ^{+}\ell ^{-}$, where $\gamma ^{*}$ is a virtual photon
whose mass, $M$, is equal to the invariant mass of the dilepton pair.

The matrix element of the physical process ${\sf M}\rightarrow {\sf M}%
^{\prime }\gamma $ for a real photon $\gamma $ has the form 
\begin{eqnarray}
{\cal M}={\cal M}_{\mu }\varepsilon _{\mu }^{*}(k)  \label{II.1}
\end{eqnarray}
where $\varepsilon _{\mu }(k)$ is a photon polarization vector. The matrix
element ${\cal M}_{\mu }$ is defined also at $k^{2}=M^{2}\neq 0$ for virtual
photons $\gamma ^{*}$. As a consequence of the gauge invariance, it is
transverse with respect to the photon momentum 
\begin{eqnarray}
{\cal M}_{\mu }k_{\mu }=0.  \label{II.2}
\end{eqnarray}

The decay rate ${\sf M}\rightarrow {\sf M}^{\prime }\gamma ^{*}$ can
formally be calculated as 
\begin{eqnarray}
d\Gamma ({\sf M}\rightarrow {\sf M}^{\prime }\gamma^{*})=\frac{1}{2\sqrt{s}}%
\sum_{f}\overline{{\cal M}_{\mu }{}{\cal M}_{\nu }{}^{*}}(-g_{\mu \nu })%
\frac{(2\pi )^{4}}{(2\pi )^{3n+3}}d\Phi _{n+1}  \label{II.3}
\end{eqnarray}
where $\sqrt{s}\ $is mass of the decaying meson and $n$ is number of
particles in the state ${\sf M}^{\prime }$. The phase space in Eq.(\ref{II.3}%
) is defined in the usual way 
\begin{eqnarray}
d\Phi _{k}(\sqrt{s},m_{1},...,m_{k})=\prod_{i=1}^{k}\frac{d{\bf p}_{i}}{%
2E_{i}}\delta ^{4}(P-\sum_{i=1}^{k}p_{i}).  \label{II.4}
\end{eqnarray}
Here, $P$ is the four-momentum of the meson ${\sf M}$ , $P^{2}=s,$ and $%
p_{i} $ are momenta of the particles in the final state, including the
virtual photon $\gamma ^{*}$. In Eq.(\ref{II.3}), the summation over the
final states and averaging over the initial states of the decaying meson is
performed. The limit $M^{2}\rightarrow 0$ gives the decay rate of the
physical process ${\sf M}\rightarrow {\sf M}^{\prime }\gamma $.

The ${\sf M}\rightarrow {\sf M}^{\prime }\ell ^{+}\ell ^{-}$ decay rate is
given by 
\begin{eqnarray}
d\Gamma ({\sf M}\rightarrow {\sf M}^{\prime }\ell ^{+}\ell ^{-})=\frac{1}{2%
\sqrt{s}}\sum_{f}\overline{{\cal M}_{\mu }{}{\cal M}_{\nu }{}^{*}}j_{\mu
}j_{\nu }{}^{*}\frac{1}{M^{4}}\frac{(2\pi )^{4}}{(2\pi )^{3n+6}}d\Phi _{n+2}
\label{II.5}
\end{eqnarray}
where $j_{\mu }\ $is the lepton current. The term $1/M^{4}$ comes from the
photon propagator, and $d\Phi _{n+2}$ is the phase space of $n$ particles in
the state ${\sf M}^{\prime }$ and of the $\ell ^{+}\ell ^{-}$ pair.

The value $\Gamma ({\sf M}\rightarrow {\sf M}^{\prime }\ell ^{+}\ell ^{-})$
can be related to the decay rates $\Gamma ({\sf M}\rightarrow {\sf M}%
^{\prime }\gamma ^{*})$ and $\Gamma (\gamma ^{*}\rightarrow \ell ^{+}\ell
^{-})$. In the analogy with massive vector particles, the width of a virtual
photon $\gamma ^{*}$ can formally be evaluated as 
\begin{eqnarray}
M\Gamma (\gamma ^{*}\rightarrow \ell ^{+}\ell ^{-})=\frac{\alpha }{3}%
(M^{2}+2m_{\ell }^{2})\sqrt{1-\frac{4m_{\ell }^{2}}{M^{2}}}  \label{II.6}
\end{eqnarray}
where $m_{\ell }$ is the lepton mass. The expression for the product of two
dilepton currents, summed up over the final states of the $\ell ^{+}\ell
^{-} $ pair, has the form 
\begin{eqnarray}
\sum_{f}j_{\mu }j_{\nu }{}^{*}=\frac{16\pi \alpha }{3}(M^{2}+2m_{\ell
}^{2})(-g_{\mu \nu }+\frac{k_{\mu }k_{\nu }}{M^{2}})  \label{II.7}
\end{eqnarray}
where $\alpha $ is the fine-structure constant and $k$ is the total momentum
of the pair. Factorizing the $n$-body invariant phase space, 
\begin{eqnarray}
d\Phi _{k}(\sqrt{s},m_{1},...,m_{k})=d\Phi _{k-1}(\sqrt{s}%
,m_{1},...,m_{k-2},M)dM^{2}\Phi _{2}(M,m_{k-1},m_{k}),  \label{II.8}
\end{eqnarray}
which can be proved by inserting the unity decomposition 
\begin{eqnarray}
1=\int d^{4}qdM^{2}\delta (q^{2}-M^{2})\delta ^{4}(q-p_{k-1}-p_{k})
\label{II.9}
\end{eqnarray}
into Eq.(\ref{II.4}), one obtains from Eqs.(\ref{II.3}), (\ref{II.5}), and (%
\ref{II.6}) with the help of Eqs.(\ref{II.7}) and (\ref{II.8}) the following
expression 
\begin{eqnarray}
d\Gamma ({\sf M}\rightarrow {\sf M}^{\prime }\ell ^{+}\ell ^{-})=d\Gamma (%
{\sf M}\rightarrow {\sf M}^{\prime }\gamma ^{*})M\Gamma (\gamma
^{*}\rightarrow \ell ^{+}\ell ^{-})\frac{dM^{2}}{\pi M^{4}}.  \label{II.10}
\end{eqnarray}
The factor $dM^{2}/(\pi M^{4})$ has the form of a properly normalized
Breit-Wigner distribution for a zero-mass resonance.

The two-body phase space in Eq.(\ref{II.8}) has the form 
\begin{eqnarray}
\Phi _{2}(\sqrt{s},m_{1},m_{2})=\frac{\pi p^{*}(\sqrt{s},m_{1},m_{2})}{\sqrt{%
s}}  \label{II.11}
\end{eqnarray}
where 
\begin{eqnarray}
p^{*}(\sqrt{s},m_{1},m_{2})=\frac{\sqrt{%
(s-(m_{1}+m_{2})^{2})(s-(m_{1}-m_{2})^{2})}}{2\sqrt{s}}  \label{II.12}
\end{eqnarray}
is momentum of the particles $1$ and $2$ in the c. m. frame.

In the following, we will work with the matrix elements of the processes $%
{\sf M}\rightarrow {\sf M}^{\prime }\gamma ^{*}$ and use Eq.(\ref{II.3}) to
derive results for the decay rates ${\sf M}\rightarrow {\sf M}^{\prime
}\gamma $ with real photons and Eq.(\ref{II.10}) to get results for decay
rates ${\sf M}\rightarrow {\sf M}^{\prime }\ell ^{+}\ell ^{-}$ with
dileptons in the final states. Now, we will consider the processes which are
interesting for the study of dilepton spectra in heavy-ion collisions. 

\section{Decays of the $\rho $-$,$ $\omega $-$,$ and $\phi $-mesons to $\ell
^{+}\ell ^{-}$ pairs}

\setcounter{equation}{0}

\begin{figure}[h]
\begin{center}
\leavevmode
\epsfxsize = 11cm
\epsffile[85 400 460 690]{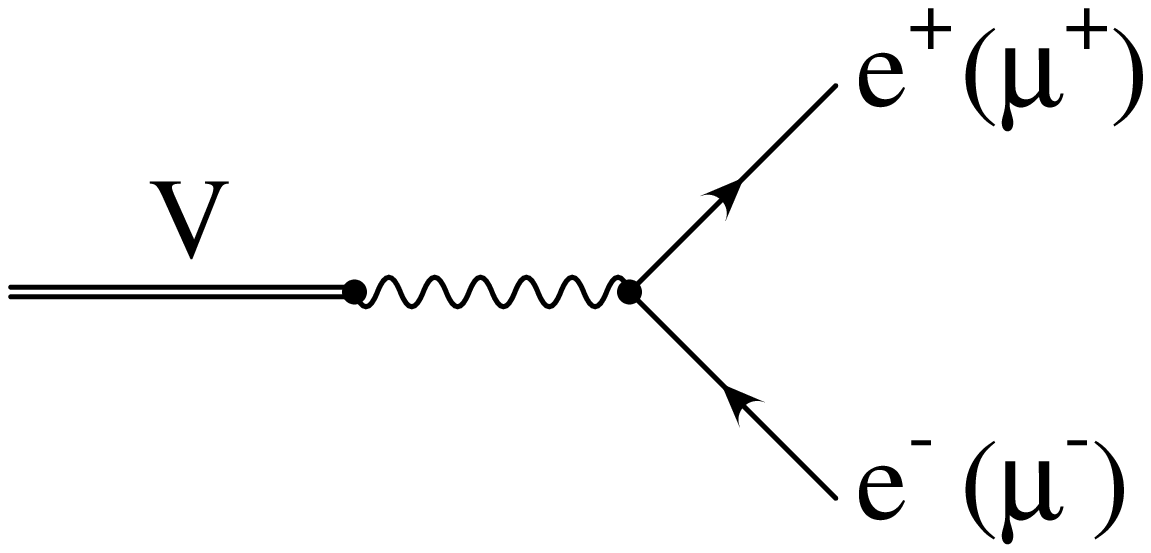} 
\end{center}
\caption{}
\label{fig1}
\end{figure}

The diagram for the $V\rightarrow \ell ^{+}\ell ^{-}$ decays with $V=\rho ,$ 
$\omega ,$ and $\phi $ is shown in Fig. \ref{fig1}. In terms of the vector
meson fields, $V_{\mu }$, the electromagnetic current has the form \cite{SAK}
\begin{eqnarray}
j_{\mu }=-e\sum_{V}\frac{m_{V}^{2}}{g_{V}}V_{\mu }  \label{III.1}
\end{eqnarray}
where $m_{V}$ are the vector meson masses and $e=-|e|$ is the electron
charge. The $SU(3)$ predictions for the coupling constants, $g_{\rho
}:g_{\omega }:g_{\phi }=1:3:\frac{-3}{\sqrt{2}},$ are in good agreement with
the ratios between the values $g_{\rho }=5.03,$ $g_{\omega }=17.1,$ and $%
g_{\phi }=-12.9$ extracted from the $e^{+}e^{-}$ decay widths of the $\rho $-%
$,$ $\omega $-$,$ and $\phi $-mesons with the use of the well known
expression 
\begin{eqnarray}
\Gamma (V\rightarrow \ell ^{+}\ell ^{-})=\frac{8\pi \alpha ^{2}}{3g_{V}^{2}}%
(1+2\frac{m_{\ell }^{2}}{m_{V}^{2}})p^{*}(m_{V},m_{\ell },m_{\ell }),
\label{III.2}
\end{eqnarray}
with $p^{*}$ defined in Eq.(\ref{II.12}).

\section{Meson decays to photons and $\ell ^{+}\ell ^{-}$ pairs}

\setcounter{equation}{0}

In this Sect., we discuss meson decays $P\rightarrow \gamma \ell ^{+}\ell
^{-}$ (Fig. 2) and $S\rightarrow \gamma \ell ^{+}\ell ^{-}$ (Fig. 3) for $%
P=\pi ^{0},\eta ,$ and $\eta ^{\prime }$ and $S=f_{0}(980)$ and $%
a_{0}^{0}(980)$. As we shall see, these decays are the dominant $e^{+}e^{-}$
modes for $\pi ^{0}$-, $\eta $-mesons, for $\eta ^{\prime }$-meson at $M$ $%
\gtrsim 250$ MeV, $f_{0}$-meson at $M$ $\gtrsim 500$ MeV and in some other
cases. The $\mu ^{+}\mu ^{-}$ modes are also discussed.

These decays are related to the experimentally measured two photon decays.
The uncertainties in the estimates originate only from the purely known
transition form factors in the time-like region. The $\eta \gamma \gamma
^{*} $ transition form factor is in reasonable agreement with the one-pole
VMD model predictions \cite{LGL}. The experimental errors in the $\eta
^{\prime } $ transition form factor are large \cite{Bayu}. The one-pole VMD
approximation for the $P\gamma \gamma ^{*}$ transition form factors is in
agreement with the quark counting rules which predict for these form factors
a $\sim 1/t$ asymptotics \cite{QCR}.

The nature of the scalar mesons has been a subject of intensive discussions
for a long time. The 4-quark content of the scalar mesons would imply a $%
\sim 1/t^{2}$ asymptotics for the $S\gamma \gamma ^{*}$ transition form
factors. In order to provide the correct asymptotics for the 4-quark meson
transition form factors, the VMD model should be extended to include
contributions from higher vector meson resonances. The $\omega \pi \gamma $
transition form factor has also $\sim 1/t^{2}$ asymptotic behavior \cite{VZ}%
. This form factor is measured in the time-like region \cite{LGL} and the
data show deviations from the naive one-pole approximation. The inclusion of
higher vector meson resonances improves the agreement and provides the
correct asymptotics.

The results of a recent measurement of the $\phi \rightarrow \gamma f_{0}$
branching ratio are interpreted as evidence for the dominance of a 4-quark
MIT bag component in the $f_{0}$-meson wave function \cite{MNA,VEPP}. We
thus calculate branching ratios $S\rightarrow \gamma \ell ^{+}\ell ^{-}$
assuming the 4-quark nature of the scalar mesons and imposing constraints
from the quark counting rules to the $S\gamma \gamma ^{*}$ transition form
factors.


\subsection{Decay modes $\pi ^{0}\rightarrow \gamma e^{+}e^{-},$ $\eta
\rightarrow \gamma \ell ^{+}\ell ^{-},$ and $\eta ^{\prime }\rightarrow
\gamma \ell ^{+}\ell ^{-}$}


The effective vertex for the $P\rightarrow \gamma \gamma $ decays has the
form 
\begin{eqnarray}
\delta {\cal L}_{P\gamma \gamma }=f_{P\gamma \gamma }\epsilon _{\tau \sigma
\mu \nu }\partial _{\sigma }A_{\tau }\partial _{\nu }A_{\mu }P  \label{IV.1}
\end{eqnarray}
where $P=\pi ^0,\eta ,$ and $\eta ^{\prime }$ and $A_\mu $ is the photon
field. The matrix element for the decay $P\rightarrow $ $\gamma \gamma ^{*}$
with a virtual photon $\gamma ^{*}$ has the form 
\begin{eqnarray}
{\cal M}=-if_{P\gamma \gamma }F_{P\gamma \gamma }(M^{2})\epsilon _{\tau
\sigma \mu \nu }k_{\tau }\varepsilon _{\sigma }^{*}(k)k_{1\mu }\varepsilon
_{\nu }^{*}(k_{1})  \label{IV.2}
\end{eqnarray}
where $k$ is the virtual photon momentum ($k^2=M^2$), $k_1$ is the real
photon momentum ($k_1^2=0$), and $F_{P\gamma \gamma }(t)$ is transition form
factor $P\gamma \gamma ^{*}$. The comparison of the $P\rightarrow \gamma
\ell ^{+}\ell ^{-}$ decay width with the decay width of a physical process $%
P\rightarrow $ $\gamma \gamma $ allows to write \cite{LGL} 
\begin{eqnarray}
\frac{d\Gamma (P\rightarrow \gamma \ell ^{+}\ell ^{-})}{\Gamma (P\rightarrow
\gamma \gamma )}=2\left( \frac{p^{*}(\sqrt{s},0,M)}{p^{*}(\sqrt{s},0,0)}%
\right) ^{3}\left| F_{P\gamma \gamma }(M^{2})\right| ^{2}M\Gamma (\gamma
^{*}\rightarrow \ell ^{+}\ell ^{-})\frac{dM^{2}}{\pi M^{4}}  \label{IV.3}
\end{eqnarray}
where $\sqrt{s}=\mu _P$ is the pseudoscalar meson mass. The value $M\Gamma
(\gamma ^{*}\rightarrow \ell ^{+}\ell ^{-})$ is given by Eq.(\ref{II.6}).
The factor $2$ in Eq.(\ref{IV.3}) occurs due the identity of photons in the
decay $P\rightarrow \gamma \gamma $. The product of the qubic term and the
absolute square of the form factor gives the ratio between squares of the
matrix element (\ref{IV.2}) at $k^2=M^2$ and $k^2=0$, multiplied by the
ratio between the two-particle phase spaces. The quark counting rules \cite
{QCR} imply that the form factor $F_{P\gamma \gamma }(t)$ behaves as $\sim
1/t$ at $t\rightarrow \infty $. The experimental data are described
reasonably well by the monopole formula 
\begin{eqnarray}
F_{P\gamma \gamma }(t)=\frac{\Lambda _{P}^{2}}{\Lambda _{P}^{2}-t}
\label{IV.4}
\end{eqnarray}
with $\Lambda _P=0.75\pm 0.03,$ $0.77\pm 0.04,$ and $0.81\pm 0.04$,
respectively, for the $\pi ^0$-$,$ $\eta $-$,$ and $\eta ^{\prime }$-mesons 
\cite{HJB}, which reproduces the correct asymptotics. Such a monopole fit
can naturally be interpreted in terms of the vector meson dominance. The
values of the $\Lambda _P$'s are close to the $\rho $- and $\omega $-meson
masses. In a QCD interpolation formula, Ref. \cite{BL}, the pole masses are
related to the PCAC coupling constants of the pseudoscalar mesons, $\Lambda
_P=2\pi f_P$. This expression gives for the pole masses similar numbers.

In the $\eta ^{\prime }\rightarrow \gamma e^{+}e^{-}$ and $\eta ^{\prime
}\rightarrow \gamma \mu ^{+}\mu ^{-}$ decays, the $\rho ^{0}$ and $\omega $
poles occur in the physical region allowed for the spectrum of the dilepton
pairs. The $SU(3)$ symmetry with the $\eta -\eta ^{\prime }$ mixing angle $%
\theta =\arcsin (-\frac{1}{3})=-19.5^{\circ }$ which is quite close to the
experimental value $\theta ^{\exp }=-15.5\pm 1.3$ (see {\it e.g.} \cite{BES}%
) predicts $f_{\rho \gamma \eta ^{\prime }}=3f_{\omega \gamma \eta ^{\prime
}}=\frac{1}{\sqrt{3}}f_{\omega \gamma \pi }$. The experimental branching
ratios $B^{\exp }(\eta ^{\prime }\rightarrow \gamma \rho ^{0})=(30.2\pm
1.3)\%$ and $B^{\exp }(\eta ^{\prime }\rightarrow \gamma \omega )=(3.02\pm
0.30)\%$ indicate that the ratio between the $\rho ^{0}$- and $\omega $%
-meson couplings with the $\eta ^{\prime }$-meson equals $f_{\rho \gamma
\eta ^{\prime }}^{2}/f_{\omega \gamma \eta ^{\prime }}^{2}=10$, in good
agreement with the $SU(3)$ predictions. In the timelike resonance region,
the transition form factor (\ref{IV.4}) should be modified: 
\begin{eqnarray}
F_{\eta ^{\prime }\gamma \gamma }(t)=c_{\rho }\frac{m_{\rho }^{2}}{m_{\rho
}^{2}-t-im_{\rho }\Gamma _{\rho }}+c_{\omega }\frac{m_{\omega }^{2}}{%
m_{\omega }^{2}-t-im_{\omega }\Gamma _{\omega }}+c_{X}\frac{m_{X}^{2}}{%
m_{X}^{2}-t}  \label{IV.5}
\end{eqnarray}
(recall that $\Lambda _{\eta ^{\prime }}\approx m_{\rho }\approx m_{\omega }$%
), with the weight coefficients $c_{\rho } \sim f_{\rho \gamma \eta ^{\prime
}}/g_{\rho }=f_{\rho \rho \eta ^{\prime }}/g_{\rho }^{2}$ and $c_{\omega }
\sim f_{\omega \gamma \eta ^{\prime }}/g_{\omega }=f_{\omega \omega \eta
^{\prime }}/g_{\omega }^{2}.$ The $SU(3)$ symmetry predicts $f_{\rho \rho
\eta ^{\prime }}=f_{\omega \omega \eta ^{\prime }}=\frac{1}{2\sqrt{3}}%
f_{\rho \omega \pi }$. The normalization condition looks like $c_{\rho }+$ $%
c_{\omega }=1$. The slope is close to $1/\Lambda _{\eta ^{\prime }}^{2}$.
From the relation $g_{\rho }/g_{\omega }\approx 0\allowbreak .\,3$ one gets $%
c_{\rho }/c_{\omega }\approx 10.5$, the relative sign of the values $c_{\rho
}\ $and $c_{\omega }$ is fixed by $SU(3) $ symmetry. The third term in Eq.(%
\ref{IV.5}) is introduced for reasons explained below. We assume for the
moment $c_{X}=0.$ With $c_{\rho }=0.9$ and $c_{\omega }=0.1,$ we obtain $%
B(\eta ^{\prime }\rightarrow \gamma \mu ^{+}\mu ^{-})=0.90\times 10^{-4}$,
in good agreement with the measured value $B^{\exp }(\eta ^{\prime
}\rightarrow \gamma \mu ^{+}\mu ^{-})=(1.04\pm 0.26)\times 10^{-4}$. Data
for the $\eta ^{\prime }\rightarrow e^{+}e^{-}\gamma $ decay are not
available. The coupling constants we used in the calculations are summarized
in Table 1.

The product of the branching ratios $B^{\exp }(\eta ^{\prime }\rightarrow
\gamma \rho ^0)B^{\exp }(\rho ^0\rightarrow \mu ^{+}\mu ^{-})=1.4\times
10^{-5}\ $is almost one order of magnitude smaller than the value $B^{\exp
}(\eta ^{\prime }\rightarrow \gamma \mu ^{+}\mu ^{-})$. The magnitude of the
direct $\rho ^0$-meson contribution to the $\eta ^{\prime }\rightarrow
\gamma \mu ^{+}\mu ^{-}$ decay can be estimated from Eq.(\ref{IV.3}) with
the use of a narrow-width approximation for the transition form factor, 
\begin{eqnarray}
\left| F_{\eta ^{\prime }\gamma \gamma }(t)\right| ^{2}\approx \left|
c_{\rho }\right| ^{2}\frac{\pi m_{\rho }^{3}}{\Gamma _{\rho }}\delta
(t-m_{\rho }^{2}),  \label{IV.6}
\end{eqnarray}
to give $B(\eta ^{\prime }\rightarrow \gamma \rho ^0\rightarrow \gamma \mu
^{+}\mu ^{-})\approx 2\times 10^{-5}$. This value is already close to the
product $B^{\exp }(\eta ^{\prime }\rightarrow \gamma \rho ^0)B^{\exp }(\rho
^0\rightarrow \mu ^{+}\mu ^{-})$, but still higher. The relative
contributions of the $\rho ^0$- and $\omega $-mesons to the $\eta ^{\prime
}\rightarrow \gamma e^{+}e^{-}$ and $\eta ^{\prime }\rightarrow \gamma \mu
^{+}\mu ^{-}$ decay rates are inversely proportional to the vector meson
widths, as it follows from Eq.(\ref{IV.6}). Since the $\omega $-meson width
is only $8.5$ MeV, its contribution is strongly enhanced. The direct
contribution of the $\omega $-meson to the transition $\eta ^{\prime
}\rightarrow \gamma \mu ^{+}\mu ^{-}$ equals $B^{\exp }(\eta ^{\prime
}\rightarrow \gamma \omega )B(\omega \rightarrow \mu ^{+}\mu ^{-})\approx
\allowbreak 2\times 10^{-6}\ $where $B(\omega \rightarrow \mu ^{+}\mu
^{-})\approx B^{\exp }(\omega \rightarrow e^{+}e^{-})=(7.15\pm 0.19)\times
10^{-5}$ (the quoted experimental values are all from Ref. \cite{PDG}). The
use of the narrow-width approximation gives $B(\eta ^{\prime }\rightarrow
\gamma \omega \rightarrow \mu ^{+}\mu ^{-}\gamma )\approx \Gamma _\rho
/\Gamma _\omega \left| c_\omega /c_\rho \right| ^2B(\eta ^{\prime
}\rightarrow \gamma \rho ^0\rightarrow \gamma \mu ^{+}\mu ^{-})\approx
4\times 10^{-6}$. This value is also greater than the product $B^{\exp
}(\eta ^{\prime }\rightarrow \gamma \omega )B(\omega \rightarrow \mu ^{+}\mu
^{-})$. It can be interpreted as a noticeable contribution from the excited
vector mesons to the form factor $F_{\eta ^{\prime }\gamma \gamma }(t).$

In any case, the above estimates demonstrate that the direct $\rho ^{0}$-
and $\omega $-meson contributions to the $\eta ^{\prime }\rightarrow \gamma
\mu ^{+}\mu ^{-}$ decay are small. Thus, the value $B(\eta ^{\prime
}\rightarrow \gamma \mu ^{+}\mu ^{-})$ is determined mainly by the
background, i.e. by those values of $M^{2}$ which are not near to the $\rho
^{0}$- and $\omega $-meson poles. Since we are interested in the spectrum of
the dilepton pairs, it is important, however, to fix the relative weights of
the vector meson contributions. A $10-20\%$ decrease of the residues $%
c_{\rho }\ $and $c_{\omega }$ due to the admixture of an excited vector
meson $X$ yields for the measured decays $\eta ^{\prime }\rightarrow \gamma
\mu ^{+}\mu ^{-}$, $\eta ^{\prime }\rightarrow \gamma \rho ^{0},$ and $\eta
^{\prime }\rightarrow \gamma \omega $ a consistent interpretation. With $%
c_{\rho }=0.8,$ $c_{\omega }=0.08,$ and $c_{X}=0.12$ we get $B(\eta ^{\prime
}\rightarrow \gamma \rho ^{0}\rightarrow \gamma \mu ^{+}\mu ^{-})\approx
1.5\times 10^{-5}$ and $B(\eta ^{\prime }\rightarrow \gamma \omega
\rightarrow \gamma \mu ^{+}\mu ^{-})\approx 2\times 10^{-6}$, in good
agreement with the products of the branching ratios $B^{\exp }(\eta ^{\prime
}\rightarrow \gamma \rho ^{0})B^{\exp }(\rho ^{0}\rightarrow \mu ^{+}\mu
^{-})$ and $B^{\exp }(\eta ^{\prime }\rightarrow \gamma \omega )B(\omega
\rightarrow \mu ^{+}\mu ^{-}).$ The mass of the radially excited vector
meson $X$ is not well fixed (see discussion in Sect. 5). We set $m_{X}=1.2$
GeV. Since it is out of the physical region, the width $\Gamma _{X}$ is set
equal to zero.


\subsection{Decay modes $f_{0}(980)\rightarrow $ $\gamma \ell ^{+}\ell ^{-}$
and $a_{0}^{0}(980)\rightarrow $ $\gamma \ell ^{+}\ell ^{-}$}


The isoscalar $f_{0}(980)$-meson and the isotriplet $a_{0}(980)$-meson have
quantum numbers $I^{G}(J^{PC})=0^{+}(0^{++})$ and $1^{-}(0^{++}).$ The $%
f_{0} $- and the neutral $a_{0}$-mesons decay to two photons. The effective
vertex for the $S\rightarrow $ $\gamma \gamma $ decay has the form 
\begin{eqnarray}
\delta {\cal L}_{S\gamma \gamma }=f_{S\gamma \gamma }F_{\tau \mu }F_{\tau
\mu }S  \label{IV.7}
\end{eqnarray}
where $F_{\tau \mu }=\partial _{\mu }A_{\tau }-\partial _{\tau }A_{\mu }$.
The matrix element for the process $S\rightarrow $ $\gamma \gamma ^{*}$ is
given by 
\begin{eqnarray}
{\cal M}=-if_{S\gamma \gamma }F_{S\gamma \gamma }(M^{2})(g_{\tau \sigma
}k_{1\lambda }-g_{\tau \lambda }k_{1\sigma })(g_{\mu \sigma }k_{\lambda
}-g_{\mu \lambda }k_{\sigma })\varepsilon _{\tau }^{*}(k_{1})\varepsilon
_{\mu }^{*}(k).  \label{IV.8}
\end{eqnarray}
Here, $k_{1}$ and $k$ are real ($k_{1}^{2}=0$) and virtual ($k^{2}=M^{2}$)
photon momenta, $F_{S\gamma \gamma }(t)$ is transition form factor of the
decay $S\rightarrow \gamma \gamma ^{*}$. The square of the matrix element
summed up over the photon polarizations equals $8sp^{*2}(\sqrt{s},0,M)$,
with $\sqrt{s}=m_{S}$ being the scalar meson mass. The width of the $%
S\rightarrow \gamma e^{+}e^{-}$ decay can be written as follows 
\begin{eqnarray}
\frac{d\Gamma (S\rightarrow \gamma \ell ^{+}\ell ^{-})}{\Gamma (S\rightarrow
\gamma \gamma )}=2\frac{p^{*3}(\sqrt{s},0,M)}{p^{*3}(\sqrt{s},0,0)}\left|
F_{S\gamma \gamma }(M^{2})\right| ^{2}M\Gamma (\gamma ^{*}\rightarrow \ell
^{+}\ell ^{-})\frac{dM^{2}}{\pi M^{4}}.  \label{IV.9}
\end{eqnarray}

The value of $\Gamma (f_0 \rightarrow\gamma \gamma )$ is given by PDG \cite
{PDG}. In case of the $a_0$-meson, PDG gives the quantity $\Gamma (a_0
\rightarrow\gamma \gamma ) \Gamma (a_0 \rightarrow\pi^0 \eta )/\Gamma_{tot}
(a_0)$. Since the mode $a_0 \rightarrow\pi^0 \eta$ is the dominant one, the
two-photon width can also be estimated.

The transition form factor $F_{S\gamma \gamma }(t)$ depends on the nature of
the scalar meson $S$. The asymptotics of the form factor according to the
quark counting rules \cite{QCR} is $\sim 1/t$ for a $2$-quark model \cite
{NAT} and $\sim 1/t^{2}$ for a 4-quark MIT bag \cite{RLJ} and a $K\bar{K}$
molecular models of the $f_{0}$- and $a_{0}$-mesons \cite{CLO}. As pointed
out by N. Achasov and Ivanchenko \cite{AIva}, the decay $\phi \rightarrow
\pi ^{0}\pi ^{0}\gamma $ can provide an important information on the
structure of the $f_{0}$-meson. In the SND experiment at the VEPP-2M $%
e^{+}e^{-}$ collider the branching ratio $\phi \rightarrow \pi ^{0}\pi
^{0}\gamma $ was measured \cite{MNA,VEPP}. The decay goes mainly through the 
$\phi \rightarrow \gamma f_{0}$ decay mode. The results are consistent with
the hypothesis of a 4-quark MIT bag nature of the $f_{0}$-meson. We thus
calculate dilepton branching ratios $S\rightarrow \gamma \ell ^{+}\ell ^{-}$
assuming the 4-quark nature of the scalar mesons.

In the framework of the VMD model, the $S\gamma \gamma ^{*}$ transition form
factor can be reproduced assuming contributions from ground-state and
excited vector mesons with masses $m_{i}$%
\begin{eqnarray}
F_{S\gamma \gamma }(t)=\sum_{i}\frac{c_{i}m_{i}^{2}}{m_{i}^{2}-t}.
\label{IV.10}
\end{eqnarray}
The normalization $F_{S\gamma \gamma }(0)=1$ and the asymptotic condition $%
F_{S\gamma \gamma }(t)\sim 1/t^{2}$ at $t\rightarrow \infty $ give
constraints to the residues $c_{i}$: 
\begin{eqnarray}
1 &=&\sum_{i}c_{i},  \label{IV.11} \\
0 &=&\sum_{i}c_{i}m_{i}^{2}.  \label{IV.12}
\end{eqnarray}

In case of the form factor $f_{0}\gamma \gamma ^{*}$, the OZI rule implies
that $\omega \omega $, $\rho \rho $ and $\phi \phi $ contributions are
small, while the contributions $\omega \rho $, $\phi \rho $ are forbidden by
isospin conservation. The relative weights of the $\omega $- and $\phi $
-mesons in the form factor are fixed and thus at least three vector mesons
should be considered to fit the asymptotic behavior: 
\begin{eqnarray}
F_{f_{0}\gamma \gamma }(t)\sim \frac{m_{\omega }^{2}}{m_{\omega }^{2}-t}+%
\frac{m_{\phi }^{2}}{m_{\phi }^{2}-t}+c_{X}\frac{m_{X}^{2}}{m_{X}^{2}-t}.
\label{IV.13}
\end{eqnarray}
The overall normalization factor can be derived from the condition $%
F_{f_{0}\gamma \gamma }(0)=1$, the value $c_{X}$ is fixed from the
requirement $F_{f_{0}\gamma \gamma }(t)\sim 1/t^{2}$ as $t\rightarrow \infty 
$. The third meson mass $m_{X}$ is not well determined. The resulting form
factor is given by 
\begin{eqnarray}
F_{f_{0}\gamma \gamma }(t)=\frac{m_{\omega }^{2}m_{\phi }^{2}m_{X}^{2}(1+Ct)%
}{(m_{\omega }^{2}-t)(m_{\phi }^{2}-t)(m_{X}^{2}-t)}  \label{IV.14}
\end{eqnarray}
where 
\begin{eqnarray}
C=-\frac{m_{\omega }^{2}(m_{X}^{2}-m_{\omega }^{2})+m_{\phi
}^{2}(m_{X}^{2}-m_{\phi }^{2})}{m_{\omega }^{2}m_{\phi
}^{2}(2m_{X}^{2}-m_{\omega }^{2}-m_{\phi }^{2})}.  \label{IV.15}
\end{eqnarray}
The transition form factor $a_{0}^{0}\gamma \gamma ^{*}$ has the same
structure with the replacement $\omega \leftrightarrow \rho .$

The virtual photon from the scalar meson decay lies in the physical region
of the $\omega $- and $\rho $-mesons, so the $\omega $- and $\rho $-meson
propagators should be modified by introducing the finite meson widths. We
use $m_{X}=1.2$ GeV. The pole $t=m_{X}^{2}$ is outside of the physical
region, so we set $\Gamma _{X}=0$.


\section{Meson decays to one meson and a $\ell ^{+}\ell ^{-}$ pair}

\setcounter{equation}{0}

The radiative decays $V\rightarrow P\gamma $ and $P\rightarrow V\gamma $ are
well studied experimentally. The dilepton modes are measured only for decays 
$\omega \rightarrow \pi ^{0}\ell ^{+}\ell ^{-}$ and $\phi \rightarrow \eta
e^{+}e^{-}$. We use data on the radiative decays to fix coupling constants
entering the $VP\gamma $ vertexes. These values are in good agreement with
the $SU(3)$ symmetry relations. We use the framework of the extended VMD
model and impose constraints to the residues of the transition form factors
from the quark counting rules. The calculation of the dilepton spectra is
then a straightforward task.

\subsection{Decay modes $\omega \rightarrow \pi ^{0}e^{+}e^{-}$, $\rho
\rightarrow \pi e^{+}e^{-},$ and $\phi \rightarrow \pi ^{0}e^{+}e^{-}$}


The effective vertex for the $V\rightarrow P\gamma $ radiative decay with $%
V=\omega $, $\rho ^{0}$, and $\phi $ and $P=\pi ^{0}$, $\eta $, and $\eta
^{\prime }$ has the form 
\begin{eqnarray}
{\cal \delta L}_{V\gamma P}=-ef_{V\gamma P}\epsilon _{\tau \sigma \mu \nu
}\partial _{\sigma }V_{\tau }\partial _{\nu }A_{\mu }P.  \label{V.1}
\end{eqnarray}

The matrix element for the $V\rightarrow P\gamma ^{*}$ transition can be
written as 
\begin{eqnarray}
{\cal M}=-ief_{V\gamma P}F_{V\gamma P}(M^{2})\epsilon _{\tau \sigma \mu \nu
}\epsilon _{\tau }(P)P_{\sigma }\varepsilon _{\mu }^{*}(k)k_{\nu }.
\label{V.2}
\end{eqnarray}
The diagram for this transition is shown in Fig. \ref{fig4}. The vertex form
factor $F_{V\gamma P}(t)$ normalized at $t=0$ to unity depends on the square
of the photon four-momentum, $t=M^{2}$. The width of the decay $V\rightarrow 
$ $P\gamma ^{*}$ has the form (see. {\it e.g.} \cite{PJain,Ulf}) 
\begin{eqnarray}
\Gamma (V\rightarrow P\gamma ^{*})=\frac{\alpha }{3}f_{V\gamma P}^{2}\left|
F_{V\gamma P}(k^{2})\right| ^{2}p^{*3}(\sqrt{s},\mu _{P},M)  \label{V.3}
\end{eqnarray}
where $\sqrt{s}=m_{V}$ is the vector meson mass and $\mu _{P}$ is the
pseudoscalar meson mass. The limit $M^{2}=0$ describes the radiative decay $%
V\rightarrow $ $P\gamma $. Using the experimental value of the $\omega
\rightarrow \pi ^{0}\gamma $ width, one gets $f_{\omega \gamma \pi }=2.3\;$%
GeV$^{-1}$. The decay width $V\rightarrow P\ell ^{+}\ell ^{-}$ is connected
to the radiative width $V\rightarrow $ $P\gamma $ \cite{CHL}: 
\begin{eqnarray}
\frac{d\Gamma (V\rightarrow P\ell ^{+}\ell ^{-})}{\Gamma (V\rightarrow
P\gamma )}=\left( \frac{p^{*}(\sqrt{s},\mu _{P},M)}{p^{*}(\sqrt{s},\mu
_{P},0)}\right) ^{3}\left| F_{V\gamma P}(M^{2})\right| ^{2}M\Gamma (\gamma
^{*}\rightarrow \ell ^{+}\ell ^{-})\frac{dM^{2}}{\pi M^{4}}  \label{V.4}
\end{eqnarray}
where $M\Gamma (\gamma ^{*}\rightarrow \ell ^{+}\ell ^{-})$ is given by Eq.(%
\ref{II.6}). The first factor is the ratio between squares of the matrix
element (\ref{V.2}) at $k^{2}=M^{2}$ and $k^{2}=0$, multiplied by the ratio
of the two-particle phase spaces $\Phi _{2}(\sqrt{s},\mu _{P},M)$ and $\Phi
_{2}(\sqrt{s},\mu _{P},0)$.

The vertex form factor $F_{V\gamma P}(t)$ falls off asymptotically as $%
1/t^{2}$ \cite{VZ}. The additional power $1/t$ as compared to the
asymptotics of the pion form factor occurs because of suppression of the
quark spin-flip amplitude due to conservation of the quark helicity in
interactions of quarks with gluons.

The $\omega \gamma \pi $ transition form factor can be reproduced taking the
contribution of the ground-state $\rho $-meson and at least one excited $%
\rho $-meson into account. The extended VMD model with $m_{\rho }=769$ MeV
and $m_{\rho ^{\prime }}=1450$ MeV provides the correct asymptotics and
yields a better description of the experimental data \cite{ARGUS,DOL} than
the one-pole model with only the ground-state $\rho $-meson. However, it
still underestimates the slope of the form factor at $t=0.$ The experimental
value $F_{\omega \gamma \pi }^{\prime }(0)=2.4\pm 0.2$ GeV$^{-2}$ \cite{LGL}
prefers a value $m_{\rho ^{\prime }}\approx 1.2$ GeV.

The possible existence of vector mesons $I=0,$ $1$ with masses around $1.2$
GeV was under discussion for a long time. A quite strong evidence for
significant contributions to the spectral functions of the nucleon form
factors at values $t$ lower than $1.45$ GeV comes from the fact that the
experimental data for the Sachs form factors are reasonably described by the
dipole formula 
\begin{eqnarray}
G_{Ep}(t)\approx G_{Mp}(t)/\mu _{p}\approx G_{Mn}(t)/\mu _{n}\approx \frac{1%
}{(1-t/0.71)^{2}}  \label{V.5}
\end{eqnarray}
where $t$ is in units GeV$^{2}.$ This formula has a double pole. The
spectral functions are proportional to 
\begin{eqnarray*}
\delta^{\prime}(t - 0.71) \approx \frac{ \delta(t - m^2_{\rho}) - \delta(t -
m^2_X)}{-m^2_{\rho} + m^2_X}.
\end{eqnarray*}
The double pole in Eq.(V.5) indicates an enhancement in the spectral
functions at $m_{X}$ close to $m_{\rho}$, whose nature is not yet clear as
long as the existence of the non-strange vector mesons at $1.2$ GeV is not
established. In any case, it is desirable to shift masses of the first
excited non-strange ''vector mesons'' to lower values. This has been done in
some popular fits of the nucleon form factors (see {\it e.g.} \cite{HPM}).
The more recent fits use, however, higher values of the vector meson masses 
\cite{NFF}. In our calculations, the value $m_{X}$ is assumed to be $1.2$
GeV, as extracted form the slope of the $F_{\omega \gamma \pi }(t)$ at $t=0$.

The multiplicative representation of the transition form factor $\omega
\gamma \pi $, which in the zero-width limit is completely equivalent to the
additive representation, is given by 
\begin{eqnarray}
F_{\omega \gamma \pi }(t)=\frac{m_{\rho }^{2}m_{X}^{2}}{(m_{\rho
}^{2}-t)(m_{X}^{2}-t)}.  \label{V.6}
\end{eqnarray}

The experimental data on the transition form factors are available for the $%
\eta ^{\prime }\gamma \gamma ^{*}$, $\eta \gamma \gamma ^{*}$, and $\omega
\pi \gamma ^{*}$ transitions. The parametrization (IV.5) for the $\eta
^{\prime }$ form factor is in good agreement with the data from Ref. \cite
{LGL}, Fig. 26. For the $\eta $-meson form factor (IV.4), our curve
coincides with the VMD curve from Fig. 24, Ref. \cite{LGL}, since the
parameter $\Lambda $ = 0.77 GeV in Eq.(IV.4) is close to the $\rho $-meson
mass.

The one-pole VMD model is known to be in rather pure agreement with the data
on the $\omega \pi \gamma ^{*}$ transition form factor \cite{Dzh}. The
two-pole extended VMD model, Eq.(V.6), improves the agreement without
introducing new parameters. It fits well the experimental points at $t<0.2$
GeV$^{2}$, but underestimates the last three ones from Ref. \cite{Dzh} below 
$0.4$ GeV$^{2}$. By price of introducing a new vector meson, $\rho ^{\prime
\prime }$, and one fitting parameter, the last three points can be described
better. The modification consists in inserting an additional multiplier $%
m_{\rho ^{\prime \prime }}^{2}(1+Ct)/(m_{\rho ^{\prime \prime }}^{2}-t)$
into the right side of Eq.(V.6), with $C$ being a free parameter. This model
again satisfies the quark counting rules. However, it overestimates the form
factors values at $t\approx 0.2$ GeV$^{2}$. For the present calculations, we
stay in the framework of the simplest two-pole VMD model, Eq.(V.6). It is
clear, however, that if one needs to calculate the dilepton production for
the DLS and/or future HADES data, the sensitivity of the results on the $%
\omega \pi \gamma ^{*}$ form factor values must be investigated.

The $\phi $-meson decay width is given by Eq.(\ref{V.4}). In the $\phi $%
-meson decay, the emitted photon has isospin $I=1$, so the transition form
factor has the same form as for the $\omega $-meson. The $\ell ^{+}\ell ^{-}$
pairs appear in the physical region of the decay $\phi \rightarrow \pi
^{0}\rho ^{0}$. In order to include the $\rho $-meson direct contribution,
the form factor (\ref{V.6}) should be modified by introducing the finite $%
\rho $-meson width. In the cases of the $\omega \rightarrow \pi
^{0}e^{+}e^{-}$and $\rho \rightarrow \pi e^{+}e^{-}$ decays, vector mesons
appear in the unphysical regions and their widths are not so important.

The direct contribution to the $\phi \rightarrow \pi ^{0}\rho ^{0}$ decay
can be estimated with the use of the narrow-width approximation for the
transition form factor in analogy with the decays $\eta ^{\prime
}\rightarrow \gamma \rho ^{0}\rightarrow \gamma \mu ^{+}\mu ^{-}$ and $\eta
^{\prime }\rightarrow \gamma \omega \rightarrow \gamma \mu ^{+}\mu ^{-}$.
The form factor is proportional to square of the residue $c_{\rho }$ ({\it cf%
}. Eq.(\ref{IV.6})) which is fixed by Eq.(\ref{V.6}). The results of the
calculations $B(\phi \rightarrow \pi ^{0}\rho ^{0}\rightarrow \pi
^{0}e^{+}e^{-})=2.3\times 10^{-6}$ are in reasonable agreement with the
product $B^{\exp }(\phi \rightarrow \pi ^{0}\rho ^{0})B^{\exp }(\rho
^{0}\rightarrow e^{+}e^{-})=(1.9\pm 0.2)\times 10^{-6}$, and similarly for
the $\mu ^{+}\mu ^{-}$ decay mode.

In case of the $\rho ^{\pm }$- and $\rho ^{0}$-meson decays, the effective
vertex has the form 
\begin{eqnarray}
{\cal \delta L}_{\rho \gamma \pi } &=&-ef_{\rho \gamma \pi }\epsilon _{\tau
\sigma \mu \nu }\partial _{\sigma }\rho _{\tau }^{\alpha }\partial _{\nu
}A_{\mu }\pi ^{\alpha }  \nonumber \\
&=&-ef_{\rho \gamma \pi }\epsilon _{\tau \sigma \mu \nu }(\partial _{\sigma
}\rho _{\tau }^{0}\partial _{\nu }A_{\mu }\pi ^{0}+\partial _{\sigma }\rho
_{\tau }^{+}\partial _{\nu }A_{\mu }\pi ^{-}+\partial _{\sigma }\rho _{\tau
}^{-}\partial _{\nu }A_{\mu }\pi ^{+}).  \label{V.7}
\end{eqnarray}
The vertex (\ref{V.1}) for $V=\rho ^{0}$ and $P=\pi ^{0}$ is a part of the
vertex (\ref{V.7}). Eq.(\ref{V.4}) remains valid for $V=\rho ^{\pm }$, $\rho
^{0}$ and $P=\pi ^{\pm }$, $\pi ^{0}$. The photon emitted in the decay $\rho
\rightarrow \pi \gamma ^{*}$ is in the isoscalar state, respectively, the
transition form factor $F_{\rho \gamma \pi }(t)$ receives contributions from
the $\omega $- and $\phi $-mesons. The effect of the $\phi $-meson is small
due to the OZI rule. The small difference between masses of the $\rho $- and 
$\omega $-mesons is beyond the accuracy of the VMD model. Away from the $%
\rho $- and $\omega $-mesons poles $F_{\omega \gamma \pi }(t)\approx F_{\rho
\gamma \pi }(t)$, whereas $F_{\omega \gamma \pi }(t)=F_{\phi \gamma \pi }(t)$
everywhere.%

\subsection{Decay modes $\omega \rightarrow \eta \ell ^{+}\ell ^{-}$, $\rho
^{0}\rightarrow \eta \ell ^{+}\ell ^{-},$ and $\phi \rightarrow \eta \ell
^{+}\ell ^{-}$}


These decays are treated in the same way as the decays discussed above. The
effective vertex for the $V\rightarrow \eta \gamma $ transitions with $%
V=\omega $, $\rho ^{0}$, and $\phi $ has the form of Eq.(\ref{V.1}) where
one should put $P=\eta $. The decay widths are given by Eq.(\ref{V.4}) with $%
\mu _{P}$ being the $\eta $-meson mass. In the simplest version of the VMD
model, the transition form factor $F_{V\gamma \eta }(t)$ has the form of Eq.(%
\ref{V.6}). For the $\omega $- and $\phi $-meson decays, the form factor is
determined by the $\rho $-meson mass, while for the $\rho ^{0}$-meson decay
it is determined by the $\omega $-meson mass.

\subsection{Decay modes $\eta ^{\prime }\rightarrow \omega e^{+}e^{-}$ and $%
\eta ^{\prime }\rightarrow \rho ^{0}e^{+}e^{-}$}


These decays are of the same nature. The effective vertex for the $\eta
^{\prime }\rightarrow V\gamma ^{*}$ transitions with $V=\omega $ and $\rho
^{0}$ has the form of Eq. (\ref{V.1}). The spectrum of the $\ell ^{+}\ell
^{-}$ pair in $P\rightarrow V\ell ^{+}\ell ^{-}$ can be obtained from Eq. (%
\ref{V.4}) with the replacements $P\leftrightarrow V$ and $\mu
_{P}\leftrightarrow m_{V}.$ In our case, $V=\omega $ and $\rho ^{0}$, $%
P=\eta ^{\prime }$, the value $\sqrt{s}=\mu _{P}$ stands for the $\eta
^{\prime }$-mesons mass. The transition form factor $F_{\eta ^{\prime
}\gamma V}(t)\ $has the asymptotics $1/t^{2}$ for $t\rightarrow \infty $. It
is described within the VMD model by the vector meson contributions from
isovector and isoscalar channels, respectively.

\subsection{Decay mode $a_{1}(1260)\rightarrow $ $\pi \ell ^{+}\ell ^{-}$}


The axial vector meson $a_1(1260)$ has quantum numbers $%
I^G(J^{PC})=1^{-}(1^{++})$. Its mass is above the $\phi (1020)$-meson mass,
so we do not expect a noticeable effect from the production and the decay of
such a meson at GSI energies. This resonance becomes, however, important at
higher energies (see \cite{BC}, \cite{LiGale}).
 
The $a_{1}\rightarrow \pi \gamma ^{*}$ transition is an electric dipole $E1$
transition. The effective vertex has the form 
\begin{eqnarray}
\delta {\cal L}_{a_{1}\gamma \pi }=f_{a_{1}\gamma \pi }(\partial _{\nu
}a_{1\mu }^{\alpha }-\partial _{\mu }a_{1\nu }^{\alpha })F_{\mu \nu }\pi
^{\alpha }  \label{V.8}
\end{eqnarray}
where $F_{\mu \nu }$ is the electromagnetic tensor. The matrix element for
the process $a_{1}\rightarrow $ $\pi \gamma ^{*}$ can be written as follows 
\begin{eqnarray}
{\cal M}=-ief_{a_{1}\gamma \pi }F_{a_{1}\gamma \pi }(M^{2})\epsilon _{\tau
}(P)(g_{\tau \sigma }P_{\lambda }-g_{\tau \lambda }P_{\sigma })(g_{\mu
\sigma }k_{\lambda }-g_{\mu \lambda }k_{\sigma })\varepsilon _{\mu }^{*}(k)
\label{V.9}
\end{eqnarray}
where $P$ is the momentum of the decaying meson, $k$ is the photon momentum, 
$F_{a_{1}\gamma \pi }(t)$ is the transition form factor, $F_{a_{1}\gamma \pi
}(0)=1$. The square of the matrix element summed up over photon
polarizations and averaged over polarizations of the $a_{1}$-meson is
proportional to $p^{*2}(\sqrt{s},\mu ,M)+\frac{3}{2}M^{2}$ where $\sqrt{s}%
=m_{a_{1}}$. The decay width $a_{1}\rightarrow \pi \ell ^{+}\ell ^{-}$ takes
the form 
\begin{eqnarray}
\frac{d\Gamma (a_{1}\rightarrow \pi \ell ^{+}\ell ^{-})}{\Gamma
(a_{1}\rightarrow \pi \gamma )}=\frac{p^{*2}(\sqrt{s},\mu ,M)+\frac{3}{2}%
M^{2}}{p^{*2}(\sqrt{s},\mu ,0)}\frac{p^{*}(\sqrt{s},\mu ,M)}{p^{*}(\sqrt{s}%
,\mu ,0)}  \nonumber \\
\times \left| F_{a_{1}\gamma \pi }(M^{2})\right| ^{2}M\Gamma (\gamma
^{*}\rightarrow \ell ^{+}\ell ^{-})\frac{dM^{2}}{\pi M^{4}}.  \label{V.10}
\end{eqnarray}
This expression differs from the analogous expression from Ref. \cite{BC},
Eq. (6.18), by the extra term $\frac{3}{2}M^{2}$ in the ratio for squares of 
the matrix elements. Such a term is needed to ensure the correct threshold 
behavior of the $s$-wave $a_{1}\rightarrow \pi \gamma ^{*}$ decay. 


\section{Meson decays to two mesons and $\ell ^{+}\ell ^{-}$ pair}

\setcounter{equation}{0} 

The four-body meson decays proceed either through a two-step mechanism
similar to that discussed a long time ago by Gell-Mann, Sharp and Wagner 
\cite{GMSW} in connection to the $\omega \rightarrow \pi ^{+}\pi ^{-}\pi
^{0} $ decay or through a bremsstrahlung mechanism when a virtual photon is
emitted form an external meson line. In the first case, the matrix element
is a smooth function of the photon mass. At small $M\;$such that $%
2m_{e}\lesssim M$, the differential widths behave like $1/M$, according to
Eq.(\ref{II.10}). In the second case, the matrix element is singular at
small $M$, as a result of which the differential width increases faster than 
$1/M$.

In this Sect., we derive analytical expressions for the double differential
widths $d^{2}\Gamma /ds_{12}dM$ of the $\eta \rightarrow $ $\pi ^{+}\pi
^{-}\ell ^{+}\ell ^{-},$ $\eta ^{\prime }\rightarrow $ $\pi ^{+}\pi ^{-}\ell
^{+}\ell ^{-},\rho ^{0}\rightarrow $ $\pi ^{+}\pi ^{-}\ell ^{+}\ell ^{-},$ $%
\rho ^{\pm }\rightarrow $ $\pi ^{\pm }\pi ^{0}\ell ^{+}\ell ^{-},$ $%
f_{0}\rightarrow $ $\pi ^{+}\pi ^{-}\ell ^{+}\ell ^{-},$ and $a_{0}^{\pm
}\rightarrow $ $\pi ^{\pm }\eta \ell ^{+}\ell ^{-}\;$decays, with $s_{12}$
being the invariant mass of two outgoing mesons. The widths of the other
decays are calculated numerically.

The experimental data are available for the $\eta \rightarrow $ $\pi ^{+}\pi
^{-}e^{+}e^{-}$ decay only, with large errors \cite{RAG66}. Nevertheless,
the dilepton widths can be predicted using the conventional assumptions on a
two-step decay mechanism and/or a bremsstrahlung decay mechanism.

\subsection{Decay modes $\eta \rightarrow $ $\pi ^{+}\pi ^{-}\ell ^{+}\ell
^{-}$ and $\eta ^{\prime }\rightarrow $ $\pi ^{+}\pi ^{-}\ell ^{+}\ell ^{-}$}


The diagram contributing to the decay $\eta \rightarrow $ $\pi ^{+}\pi
^{-}\gamma ^{*}$ is shown in Fig. \ref{fig5}. The $\rho ^{0}\gamma \eta $
vertex is given by Eq.(\ref{V.1}). The vertex for the $\rho \rightarrow \pi
\pi $ decay has the form 
\begin{eqnarray}
{\cal \delta L}_{\rho \pi \pi } &=& \frac{1}{2}f_{\rho \pi \pi }\epsilon
_{\alpha \beta \gamma }\rho _{\mu }^{\alpha }(\pi ^{\beta }\stackrel{%
\leftrightarrow }{\partial }_{\mu }\pi ^{\gamma })  \nonumber \\
&=& f_{\rho \pi \pi }(\rho _{\mu }^{0}\pi ^{+}i\stackrel{\leftrightarrow }{%
\partial }_{\mu }\pi ^{-}+\rho _{\mu }^{+}\pi ^{-}i\stackrel{\leftrightarrow 
}{\partial }_{\mu }\pi ^{0}+\rho _{\mu }^{-}\pi ^{0}i\stackrel{%
\leftrightarrow }{\partial }_{\mu }\pi ^{+})  \label{VI.1}
\end{eqnarray}
where $\alpha ,$ $\beta ,$ and $\gamma $ are isotopic indices of the $\rho $%
- and $\pi $-mesons. The coupling constant $f_{\rho \pi \pi }=6.0$ is
determined from the $\rho $-meson width 
\begin{eqnarray}
\Gamma (\rho ^{0}\rightarrow \pi ^{+}\pi ^{-})=\frac{1}{6\pi s}f_{\rho \pi
\pi }^{2}p^{*3}(\sqrt{s},\mu ,\mu )  \label{VI.2}
\end{eqnarray}
where $\sqrt{s}=m_{\rho }$ is the $\rho $-meson mass. The value $f_{\rho \pi
\pi }$ is in good agreement with the relation $f_{\rho \pi \pi }/g_{\rho }=1$
which follows from the $\rho ^{0}$-meson dominance in the pion form factor.
In the FFGS model \cite{FF} the $t$-dependence of the pion form factor $%
F_{\pi }(t)$ is attributed to the $\rho $-meson propagator. It means that
the constant $f_{\rho \pi \pi }$ does not depend on the value $t$ when the
two pions are on the mass shell.

The radiative decay $\eta \rightarrow $ $\pi ^{+}\pi ^{-}\gamma $ has been
measured in Ref. \cite{JGL}. The $\eta ^{\prime }\rightarrow $ $\pi ^{+}\pi
^{-}\gamma $ decay is studied experimentally much better (see \cite{HA87}-%
\cite{SIB} and references therein). The last decay is more complicated,
since two pions occur in the physical region of the $\rho $-meson. It was
shown that an attempt to fit the experimental data within the framework of
the cascade model $\eta ^{\prime }\rightarrow $ $\rho ^0\gamma \rightarrow $ 
$\pi ^{+}\pi ^{-}\gamma $ results in overestimating the events with a $\pi
\pi $ invariant mass below the $\rho $-meson mass. In Ref. \cite{SIB}, a
nonresonant contribution is added to the amplitude and its parameters are
fixed by fitting the experimental data. One of the modifications considered
for the $\rho $-meson propagator is the following one 
\begin{eqnarray}
(s_{12}-m_{\rho }^{2}+im_{\rho }\Gamma _{\rho })^{-1}\rightarrow
(s_{12}-m_{\rho }^{2}+im_{\rho }\Gamma _{\rho })^{-1}+C_{\rho }  \label{VI.3}
\end{eqnarray}
with a complex value of $C_{\rho }$.

It was proposed that the existence of a non-resonant term is connected to
the chiral box anomaly \cite{CPIC}. The anomaly results to a modification of
the $\rho $-meson propagator similar to that of Eq.(\ref{VI.3}) with $C_\rho
=1/(3m_\rho ^2)$. The positive sign of the term added guarantees a reduction
of the number of events on the left hand side of the $\rho $-meson peak. The
contact term describes a production of two mesons in the relative $p$-wave
where a strong resonant pion-pion interaction due to existence of the $\rho $%
-meson is present. The final state interaction can be taken into account by
dividing the bare vertex by the Jost function determined by the $\pi $-meson 
$p$-wave phase shift. In this way, the $\rho $-meson propagator reoccurs,
and so we get from the contact term the same old $\rho $-meson pole term of
the earlier models. The net effect of the box anomaly is a redefinition of
the coupling constant for those models which include only the single diagram
of Fig. \ref{fig5}.

We keep the constant term $C_\rho $ in the form suggested by Ref. \cite{CPIC}
and interpret it purely phenomenologically as a modification of the $\rho $%
-meson propagator needed to describe the distribution of the two-pion events
in the $\eta ^{\prime }\rightarrow $ $\pi ^{+}\pi ^{-}\gamma $ decay. The
contact term $C_\rho $ for the $\eta $-meson is assumed to be of the same
magnitude as for the $\eta ^{\prime }$-meson.

The matrix element of the $\eta \rightarrow $ $\pi ^{+}\pi ^{-}\gamma ^{*}$
decay has the form 
\begin{eqnarray}
{\cal M}=-ief_{\rho \pi \pi }f_{\rho \gamma \eta }{\cal M}_{\mu }\varepsilon
_{\mu }^{*}(k)  \label{VI.4}
\end{eqnarray}
where 
\begin{eqnarray}
{\cal M}_{\mu }=\epsilon _{\tau \sigma \mu \nu }(p_{1}-p_{2})_{\tau
}(p_{1}+p_{2})_{\sigma }k_{\nu }(\frac{1}{s_{12}-m_{\rho }^{2}+im_{\rho
}\Gamma _{\rho }}+C_{\rho }).  \label{VI.5}
\end{eqnarray}
Here, $p_{1}$ and $p_{2}$ are the pion momenta, $s_{12}=(p_{1}+p_{2})^{2}$, $%
k$ is the photon momentum, and $m_{\rho }$ and $\Gamma _{\rho }$ are the $%
\rho $-meson mass and width. Energy-momentum conservation implies $%
P=p_{1}+p_{2}+k$. After summation of the squared matrix element over the
photon polarizations and averaging over the direction of the two pion
momenta their in c. m., we get 
\begin{eqnarray}
{\cal R} &\equiv &\int \frac{d\Omega _{12}}{4\pi }{\cal M}_{\mu }{\cal M}%
_{\nu }{}^{*}(-g_{\mu \nu })  \nonumber \\
&=&\frac{8}{3}sp^{*2}(\sqrt{s},\sqrt{s_{12}},M)p^{*2}(\sqrt{s_{12}},\mu ,\mu
)\left| \frac{1}{s_{12}-m_{\rho }^{2}+im_{\rho }\Gamma _{\rho }}+C_{\rho
}\right| ^{2}  \label{VI.6}
\end{eqnarray}
with $d\Omega _{12}=2\pi \sin \theta d\theta $, and $\theta $ being the
angle between the momenta ${\bf p}_{1}$ and ${\bf k}$ in the c. m. frame of
two pions.

The decay width $\eta \rightarrow $ $\pi ^{+}\pi ^{-}\gamma ^{*}$ becomes 
\begin{eqnarray}
\Gamma (\eta \rightarrow \pi ^{+}\pi ^{-}\gamma ^{*}) &=& \frac{\alpha }{%
16\pi ^{2}s}f_{\rho \pi \pi }^{2}f_{\rho \gamma \eta }^{2} \left| F_{\rho
\gamma \eta }(M^{2})\right| ^{2}  \nonumber \\
&\times& \int_{4\mu ^{2}}^{(\sqrt{s}-M)^{2}}{\cal R\;}\frac{p^{*}(\sqrt{s},%
\sqrt{s_{12}},M)p^{*}(\sqrt{s_{12}},\mu ,\mu )}{\sqrt{s_{12}}}ds_{12}
\label{VI.7}
\end{eqnarray}
We neglect thereby for the $\rho $-meson off-shellness in the form factor $%
F_{\rho \gamma \eta }(t)$. The value $f_{\rho \gamma \eta }=1.86$ GeV$^{-1}$
($=\sqrt{\frac{2}{3}}f_{\omega \gamma \pi }$ due to $SU(3)$ symmetry)\ is
extracted from the $\rho ^{0}\rightarrow \eta \gamma $ width with the use of
Eq.(\ref{V.3}).

The dilepton spectrum in the decay $\eta \rightarrow \pi ^{+}\pi
^{-}e^{+}e^{-}$ is described by expression 
\begin{eqnarray}
d\Gamma (\eta \rightarrow \pi ^{+}\pi ^{-}\ell ^{+}\ell ^{-})=\Gamma (\eta
\rightarrow \pi ^{+}\pi ^{-}\gamma ^{*})M\Gamma (\gamma ^{*}\rightarrow \ell
^{+}\ell ^{-})\frac{dM^{2}}{\pi M^{4}}.  \label{VI.8}
\end{eqnarray}

The model for the decay $\eta ^{\prime }\rightarrow $ $\pi ^{+}\pi
^{-}\gamma ^{*}$ is a direct extension of the model for the $\eta $-meson
decay. The coupling constant $f_{\rho \gamma \eta ^{\prime }}=1.3$ GeV$^{-1}$
($=\frac{1}{\sqrt{3}}f_{\omega \gamma \pi }$ due to $SU(3)$ symmetry)\ can
be extracted from the radiative $\eta ^{\prime }$-meson decay width. The
two-step decay $\eta (\eta ^{\prime })\rightarrow \rho ^{\pm }\pi ^{\mp
}\rightarrow \pi ^{+}\pi ^{-}\gamma ^{*}$ is forbidden by $C$-parity
conservation. Constructing the dilepton spectra, one should not take into
account {\it e.g.} channels $\eta ^{\prime }\rightarrow $ $\pi ^{+}\pi
^{-}\gamma ^{*}$ and $\eta ^{\prime }\rightarrow $ $\rho ^{0}\gamma ^{*}$
simultaneously. The second one is already contained in the first one.


\subsection{Decay mode $\rho ^{0}\rightarrow $ $\pi ^{+}\pi ^{-}\ell
^{+}\ell ^{-}$}


The diagrams contributing to the decay $\rho ^0\rightarrow $ $\pi ^{+}\pi
^{-}\gamma ^{*}$ are depicted in Fig. \ref{fig6}. The first two diagrams
describe the photon bremsstrahlung. The third contact diagram is added to
restore the gauge invariance.

The vertex $\rho \pi \pi $ is given by Eq.(\ref{VI.1}). The matrix element
of the process $\rho ^{0}\rightarrow $ $\pi ^{+}\pi ^{-}\gamma ^{*}$ has the
form 
\begin{eqnarray}
{\cal M}=ief_{\rho \pi \pi }F_{\pi }(M^{2})\epsilon _{\tau }(P){\cal M}%
_{\tau \mu }\varepsilon _{\mu }^{*}(k)  \label{VI.9}
\end{eqnarray}
where 
\begin{eqnarray}
{\cal M}_{\tau \mu }=(2Q_{1}-P)_{\tau }\frac{(2Q_{1}-k)_{\mu }}{%
Q_{1}^{2}-\mu ^{2}}+(2Q_{2}-P)_{\tau }\frac{(2Q_{2}-k)_{\mu }}{Q_{2}^{2}-\mu
^{2}}-2g_{\tau \mu }.  \label{VI.10}
\end{eqnarray}
Here, $Q_{1}=p_{1}+k$, $Q_{2}=p_{2}+k$, $p_{1}$ and $p_{2}$ are the $\pi
^{+} $ and $\pi ^{-}$ momenta, $\mu $ is the pion mass, and $F_{\pi }(t)$ is
the pion electromagnetic form factor. The tensor part of the matrix element, 
${\cal M}_{\tau \mu }$, is transverse with respect to the photon momentum, $%
k,$ and the $\rho $-meson momentum, $P$: 
\begin{eqnarray}
{\cal M}_{\tau \mu }k_{\mu } &=&0,  \label{VI.11} \\
{\cal M}_{\tau \mu }P_{\tau } &=&0.  \label{VI.12}
\end{eqnarray}
The role of the last equation is the elimination of the coupling of the
spurious spin-zero component of the $\rho $-meson wave function from the
physical sector. Gauge invariance requires that the form factor entering the
third diagram be equal to the pion form factor.

The square of the matrix element summed up over the photon polarizations,
averaged over the initial $\rho $-meson polarizations and over directions of
the pion momenta in the c. m. frame of the two pions, has a compact form 
\begin{eqnarray}
{\cal R} &\equiv& \int \frac{d\Omega _{12}}{4\pi }{\cal M}_{\tau \mu }{\cal M%
}_{\sigma\nu}^{*}\frac{1}{3}(-g_{\tau \sigma}+\frac{P_{\tau }P_{\sigma}}{s}%
)(-g_{\mu \nu})  \nonumber \\
&=&\frac{2}{3B_{\pi }^{2}} \left(4B_{\pi }^{2}+(M^{2}-4\mu
^{2})(s-4\mu^{2})F(\xi) \right.  \nonumber \\
&+&\left.((s-4\mu ^{2})(M^{2}-4\mu ^{2}+2s_{12})+2s_{12}M^{2})L(\xi)\right)
\label{II.13}
\end{eqnarray}
where $s=P^2=m_\rho ^2$ and $s_{12}=(p_1+p_2)^2$. We have also 
\begin{eqnarray}
F(\xi ) &=&\frac{1}{1-\xi ^{2}},  \label{VI.14} \\
L(\xi ) &=&\frac{1}{2\xi }\ln (\frac{1+\xi }{1-\xi })  \label{VI.15}
\end{eqnarray}
and 
\begin{eqnarray}
\xi &=&\frac{2}{B_{\pi }}\sqrt{\frac{s}{s_{12}}}p^{*}(\sqrt{s_{12}},\mu ,\mu
)p^{*}(\sqrt{s},\sqrt{s_{12}},M),  \label{VI.16} \\
B_{\pi } &=&\frac{1}{2}(s+M^{2}-s_{12}).  \label{VI.17}
\end{eqnarray}
The decay width $\Gamma (\rho ^0\rightarrow \pi ^{+}\pi ^{-}\gamma ^{*})$ is
given by 
\begin{eqnarray}
\Gamma (\rho ^{0}\rightarrow \pi ^{+}\pi ^{-}\gamma ^{*})&=&\frac{\alpha }{%
16\pi ^{2}s}f_{\rho \pi \pi }^{2}\left| F_{\pi }(M^{2})\right| ^{2} 
\nonumber \\
&\times&\int_{4\mu ^{2}}^{(\sqrt{s}-M)^{2}}{\cal R\;}\frac{p^{*}(\sqrt{s},%
\sqrt{s_{12}},M)p^{*}(\sqrt{s_{12}},\mu ,\mu )}{\sqrt{s_{12}}}ds_{12}.
\label{VI.18}
\end{eqnarray}
The integral over the $s_{12}$ runs from $4\mu ^2\,$to $(\sqrt{s}-M)^2$. In
the case of real photon emission ($M^2=0$), we reproduce the corresponding
expression from Ref. \cite{PS}. The pion form factor can be taken as 
\begin{eqnarray}
F_{\pi }(t)=\frac{m_{\rho }^{2}}{m_{\rho }^{2}-t-im_{\rho }\Gamma _{\rho }}.
\label{VI.19}
\end{eqnarray}
The dilepton spectrum can be obtained with the use of Eq.(\ref{II.10}) as 
\begin{eqnarray}
d\Gamma (\rho ^{0}\rightarrow \pi ^{+}\pi ^{-}\ell ^{+}\ell ^{-})=\Gamma
(\rho ^{0}\rightarrow \pi ^{+}\pi ^{-}\gamma ^{*})M\Gamma (\gamma
^{*}\rightarrow \ell ^{+}\ell ^{-})\frac{dM^{2}}{\pi M^{4}}.  \label{VI.20}
\end{eqnarray}

The square of the matrix element at small invariant masses $2m_e\lesssim M$
and small momenta of the virtual photon is proportional to $1/\omega ^{*2}$
where $\omega ^{*}=B_\pi /\sqrt{s}$ is the photon energy in the c. m. frame
of the decaying meson. The contribution of the region $s_{12}\lesssim (\sqrt{%
s}-M)^2$ to the integral over the $s_{12}$ is of the order $\log ^3(1/M),$
and so $dB/dM\sim \log ^3(1/M)/M$. This is a general feature of the
bremsstrahlung mechanism. The square of the matrix element in a two-step
decay contains no a large parameter. In such a case, $dB/dM\sim 1/M$.


\subsection{Decay modes $\rho ^{0}\rightarrow $ $\pi ^{0}\pi ^{0}\ell
^{+}\ell ^{-}$, $\omega \rightarrow $ $\pi ^{0}\pi ^{0}\ell ^{+}\ell ^{-}$,
and $\omega \rightarrow $ $\pi ^{+}\pi ^{-}\ell ^{+}\ell ^{-}$}


The diagram contributing to the decay $\rho ^{0}\rightarrow $ $\pi ^{0}\pi
^{0}\gamma ^{*}$ is shown in Fig. \ref{fig7}. The vertex $\rho \omega \pi $
has the form 
\begin{eqnarray}
{\cal \delta L}_{\rho \omega \pi } &=&f_{\rho \omega \pi }\epsilon _{\tau
\sigma \mu \nu }\partial _{\sigma }\omega _{\tau }\partial _{\nu }\rho _{\mu
}^{\alpha }\pi ^{\alpha }  \nonumber \\
&=&f_{\rho \omega \pi }\epsilon _{\tau \sigma \mu \nu }(\partial _{\sigma
}\omega _{\tau }\partial _{\nu }\rho _{\mu }^{0}\pi ^{0}+\partial _{\sigma
}\omega _{\tau }\partial _{\nu }\rho _{\mu }^{+}\pi ^{-}+\partial _{\sigma
}\omega _{\tau }\partial _{\nu }\rho _{\mu }^{-}\pi ^{+}).  \label{VI.21}
\end{eqnarray}

The coupling constant $f_{\rho \omega \pi }\approx 16$ GeV$^{-1}$ was
determined by Gell-Mann, Sharp and Wagner \cite{GMSW} from the $\omega
\rightarrow \pi ^{+}\pi ^{-}\pi ^{0}$ decay assuming a two-step mechanism $%
\omega \rightarrow \rho \pi \rightarrow \pi ^{+}\pi ^{-}\pi ^{0}$.

In Refs. \cite{Ulf} a direct contact term $\omega \rightarrow \pi ^{+}\pi
^{-}\pi ^{0}$ originating from the chiral anomaly is taken into account. The
analysis of the $\omega \rightarrow \pi ^{+}\pi ^{-}\pi ^{0}$ and $\phi
\rightarrow \rho \pi $, $\pi ^{+}\pi ^{-}\pi ^{0}$ decays with inclusion of
the contact vertex gives a value $f_{\rho \omega \pi }\approx 12$ GeV$^{-1}$%
. In the contact vertex, one can always select a pion pair with quantum
numbers of the $\rho $-meson. The final-state interaction between these
pions, as usually, can be taken into account dividing the bare vertex by the
Jost function determined from the isovector $p$-wave $\pi \pi $-scattering
phase shift. In the two-pion channels with the $\rho $-meson quantum
numbers, the ordinary $\rho $-meson propagator therefore occurs, as a result
of which the contact vertex reduces to the ordinary pole diagrams of the
same old two-step mechanism.

The coupling constant $f_{\rho \omega \pi }$ can be extracted from the
decays $\rho \rightarrow \pi \gamma $ and $\omega \rightarrow \pi \gamma .$
The one-pole VMD approximation for the transition form factors gives $%
f_{\rho \omega \pi }=f_{\rho \gamma \pi }g_\omega \approx 12.6$ GeV$^{-1}$
and $f_{\rho \omega \pi }=f_{\omega \gamma \pi }g_\rho \approx 11.7$ GeV$%
^{-1}$, respectively. In the two-pole VMD model which provides the
transition form factors with the correct asymptotics these numbers should be
multiplied by a factor $m_X^2/(m_X^2-m_\rho ^2)$. For $m_X=1.2$ GeV, one
gets, respectively, $f_{\rho \omega \pi }\approx 21$ GeV$^{-1}$ and $f_{\rho
\omega \pi }\approx 20$ GeV$^{-1}$. The QCD sum rules \cite{Lubl} give $%
f_{\rho \omega \pi }\approx 16$ GeV$^{-1}$. In Ref. \cite{ABram}, the value $%
f_{\rho \omega \pi }\approx 14$ GeV$^{-1}$ is used.

We consider contributions to the strong part of the decay amplitudes from
the ground-state vector mesons only. The two-step mechanism of the $\omega
\rightarrow \pi ^{+}\pi ^{-}\pi ^{0}$ decay with a ground-state $\rho $%
-meson in the intermediate state is essentially identical to the mechanism
of the decays discussed in this subsection. In our calculations, we use $%
f_{\rho \omega \pi }=16$ GeV$^{-1}$.

The vertex $\omega \gamma \pi $ is defined by Eq.(\ref{V.1}). The matrix
element of the decay $\rho ^{0}\rightarrow $ $\pi ^{0}\pi ^{0}\gamma ^{*}$
has the form 
\begin{eqnarray}
{\cal M}=-ief_{\rho \omega \pi }f_{\omega \gamma \pi }F_{\omega \gamma \pi
}(M^{2})\epsilon _{\tau }(P){\cal M}_{\tau \mu }\varepsilon _{\mu }^{*}(k)
\label{VI.22}
\end{eqnarray}
with

\begin{eqnarray}
{\cal M}_{\tau \mu }=\epsilon _{\tau \sigma \rho \lambda }P_{\sigma
}Q_{1\lambda }\epsilon _{\rho \kappa \mu \nu }Q_{1\kappa }k_{\nu }\frac{1}{%
Q_{1}^{2}-m_{\omega }^{2}}+\epsilon _{\tau \sigma \rho \lambda }P_{\sigma
}Q_{2\lambda }\epsilon _{\rho \kappa \mu \nu }Q_{2\kappa }k_{\nu }\frac{1}{%
Q_{2}^{2}-m_{\omega }^{2}}  \label{VI.23}
\end{eqnarray}
where $Q_{1}=p_{1}+k$ and $Q_{2}=p_{2}+k$.

The square of the matrix element 
\begin{eqnarray}
{\cal R}\equiv \int \frac{d\Omega _{12}}{4\pi }{\cal M}_{\tau \mu }{\cal M}%
_{\sigma\nu}{}^{*}\frac{1}{3}(-g_{\tau \sigma}+\frac{P_{\tau }P_{\sigma}}{s}%
)(-g_{\mu \nu})\text{,}  \label{VI.24}
\end{eqnarray}
is averaged numerically over the directions of the photon momentum ${\bf k}$
in the c. m. frame of the pion $1$ and the photon.

The decay widths can be written as 
\begin{eqnarray}
\Gamma (\rho ^{0}\rightarrow \pi ^{0}\pi ^{0}\gamma ^{*})&=&\frac{1}{32\pi
^{2}s}f_{\rho \omega \pi }^{2}f_{\omega \gamma \pi }^{2}\left| F_{\omega
\gamma \pi }(M^{2})\right| ^{2}  \nonumber \\
&\times&\int_{(\mu +M)^{2}}^{(\sqrt{s}-\mu )^{2}}{\cal R\;}\frac{p^{*}(\sqrt{%
s},\sqrt{s_{12}},M)p^{*}(\sqrt{s_{12}},\mu ,\mu )}{\sqrt{s_{12}}}ds_{12}.
\label{VI.26}
\end{eqnarray}
The integral over the $s_{12}$ runs from $(\mu +M)^{2}$ to $(\sqrt{s}-\mu
)^{2}$. The additional factor $2$ in the denominator occurs due to identity
of two pions. The dilepton spectrum in the decay $\rho ^{0}\rightarrow \pi
^{0}\pi ^{0}\ell ^{+}\ell ^{-}$ can be constructed with the use of Eq.(\ref
{II.10}): 
\begin{eqnarray}
d\Gamma (\rho ^{0}\rightarrow \pi ^{0}\pi ^{0}\ell ^{+}\ell ^{-})=\Gamma
(\rho ^{0}\rightarrow \pi ^{0}\pi ^{0}\gamma ^{*})M\Gamma (\gamma
^{*}\rightarrow \ell ^{+}\ell ^{-})\frac{dM^{2}}{\pi M^{4}}.  \label{VI.25}
\end{eqnarray}

The $\omega \rightarrow $ $\pi ^{0}\pi ^{0}\gamma ^{*}$ decay is analogous
to the $\rho ^{0}\rightarrow $ $\pi ^{0}\pi ^{0}\gamma ^{*}$ decay. The
corresponding diagrams are shown in Fig. \ref{fig8}. The matrix element has
the form of Eqs.(\ref{VI.22}) and (\ref{VI.23}) with interchanged $V\gamma P$
coupling constants, $f_{\omega \gamma \pi }$ $\leftrightarrow $ $f_{\rho
\gamma \pi }$, and vector meson masses, $m_{\omega }$ $\leftrightarrow $ $%
m_{\rho }$.

The diagrams contributing to the $\omega \rightarrow $ $\pi ^{+}\pi
^{-}\gamma ^{*}$ decay are shown in Fig. \ref{fig9}. The first two diagrams
(a) and (b) correspond to the common two-step mechanism. The last three
diagrams occur due to the $\omega \rho $ mixing and correspond to the
bremsstrahlung. The diagram (e) restores the gauge invariance of the
amplitude. The bremsstrahlung dominates at small values of the $e^{-}e^{+}$
invariant mass. The vertices $\omega \rho \pi $ and $\rho \gamma \pi $
entering into the matrix element of the $\omega \rightarrow $ $\pi ^{+}\pi
^{-}\gamma ^{*}$ decay have the form of Eqs.(\ref{VI.21}) and (\ref{V.7}).

The first two diagrams and the last three diagrams from Fig. \ref{fig9} are
separately gauge invariant. These two classes of diagrams contrary to the
statement of Ref. \cite{FO} interfere: Two pions in the two-step part of the
amplitude are in an isosinglet state. Due to the VMD, the photon in the
diagrams (c) and (d) is emitted from the $\rho ^0$-meson. It is equivalent
to introducing the pion form factor into the $\pi \pi \gamma $\ vertex. The
outgoing pion and the $\rho $-meson with a common vertex form an isotriplet
which, in turn, forms a new isotriplet with the second outgoing pion. The
isospin wave function of the final state looks like $[\vec \pi \times [\vec %
\pi \times \vec \rho ]]=\vec \pi (\vec \pi \cdot \vec \rho )-\vec \rho (\vec %
\pi \cdot \vec \pi ).$\ The first term corresponds to two neutral pions and
can be dropped, while the second term describes two charged pions in an
isoscalar state. Two pions in an isoscalar state from the diagrams (a) and
(b) and a photon interfere with two pions in an isoscalar state from
diagrams (c) and (d) and a photon. The final states are identical.

The interference pattern depends on the production dynamics. The physical $%
\omega $-meson wave function has the form $|\omega >=|\omega _0>+\varepsilon
|\rho _0>$ where $|\omega _0>\ $and $|\rho _0>$ are pure isospin $I=0$ and $%
1 $ states and $\varepsilon $ is a complex $\omega \rho $ mixing parameter.
The total matrix element is given by ${\cal M}={\cal M}^{[p]}+\varepsilon 
{\cal M}^{[b]}.$ The value ${\cal M}^{[p]}$ is the same as in Eq.(VI.22).
The value ${\cal M}^{[b]}$ is given by Eq.(VI.9). The $\omega \rightarrow $ $%
\pi ^{+}\pi ^{-}\gamma ^{*}$ decay width represents a sum of three terms 
\begin{eqnarray}
\Gamma (\omega \rightarrow \pi ^{+}\pi ^{-}\gamma ^{*})=\Gamma ^{[p]}(\omega
\rightarrow \pi ^{+}\pi ^{-}\gamma ^{*})+\Gamma ^{[b]}(\omega \rightarrow
\pi ^{+}\pi ^{-}\gamma ^{*})+\Gamma ^{[pb]}(\omega \rightarrow \pi ^{+}\pi
^{-}\gamma ^{*})  \label{VI.27}
\end{eqnarray}
originating from the vector-meson pole diagrams $[p]$, the bremsstrahlung
diagrams $[b]$, and from their interference $[pb]$. It was pointed out in
Ref. \cite{PS} that $\Gamma ^{[p]}(\omega \rightarrow \pi ^{+}\pi ^{-}\gamma
^{*})=2\Gamma (\omega \rightarrow \pi ^0\pi ^0\gamma ^{*}).$ The
bremsstrahlung contribution to the $\omega \rightarrow $ $\pi ^{+}\pi
^{-}\gamma ^{*}$ decay equals $\Gamma ^{[b]}(\omega \rightarrow \pi ^{+}\pi
^{-}\gamma ^{*})=|\varepsilon |^2\Gamma ^{[b]}(\rho ^0\rightarrow \pi
^{+}\pi ^{-}\gamma ^{*})$, with $|\varepsilon |=(\Gamma (\omega \rightarrow
\pi ^{+}\pi ^{-})/\Gamma (\rho ^0\rightarrow \pi ^{+}\pi
^{-}))^{1/2}=3.4\times 10^{-2}$ being the $\rho ^0$-meson admixture, as
determined from the $\omega \rightarrow \pi ^{+}\pi ^{-}$ decay.

The interference term 
\begin{eqnarray}
{\cal R}^{[pb]}\equiv 2\int \frac{d\Omega _{12}}{4\pi }{\cal M}_{\tau \mu
}^{[p]}{\cal M}_{\sigma \nu }{}^{[b]*}\frac{1}{3}(-g_{\tau \sigma }+\frac{%
P_{\tau }P_{\sigma }}{s})(-g_{\mu \nu })  \label{VI.28a}
\end{eqnarray}
can be found to be 
\begin{eqnarray}
{\cal R}^{[pb]} &=&\frac{8}{3}sp^{*2}(\sqrt{s_{12}},\mu ,\mu )p^{*2}(\sqrt{s}%
,\sqrt{s_{12}},M)  \nonumber \\
&&\times \frac{1}{B_{\pi }B_{\rho }(\xi +\zeta _{\rho })}[\xi L(\xi )+\zeta
_{\rho }L(\zeta _{\rho })+(1-L(\xi ))/\xi +(1-L(\zeta _{\rho }))/\zeta
_{\rho }]  \label{INT}
\end{eqnarray}
where $B_\rho =B_\pi +\mu ^2-m_\rho ^2$ and $\zeta _\rho =B_\pi \xi /B_\rho
. $ The value $B_\pi $ is defined by Eq.(\ref{VI.17}). The interference
contribution to the width is given by 
\begin{eqnarray}
\Gamma ^{[pb]}(\omega &\rightarrow &\pi ^{+}\pi ^{-}\gamma ^{*})=\frac{1}{%
16\pi ^{2}s}Re\{(ief_{\rho \pi \pi }F_{\pi }(M^{2}))^{*}(-ie\varepsilon
f_{\rho \omega \pi }f_{\omega \gamma \pi }F_{\omega \gamma \pi }(M^{2}))\} 
\nonumber \\
&\times& \int_{(\mu +M)^{2}}^{(\sqrt{s}-\mu )^{2}}{\cal R}^{[pb]}{\cal \;}%
\frac{p^{*}(\sqrt{s},\sqrt{s_{12}},M)p^{*}(\sqrt{s_{12}},\mu ,\mu )}{\sqrt{%
s_{12}}}ds_{12}.
\end{eqnarray}
The couplings $f_{\rho \omega \pi }\ $and $f_{\rho \pi \pi }$ are determined
up a sign. However, the relations $f_{\omega \gamma \pi }\approx f_{\rho
\omega \pi }/g_\rho $ and $f_{\rho \pi \pi }/g_\rho \approx 1$ allow to fix
the sign of the interference term. The phase $\varphi $ of the mixing
parameter $\varepsilon $ is a sum of phases $\varphi _{prod}$ and $\varphi
_{decay}$ originating, respectively, from production and $\pi \pi $ decay of
the $\omega $-meson. The last one is close to $\pi /2$ (see {\it e.g.} Ref. 
\cite{ABK}). We set the value $\varphi _{prod}$ equal to zero. Since the
product $F_\pi (M^2)^{*}F_{\omega \gamma \pi }(M^2)$ is approximately real
below $1$ GeV, the interference can then be neglected.

The dilepton spectra of the $\omega \rightarrow \pi ^{+}\pi ^{-}\ell
^{+}\ell ^{-}$ decays can be obtained as usually from Eq.(II.10).


\subsection{Decay modes $\rho \rightarrow $ $\eta \pi e^{+}e^{-}$and $\omega
\rightarrow $ $\eta \pi ^{0}e^{+}e^{-}$}


These decays also proceed through the two-step mechanism. The diagrams are
shown in Figs. \ref{fig10} and \ref{fig11}. The matrix element of the decay $%
\rho \rightarrow $ $\eta \pi \gamma ^{*}$ is given by

\begin{eqnarray}
{\cal M}=ef_{\rho \omega \pi }f_{\omega \gamma \eta }\epsilon _{\tau }(P)%
{\cal M}_{\tau \mu }\varepsilon _{\mu }^{*}(k)  \label{VI.28}
\end{eqnarray}
with 
\begin{eqnarray}
{\cal M}_{\tau \mu }=\epsilon _{\tau \sigma \rho \lambda }P_{\sigma
}Q_{1\lambda }\epsilon _{\rho \kappa \mu \nu }Q_{1\kappa }k_{\nu }\frac{1}{%
Q_{1}^{2}-m_{\omega }^{2}}+\epsilon _{\tau \sigma \rho \lambda }P_{\sigma
}Q_{2\lambda }\epsilon _{\rho \kappa \mu \nu }Q_{2\kappa }k_{\nu }\frac{1}{%
Q_{2}^{2}-m_{\rho }^{2}}  \label{VI.29}
\end{eqnarray}
where $Q_{1}=p_{1}+k$, $Q_{2}=p_{2}+k$, $p_{1}$ is the pion momentum, and $%
p_{2}$ is the $\eta $-meson momentum. We take the $SU(3)$ symmetry relation $%
f_{\rho \omega \pi }f_{\omega \gamma \eta }=2f_{\rho \rho \eta }f_{\rho
\gamma \pi }$ into account.

The average of the squared matrix element, 
\begin{eqnarray}
{\cal R}\equiv \int \frac{d\Omega _{12}}{4\pi }{\cal M}_{\tau \mu }{\cal M}%
_{\sigma \nu}{}^{*}\frac{1}{3}(-g_{\tau \sigma}+\frac{P_{\tau }P_{\sigma}}{s}%
)(-g_{\mu \nu})\text{,}  \label{VI.30}
\end{eqnarray}
is calculated numerically. The integration over the directions of the
momenta of the two mesons is performed in the c. m. frame of $\pi $ and $%
\eta $. The decay widths can be written as 
\begin{eqnarray}
\Gamma (\rho \rightarrow \eta \pi \gamma ^{*})&=&\frac{\alpha }{16\pi ^{2}s}%
f_{\rho \omega \pi }^{2}f_{\omega \gamma \eta }^{2}\left| F_{\omega \gamma
\eta }(M^{2})\right| ^{2}  \nonumber \\
&\times&\int_{(\mu +\mu ^{\prime })^{2}}^{(\sqrt{s}-M)^{2}}{\cal R\;}\frac{%
p^{*}(\sqrt{s},\sqrt{s_{12}},M)\;p^{*}(\sqrt{s_{12}},\mu ,\mu ^{\prime })}{%
\sqrt{s_{12}}}ds_{12}.  \label{VI.31}
\end{eqnarray}
where $\mu ^{\prime }$ is the $\eta $-meson mass. In the form factor $%
F_{\omega \gamma \eta }(t)$, the shift of the $\omega $-meson from the mass
shell is neglected. The dilepton spectrum in the decays $\rho \rightarrow
\eta \pi e^{+}e^{-}$ can be constructed with the use of Eq.(\ref{II.10}).

The $\omega \rightarrow $ $\eta \pi ^{0}\gamma ^{*}$ decay is depicted in
Fig. \ref{fig11}. The matrix element has the form of Eq.(\ref{VI.28}), its
tensor part once the $SU(3)$ symmetry relation $f_{\rho \omega \pi }f_{\rho
\gamma \eta }=2f_{\omega \omega \eta }f_{\omega \gamma \pi }$ is taken into
account has the form of Eq.(\ref{VI.29}), and the $\omega \rightarrow $ $%
\eta \pi ^{0}\gamma ^{*}$ decay width is given by Eq.(\ref{VI.30}), with the
apparent substitutions $f_{\rho \gamma \eta }\leftrightarrow f_{\omega
\gamma \eta }$ and $m_{\omega }\leftrightarrow m_{\rho }$. The dilepton
spectrum can then be constructed, as usually, using Eq.(\ref{II.10}).\ 

\subsection{Decay mode $\rho ^{+}\rightarrow $ $\pi ^{+}\pi ^{0}\ell
^{+}\ell ^{-}$}


This decay is depicted in Fig. \ref{fig12}. The first three diagrams
describe the dilepton bremsstrahlung. The last diagram corresponds to the
two-step mechanism of the dilepton emission. These two sets of diagrams,
each of them is the gauge invariant, interfere and we calculate the
interference without further approximations.

The photon emitted from the $\rho ^{+}$-meson line interacts with the
electromagnetic current 
\begin{eqnarray}
j_{\mu }(p^{\prime },p)=\frac{1}{2m_{\rho }}\epsilon _{\sigma
}^{*}(p^{\prime })\left[ -(p^{\prime }+p)_{\mu }g_{\tau \sigma }+\lambda
(q_{\tau }g_{\sigma \mu }-q_{\sigma }g_{\tau \mu })\right] \epsilon _{\tau
}(p)  \label{VI.32}
\end{eqnarray}
where $q=p^{\prime }-p$, $\epsilon _{\tau }(p)$ is the $\rho $-meson
polarization vector. The value of $\lambda $ is connected to the $\rho ^{+}$
magnetic moment. Let's fix first the value of $\lambda $.

The magnetic moment of a particle can be calculated from expression 
\begin{eqnarray}
{\bf \mu }=\int d{\bf x}\frac{1}{2}[{\bf x}\times {\bf j}].  \label{VI.33}
\end{eqnarray}

The wave function of a vector particle in its rest frame where $p=(m,{\bf 0}%
) $ has the form $\epsilon _{\tau }(p)=(0,{\bf e})$. Substituting this
expression into Eq.(\ref{VI.32}) and Eq.(\ref{VI.32}) in turn into Eq.(\ref
{VI.33}), we obtain 
\begin{eqnarray}
{\bf \mu }=\frac{\lambda }{2m_{\rho }}e_{\gamma }^{*}{\bf s}e_{\gamma }.
\label{VI.34}
\end{eqnarray}
where ${\bf s}$ is the spin operator acting on the space-like part of the $%
\rho $-meson wave functions according to the rule $s_{\alpha }e_{\beta
}=-i\epsilon _{\alpha \beta \gamma }e_{\gamma }$ ($\alpha ,\beta ,\gamma
=1,2,3$). In the non-relativistic quark model, the $\rho ^{+}$ magnetic
moment is the difference of the magnetic moments of the $u$- and $d$-quarks: 
$\mu _{\rho ^{+}}=\mu _{u}-\mu _{d}$. The proton and neutron magnetic
moments are given by the well known expressions 
\begin{eqnarray}
\mu _{p} &=&\frac{4\mu _{u}-\mu _{d}}{3},  \label{VI.35} \\
\mu _{n} &=&\frac{4\mu _{d}-\mu _{n}}{3},  \label{VI.36}
\end{eqnarray}
and thus one can write 
\begin{eqnarray}
\frac{\lambda }{2m_{\rho }}=\mu _{u}-\mu _{d}=\frac{3}{5}(\mu _{p}-\mu _{n}).
\label{VI.37}
\end{eqnarray}
One gets $\lambda =\allowbreak 2.\,3$. The additive quark model predictions
are valid with an accuracy of $10-30\%$. The one-gluon and one-pion exchange
currents and the anomalous quark moments connected to the violation of the
OZI rule, are shown to be important in order to explain the deviations of
the naive $SU(3)$ predictions from the experimental data \cite{MIK,Bre}.

The diagram (a) in Fig. 12 contains the pion form factor $F_{\pi }(t)$. A
part of the second diagram (b) connected to the $\rho $-meson convection
current contains a $\rho $-meson form factor $F_{1\rho }(t)$ which decreases
as $\sim 1/t^{2}$ at $t\rightarrow \infty $. The second part of the same
diagram connected to the spin current contains a different the $\rho $-meson
form factor $F_{2\rho }(t)$ which decreases also like $\sim 1/t^{2}$ at $%
t\rightarrow \infty $ \cite{VZ}. It is not clear what kind of the propagator
should enter into the diagram (c). The gauge invariance requires, within the
framework of the considered model, that the two form factors $F_{1\rho }(t)$
and $F_{\pi }(t)$ should be equal. This is, however, a shortcoming of the
model. It can be overcome on the basis of a more complete theory. The second
part of the diagram (b) is gauge invariant. The form factor $F_{2\rho }(t)$
is therefore not subjected to additional constraints. The dilepton invariant
masses in the $\rho ^{+}\rightarrow $ $\pi ^{+}\pi ^{0}\ell ^{+}\ell ^{-}$
decay are not very large, so the distinction between these three form
factors does not exceed a $20\%$ effect. We set the form factor $F_{1\rho
}(t)$ and the form factor entering the diagram (c) both equal to the pion
form factor. The sum of the diagrams (a) - (c) is then gauge invariant. The
form factor $F_{2\rho }(t)$ coincides with the transition form factor $%
F_{\omega\gamma\pi }(t)$ in the framework of the extended VMD model.

The matrix element of the $\rho ^{+}\rightarrow $ $\pi ^{+}\pi ^{0}\gamma
^{*}$ decay is given then by 
\begin{eqnarray}
{\cal M}=ief_{\rho \pi \pi }F_{\pi }(M^{2})\epsilon _{\tau }(P){\cal M}%
_{\tau \mu }\varepsilon _{\mu }^{*}(k)  \label{VI.38}
\end{eqnarray}
Its tensor part consists of the $5$ pieces: 
\begin{eqnarray}
{\cal M}_{\tau \mu }=\sum_{i=1}^{5}{\cal M}_{\tau \mu }^{i}  \label{VI.39}
\end{eqnarray}
where 
\begin{eqnarray}
{\cal M}_{\tau \mu }^{1} &=&\frac{(p_{1}-p_{2})_{\tau }(2P-k)_{\mu }}{%
(P-k)^{2}-m_{\rho }^{2}},  \label{VI.40} \\
{\cal M}_{\tau \mu }^{2} &=&\lambda \frac{-k\cdot (p_{1}-p_{2})g_{\tau \mu
}+k_{\tau }(p_{1}-p_{2})_{\mu }}{(P-k)^{2}-m_{\rho }^{2}},  \label{VI.41} \\
{\cal M}_{\tau \mu }^{3} &=&\frac{(p_{1}-p_{2}+k)_{\tau }(2p_{1}+k)_{\mu }}{%
(p_{1}+k)^{2}-\mu ^{2}},  \label{VI.42} \\
{\cal M}_{\tau \mu }^{4} &=&-g_{\tau \mu },  \label{VI.43} \\
{\cal M}_{\tau \mu }^{5} &=&\gamma \frac{\epsilon _{\tau \sigma \rho \lambda
}P_{\sigma }(p_{2}+k)_{\lambda }\epsilon _{\rho \kappa \mu \nu
}(p_{2}+k)_{\kappa }k_{\nu }}{(p_{2}+k)^{2}-m_{\omega }^{2}},  \label{VI.44}
\end{eqnarray}
with 
\begin{eqnarray}
\gamma =\frac{f_{\rho \omega \pi }f_{\omega \gamma \pi }}{f_{\rho \pi \pi }}%
\frac{F_{\omega \gamma \pi }(M^{2})}{F_{\pi }(M^{2})}\approx \frac{f_{\rho
\omega \pi }^{2}}{f_{\rho \pi \pi }^{2}}\frac{F_{\omega \gamma \pi }(M^{2})}{%
F_{\pi }(M^{2})}.  \label{VI.45}
\end{eqnarray}
The coupling constants $f_{\rho \omega \pi }$, $f_{\omega \gamma \pi }$, and 
$f_{\rho \pi \pi }\ $ are determined earlier (see Table 1). We used in Eq.(%
\ref{VI.45}) the relations $f_{\omega \gamma \pi }\approx f_{\rho \omega \pi
}/g_{\rho }$ and $f_{\rho \pi \pi }/g_{\rho }\approx 1.$ It is seen that,
while signs of the couplings $f_{\rho \pi \pi }$ and $f_{\rho \omega \pi }$
are a matter of convention, the value $\gamma $ is positive and thus the
sign of the interference term is fixed.

The first two terms, (\ref{VI.40}) and (\ref{VI.41}), originate from the
diagram (b) of Fig. \ref{fig12}. The third term (\ref{VI.42}) originates
from the diagram (a), the fourth term comes from the diagram (c), and the
last term originates from the diagram (d). The second and the last terms and
the sum of the rest are transverse with respect to the photon momentum: $%
{\cal M}_{\tau \mu }^{2}k_{\mu }=0,$ ${\cal M}_{\tau \mu }^{5}k_{\mu }=0,$ $%
\sum_{i=1,3,4}{\cal M}_{\tau \mu }^{i}k_{\mu }=0$. The matrix element, $%
{\cal M}_{\tau \mu },$ is not transverse, however, with respect to the $\rho 
$-meson momentum, ${\cal M}_{\tau \mu }P_{\tau }\neq 0$. A complete theory
should yield a matrix element which is transverse with respect to both, the
photon and the $\rho $-meson momenta. This is not the case in our effective
theory. In the following, we work exclusively with the transverse part of
the matrix element which is responsible for decays of the physical states,
and ignore the longitudinal part responsible for a decay of a spurious
spin-zero component of the $\rho $-meson. This is achieved by contraction of
the square of the matrix element with the polarization tensor 
\begin{eqnarray}
\Sigma _{\tau \sigma }(P)=-g_{\tau \sigma }+\frac{P_{\tau }P_{\sigma }}{s}
\label{VI.46}
\end{eqnarray}
of the massive vector particle, as we did {\it e.g.} in Eq.(VI.24). The
difference is, however, that the term $P_{\tau }P_{\sigma}/s$ in Eq.(VI.24)
can be dropped before calculations. In the case considered in this
subsection the term $P_{\tau }P_{\sigma}/s$ cannot be dropped.

The square of the matrix element averaged over the initial polarizations of
the $\rho $-meson and summed up over the photon polarizations has the form 
\begin{eqnarray}
{\cal R}\equiv \sum_{f}\overline{\left| {\cal M}\right| ^{2}}%
=\sum_{i,j=1}^{5}{\cal R}_{ij}  \label{VI.47}
\end{eqnarray}
where 
\begin{eqnarray}
{\cal R}_{ij}={\cal M}_{\tau \mu }^{i}{\cal M}_{\sigma \nu }^{j*}\frac{1}{3}%
(-g_{\tau \sigma }+\frac{P_{\tau }P_{\sigma }}{s})(-g_{\mu \nu }).
\label{VI.48}
\end{eqnarray}
Let's now consider separately the different interference terms.

We perform an additional averaging of the terms ${\cal R}_{ij}$ with $%
i,j=1\div 4$ and $i=1 \div 3$ and $j=5$ over directions of the pion momentum 
${\bf p}_{1}$ in the c. m. frame of two pions. The remaining terms ${\cal R}%
_{45}$ and ${\cal R}_{55}$ are averaged over the directions of the pion
momentum ${\bf p}_{2}$ in the c. m. frame of the pion $\pi ^{0}$ and the
photon. The decay rate will finally be given as a sum of the two integrals
over the variables $s_{12}=(p_{1}+p_{2})^{2}$ and $s_{13}=(p_{2}+k)^{2}$.
The straightforward calculations give 
\begin{eqnarray}
{\cal R}_{11} &=&-\frac{4}{9}\frac{2s_{12}+2s-M^{2}}{(s_{12}-s)^{2}}p^{*2}(%
\sqrt{s_{12}},\mu ,\mu )(3+\frac{p^{*2}(\sqrt{s},\sqrt{s_{12}},M)}{s_{12}}),
\nonumber \\
{\cal R}_{22} &=&\frac{4}{9}\frac{\lambda ^{2}}{(s_{12}-s)^{2}}p^{*2}(\sqrt{%
s_{12}},\mu ,\mu )\left[ p^{*2}(\sqrt{s},\sqrt{s_{12}},M)-3M^{2}+\frac{1}{%
4s_{12}}(s-M^{2}-s_{12})^{2}\right.  \nonumber \\
&&\left. +\frac{1}{4ss_{12}}(s+M^{2}-s_{12})(s+s_{12}-M^{2})(s-M^{2}-s_{12})%
\right] ,  \nonumber \\
{\cal R}_{33} &=&-\frac{1}{3B_{\pi }^{2}}\left[ -\frac{1}{s}B_{\pi
}^{2}(3s+s_{12}-4\mu ^{2})\right.  \nonumber \\
&&\left. -(s-4\mu ^{2})(M^{2}-4\mu ^{2})F(\xi )+2B_{\pi }(s+M^{2}-8\mu
^{2})L(\xi )\right] ,  \nonumber \\
{\cal R}_{44} &=&1,  \nonumber \\
{\cal R}_{55} &=&\frac{4}{9}\frac{\gamma ^{2}}{(s_{23}-m_{\omega }^{2})^{2}}%
ss_{23}p^{*2}(\sqrt{s},\sqrt{s_{23}},\mu )p^{*2}(\sqrt{s_{23}},\mu ,M),
\label{VI.49}
\end{eqnarray}
\begin{eqnarray}
{\cal R}_{12}+{\cal R}_{21} &=&\frac{8}{9}\frac{\lambda }{(s_{12}-s)^{2}}%
\frac{s+s_{12}-M^{2}}{s_{12}}p^{*2}(\sqrt{s_{12}},\mu ,\mu )p^{*2}(\sqrt{s},%
\sqrt{s_{12}},M),  \nonumber \\
{\cal R}_{13}+{\cal R}_{31} &=&-\frac{2}{3}\frac{1}{(s_{12}-s)B_{\pi }}%
\left[ (-4\mu ^{2}+M^{2}-s)B_{\pi }+\frac{1}{3s}B_{\pi }^{3}\xi ^{2}+\right.
\nonumber \\
&&\left. \frac{1}{2}(s+s_{12}-M^{2})(s+s_{12}+M^{2}-8\mu ^{2})L(\xi )\right]
,  \nonumber \\
{\cal R}_{14}+{\cal R}_{41} &=&0,  \nonumber \\
{\cal R}_{15}+{\cal R}_{51} &=&\frac{4\gamma }{3}\frac{sp^{*2}(\sqrt{s_{12}}%
,\mu ,\mu )p^{*2}(\sqrt{s},\sqrt{s_{12}},M)}{(s_{12}-s)B_{\omega }}\frac{%
1+(\zeta ^{2}-1)L(\zeta )}{\zeta ^{2}},  \nonumber \\
{\cal R}_{23}+{\cal R}_{32} &=&-\frac{2}{3}\frac{\lambda }{(s_{12}-s)B_{\pi }%
}\left[ \frac{B_{\pi }}{12ss_{12}}((4\mu ^{2}-s_{12})M^{4}\right.  \nonumber
\\
&&+(-8\mu ^{2}s-10ss_{12}-4s_{12}^{2}+16\mu ^{2}s_{12})M^{2}+(4\mu
^{2}-s_{12})(s-s_{12})(s+5s_{12}))  \nonumber \\
&&\left. +\frac{1}{2}M^{2}(s_{12}+M^{2}+s-8\mu ^{2})L(\xi )\right] , 
\nonumber \\
{\cal R}_{24}+{\cal R}_{42} &=&0,  \nonumber \\
{\cal R}_{25}+{\cal R}_{52} &=&\frac{2}{3}\frac{\lambda \gamma }{%
(s_{12}-s)B_{\omega }}\left[ \frac{1}{2}B_{\omega }^{3}(\frac{1}{3}\zeta
^{2}+1-L(\zeta ))+\frac{1}{2}s_{12}B_{\omega }^{2}(-1+L(\zeta ))\right. 
\nonumber \\
&&+B_{\omega }(-\frac{1}{2}sp^{*2}(\sqrt{s},\sqrt{s_{12}}%
,M)+(s+M^{2}-s_{12})p^{*2}(\sqrt{s_{12}},\mu ,\mu ))(1-L(\zeta ))  \nonumber
\\
&&\left. +sp^{*2}(\sqrt{s},\sqrt{s_{12}},M)p^{*2}(\sqrt{s_{12}},\mu ,\mu )%
\frac{1+(\zeta ^{2}-1)L(\zeta )}{\zeta ^{2}}\right] ,  \nonumber \\
{\cal R}_{34}+{\cal R}_{43} &=&\frac{2}{3}\frac{1}{B_{\pi }}\left[ (-4\mu
^{2}+s_{12})L(\xi )+B_{\pi }(-1+L(\xi ))\right] ,  \nonumber \\
{\cal R}_{35}+{\cal R}_{53} &=&\frac{8}{3}\frac{\gamma }{B_{\pi }B_{\omega }}%
\frac{1}{\xi +\zeta }\left[ (-\frac{1}{4}s_{12}+\mu ^{2})(s_{12}M^{2}-\frac{1%
}{4}(s-M^{2}-s_{12})^{2})(\xi L(\xi )+\zeta L(\zeta ))\right.  \nonumber \\
&&\left. -\frac{1}{4}s_{12}\xi B_{\pi }{}^{2}(-1+L(\xi )+\frac{\xi }{\zeta }%
(-1+L(\zeta )))\right] ,  \nonumber \\
{\cal R}_{45}+{\cal R}_{54} &=&0.  \label{VI.50}
\end{eqnarray}
where $B_{\omega }=B_{\pi }+\mu ^{2}-m_{\omega }^{2}$ and $\zeta =B_{\pi
}\xi /B_{\omega }$.

The decay width $\rho ^{+}\rightarrow \pi ^{+}\pi ^{0}\gamma ^{*}$ is given
by expression 
\begin{eqnarray}
\Gamma (\rho ^{+} &\rightarrow& \pi ^{+}\pi ^{0}\gamma ^{*}) = \frac{\alpha 
}{16\pi ^{2}s}f_{\rho \pi \pi }^{2}\left| F_{\pi }(M^{2})\right| ^{2} 
\nonumber \\
&&\times \left[ \int_{4\mu ^{2}}^{(\sqrt{s}-M)^{2}}(\sum_{i=1}^{4}%
\sum_{j=1}^{4}+2\sum_{i=1}^3\sum_{j=5}^5) {\cal R}_{ij}{\cal \;}\frac{p^{*}(%
\sqrt{s},\sqrt{s_{12}},M)p^{*}(\sqrt{s_{12}},\mu ,\mu )}{\sqrt{s_{12}}}%
ds_{12}\right.  \nonumber \\
&&\left. +\int_{(\mu +M)^{2}}^{(\sqrt{s}-\mu )^{2}}{\cal R}_{55}{\cal \;} 
\frac{p^{*}(\sqrt{s},\sqrt{s_{23}},M)p^{*}(\sqrt{s_{23}},\mu ,M)}{\sqrt{%
s_{23}}}ds_{23}\right] .  \label{VI.52}
\end{eqnarray}
The dilepton spectrum in the $\rho ^{+}\rightarrow \pi ^{+}\pi ^{0}\ell
^{+}\ell ^{-}$ decay can be calculated from equation 
\begin{eqnarray}
d\Gamma (\rho ^{+}\rightarrow \pi ^{+}\pi ^{0}\ell ^{+}\ell ^{-})=\Gamma
(\rho ^{+}\rightarrow \pi ^{+}\pi ^{0}\gamma ^{*})M\Gamma (\gamma
^{*}\rightarrow \ell ^{+}\ell ^{-})\frac{dM^{2}}{\pi M^{4}}.  \label{VI.53}
\end{eqnarray}
The decay $\rho ^{-}\rightarrow \pi ^{-}\pi ^{0}\ell ^{+}\ell ^{-}$ is
completely equivalent to the $\rho ^{+}$-meson decay.

\subsection{Decay mode $f_{0}(980)\rightarrow $ $\pi ^{+}\pi ^{-}\ell
^{+}\ell ^{-}$}


The dominant decay mode of the $f_{0}(980)$-meson is $f_{0}\rightarrow $ $%
\pi \pi $. The effective vertex for this decay looks like 
\begin{eqnarray}
\delta {\cal L}=g_{f_{0}\pi \pi }f_{0}\pi ^{\alpha }\pi ^{\alpha }.
\label{VI.54}
\end{eqnarray}
The $f_{0}\rightarrow $ $\pi ^{+}\pi ^{-}$ decay width has the form 
\begin{eqnarray}
\Gamma (f_{0}\rightarrow \pi ^{+}\pi ^{-})=\frac{1}{8\pi s}g_{f_{0}\pi \pi
}^{2}p^{*}(\sqrt{s},\mu ,\mu ),  \label{VI.55}
\end{eqnarray}
with $\sqrt{s}=m_{f_{0}}.$ The experimental width $\Gamma
^{tot}(f_{0}\rightarrow \pi \pi )=\frac{3}{2}\Gamma (f_{0}\rightarrow \pi
^{+}\pi ^{-})=3\Gamma (f_{0}\rightarrow \pi ^{0}\pi ^{0})$ lies within the
interval $50-100$ MeV \cite{PDG}. The coupling constant can be found to be $%
g_{f_0 \pi \pi } = 1.6$ GeV$\pm \; 30 \%$.

The diagrams for the $f_{0}\rightarrow $ $\pi ^{+}\pi ^{-}\gamma ^{*}$ decay
are shown on Fig. \ref{fig13}. The matrix element has the form 
\begin{eqnarray}
{\cal M}=-ieg_{f\pi \pi }{\cal M}_{\mu }\varepsilon _{\mu }^{*}(k)
\label{VI.56}
\end{eqnarray}
with 
\begin{eqnarray}
{\cal M}_{\mu }=\frac{(2p_{1}+k)_{\mu }}{(p_{1}+k)^{2}-\mu ^{2}}-\frac{%
(2p_{2}+k)_{\mu }}{(p_{2}+k)^{2}-\mu ^{2}}.  \label{VI.57}
\end{eqnarray}
The tensor ${\cal M}_{\mu }$ is transverse with respect to the photon
momentum. The square of the matrix element summed up over the photon
polarizations and averaged over the directions of the pion momentum in the
c. m. frame of two pions has the form 
\begin{eqnarray}
{\cal R} &\equiv &\int \frac{d\Omega _{12}}{4\pi }{\cal M}_{\mu } {\cal M}%
_{\nu }^{*}(-g_{\mu \nu })  \nonumber \\
&=&\frac{2}{B_{\pi }^{2}}\left[ (M^{2}-4\mu ^{2})F(\xi )+(2s_{12}-4\mu
^{2}-M^{2})L(\xi )\right] ,  \label{VI.59}
\end{eqnarray}
where $s_{12}=(p_{1}+p_{2})^{2}$. The functions $F(\xi )$ and $L(\xi )$ are
defined in Eqs. (\ref{VI.14}) and (\ref{VI.15}) in terms of the parameter $%
\xi $ which, in turn, is defined in Eq. (\ref{VI.16}). The decay width
becomes 
\begin{eqnarray}
\Gamma (f_{0}\rightarrow \pi ^{+}\pi ^{-}\gamma ^{*})&=&\frac{\alpha }{16\pi
^{2}s}g_{f_{0}\pi \pi }^{2}\left| F_{\pi }(M^{2})\right| ^{2}  \nonumber \\
&\times& \int_{4\mu ^{2}}^{(\sqrt{s}-M)^{2}}{\cal R\;}\frac{p^{*}(\sqrt{s},%
\sqrt{s_{12}},M)p^{*}(\sqrt{s_{12}},\mu ,\mu )}{\sqrt{s_{12}}}ds_{12}.
\label{VI.60}
\end{eqnarray}
The dilepton spectrum can be obtained with the use of Eq.(\ref{II.10}). 

\subsection{Decay mode $a_{0}(980)\rightarrow $ $\pi \eta \ell ^{+}\ell ^{-}$%
}


The dominant decay mode of the $a_{0}(980)$-meson is $a_{0}\rightarrow $ $%
\pi \eta .$ The effective vertex for this decay looks like 
\begin{eqnarray}
\delta {\cal L}=g_{a_{0}\pi \eta }a_{0}^{\alpha }\pi ^{\alpha }\eta .
\label{VI.61}
\end{eqnarray}
The $a_{0}\rightarrow $ $\pi \eta $ decay width has the form 
\begin{eqnarray}
\Gamma (a_{0}\rightarrow \pi \eta )=\frac{1}{8\pi s}g_{a_{0}\pi \eta
}^{2}p^{*}(\sqrt{s},\mu ,\mu ^{\prime })  \label{VI.62}
\end{eqnarray}
where $\mu $ and $\mu ^{\prime }$ are the pion and the $\eta $-meson masses.
The experimental width $\Gamma (a_{0}\rightarrow \pi \eta )$ lies within the
interval $50-100$ MeV. The value of the coupling constant $g_{a_{0}\pi \eta}$
determined from the width is given in Table 1.

Since the $\eta $-meson is neutral, photons can only be emitted from the $%
a_{0}$- and $\pi $-mesons in an isospin $I_{3}=\pm 1$ state. To the lowest
order in $\alpha$, $\Gamma (a_{0}^{0}\rightarrow \pi ^{0}\eta \gamma )=0.$
Two diagrams for the $a_{0}^{+}\rightarrow $ $\eta \pi ^{+}\gamma ^{*}$
decays are shown on Fig. \ref{fig14}. The matrix element can be written in
the form 
\begin{eqnarray}
{\cal M}=-ieg_{a\eta \pi }{\cal M}_{\mu }\varepsilon _{\mu }^{*}(k)
\label{VI.63}
\end{eqnarray}
with 
\begin{eqnarray}
{\cal M}_{\mu }=\frac{(2P-k)_{\mu }}{(P-k)^{2}-s}+\frac{(2p_{1}+k)_{\mu }}{%
(p_{1}+k)^{2}-\mu ^{2}}  \label{VI.64}
\end{eqnarray}
where $p_{1}$ is the pion momentum, $\sqrt{s}=m_{a_{0}}$ is the $a_{0}$%
-meson mass, and $P^{2}=s$. The tensor part of the matrix element, ${\cal M}%
_{\mu }$, is transverse with respect to the photon momentum. The square of
the matrix element summed up over photon polarizations and averaged over the
directions of the pion momentum in the c. m. frame of the $\pi $- and $\eta $%
-mesons has the form 
\begin{eqnarray}
{\cal R} &\equiv &\int \frac{d\Omega _{12}}{4\pi }{\cal M}_{\mu }{\cal M}%
_{\nu }{}^{*}(-g_{\mu \nu })  \nonumber \\
&=&-\frac{1}{(s_{12}-s)^{2}}\left[ 2s+2s_{12}-M^{2}\right] -\frac{1}{B_{\pi
}^{\prime 2}}\left[ (4\mu ^{2}-M^{2})F(\xi ^{\prime })+2B_{\pi }^{\prime
}L(\xi ^{\prime })\right]  \nonumber \\
&&-\frac{2}{(s_{12}-s)B_{\pi }^{\prime }}\left[ (s+s_{12}-M^{2}+2(\mu
^{2}-\mu ^{\prime 2}))L(\xi ^{\prime })+B_{\pi }^{\prime }\right]
\label{VI.65}
\end{eqnarray}
where $s_{12}=(p_{1}+p_{2})^{2},$ $p_{2}$ is the $\eta $ momentum, $%
p_{2}^{2}=\mu ^{\prime 2}$, and $F(\xi ^{\prime })$ and $L(\xi ^{\prime })$
are functions of the parameter 
\begin{eqnarray}
\xi ^{\prime }=\frac{2}{B_{\pi }^{\prime }}\sqrt{\frac{s}{s_{12}}}p^{*}(%
\sqrt{s_{12}},\mu ,\mu ^{\prime })p^{*}(\sqrt{s},\sqrt{s_{12}},M).
\label{VI.66}
\end{eqnarray}
The value $B_{\pi }^{\prime }$ is given by 
\begin{eqnarray}
B_{\pi }^{\prime }=\frac{1}{2}(s+M^{2}-s_{12})+\frac{1}{2s_{12}}(\mu
^{2}-\mu ^{\prime 2})(s-M^{2}-s_{12}).  \label{VI.67}
\end{eqnarray}

The $a_{0}^{+}\rightarrow \pi ^{+}\eta \gamma ^{*}$ decay width becomes 
\begin{eqnarray}
\Gamma (a_{0}^{+}\rightarrow \pi ^{+}\eta \gamma ^{*})&=&\frac{\alpha }{%
16\pi ^{2}s}g_{a_{0}\pi \eta }^{2}\left| F_{\pi }(M^{2})\right| ^{2} 
\nonumber \\
&\times&\int_{(\mu +\mu ^{\prime })^{2}}^{(\sqrt{s}-M)^{2}}{\cal R\;}\frac{%
p^{*}(\sqrt{s},\sqrt{s_{12}},M)p^{*}(\sqrt{s_{12}},\mu ,\mu ^{\prime })}{%
\sqrt{s_{12}}}ds_{12}.  \label{VI.68}
\end{eqnarray}
The dilepton spectrum can be calculated using Eq.(\ref{II.10}): 
\begin{eqnarray}
d\Gamma (a_{0}^{+}\rightarrow \pi ^{+}\eta e^{+}e^{-})=\Gamma
(a_{0}^{+}\rightarrow \pi ^{+}\eta \gamma ^{*})M\Gamma (\gamma
^{*}\rightarrow e^{+}e^{-})\frac{dM^{2}}{\pi M^{4}}  \label{VI.69}
\end{eqnarray}
The possible values of $M$ lie in the interval $2m\leq M\leq \sqrt{s}-\mu
-\mu ^{\prime }$. 

\section{Numerical results}

\setcounter{equation}{0}

The dilepton meson decays have essentially the same physical origin as the
radiative meson decays. Eq.(\ref{II.3}) allows a direct calculation of the
radiative widths within the framework used for the calculation of the
dilepton widths. Thus, the radiative decays provide a useful test for the
present model.

\subsection{Radiative meson decays}

The results for the radiative decays with two mesons in the final states are
shown in Table 2. The $\rho ^0\rightarrow \pi ^{+}\pi ^{-}\gamma $, $\omega
\rightarrow \pi ^0\pi ^0\gamma $, and $\eta ^{\prime }\rightarrow \pi
^{+}\pi ^{-}\gamma $ branching ratios are in the reasonable agreement with
experiment. The branching of the $\eta \rightarrow \pi ^{+}\pi ^{-}\gamma $
decay is higher than the experimental value. The $\eta $-meson width is
known with an accuracy of $10\%$. This uncertainty is also present in the
calculated branching. The coupling constant $f_{\rho \gamma \eta }^2$ to
which this branching is proportional is known with an accuracy of $20\%$. It
is, however, not possible to use these uncertainties in order to decrease
the predicted branching, since the branching for the $\eta \rightarrow \pi
^{+}\pi ^{-}e^{+}e^{-}$ decay is lower than the experimental value (see
Table 3). The measurement of the $\eta \rightarrow \pi ^{+}\pi
^{-}e^{+}e^{-} $ reaction has been done quite a long time ago and with
rather pure statistics (see Ref. \cite{RAG66}). It would be desirable to
remeasure this process. The experimental branching for the $\eta ^{\prime
}\rightarrow \pi ^{+}\pi ^{-}\gamma $ decay is taken from Ref. \cite{SIB},
others are from \cite{PDG}. The results for the $\rho ^0\rightarrow \pi
^0\pi ^0\gamma $, $\rho \rightarrow \pi ^0\eta \gamma $, $\omega \rightarrow
\pi ^0\pi ^0\gamma $, and $\omega \rightarrow \pi ^0\eta \gamma $ decays are
in agreement with the calculations by Bramon, Grau and Pancheri \cite{ABram}
where the same model for these decays is used. The branching ratios for
decays including photon bremsstrahlung are calculated for photon energies
above $50$ MeV. The data on the radiative decays with one meson in the final
state were used as input to fix the $V\gamma P$ couplings. These decays are
not listed in Table 2.

The results for the dilepton branching ratios are summarized in Table 3.

\subsection{$\rho $-meson decays}

The branching ratio $B(\rho ^{0}\rightarrow \pi ^{+}\pi ^{-}\gamma )=(9.9\pm
1.6)\times 10^{-3}$ is almost one order of magnitude greater than the
branching ratio $B(\rho ^{0}\rightarrow \pi ^{0}\gamma )=(7.9\pm 2.0)\times
10^{-4}$. Despite the fact that the first one depends on the experimental
cut in the photon energy, one can expect that the decay mode $\rho
^{0}\rightarrow $ $\pi ^{+}\pi ^{-}e^{+}e^{-}$ is important. Indeed, it is
seen from Table 3 that the $\rho ^{0}\rightarrow \pi ^{+}\pi ^{-}e^{+}e^{-}$
branching ratio is more than one order of magnitude greater than the $\rho
\rightarrow \pi e^{+}e^{-}$ branching ratio and $4$ times greater than
branching ratio of the direct $\rho ^{0}\rightarrow e^{+}e^{-}$ decay. The
dominant contribution comes, however, from the region of small invariant
masses of the $e^{+}e^{-}$ pair, which is not seen experimentally. The same
is true for the $\rho ^{\pm }\rightarrow \pi ^{\pm }\pi ^{0}e^{+}e^{-}$
decays. For invariant masses $M>100$ MeV, the number of the $e^{+}e^{-}$
pairs due to the $\rho ^{0}\rightarrow \pi ^{+}\pi ^{-}e^{+}e^{-}$ and $\rho
^{\pm }\rightarrow \pi ^{\pm }\pi ^{0}e^{+}e^{-}$ decays increases by $30\%$
as compared to the direct mode, if one assumes that the $\rho ^{0}$- and $%
\rho ^{\pm }$-mesons are produced with equal probabilities. The differential
branching ratios for the $\rho $-meson decays are shown for the $e^{+}e^{-}$
channels in Fig. \ref{fig15} (a) and for the $\mu ^{+}\mu ^{-}$ channels in
Fig. \ref{fig15} (b). At small invariant masses such that $2m_{e}\lesssim M$%
, the slope equals $d\log B^{\prime}/d\log M\approx -1$ where $%
B^{\prime}=dB/dM$, when there is no bremsstrahlung. Such a behavior appears
because of the $dM/M $ dependence of the differential branching ratio
according to Eq.(\ref{II.10}). The bremsstrahlung makes the ${\sf M}%
\rightarrow {\sf M}^{\prime }\gamma ^{*}$ widths more singular ($\sim \log
^{3}(1/M)$) and thus the slope of the other curves is larger.

In order to get a comparison with the direct channel, we plotted also a
weighted $\rho ^{0}\rightarrow e^{+}e^{-}$ distribution according to
equation 
\begin{eqnarray}
\frac{dB}{dM}=\frac{1}{\pi }\frac{2Mm_{\rho }\Gamma_{\rho} (M)}{%
(M^{2}-m_{\rho }^{2})^{2}+m_{\rho }^{2}\Gamma_{\rho}(M)^{2}}B(\rho
\rightarrow \ell ^{+}\ell ^{-}).  \label{VII.1}
\end{eqnarray}
The $\rho $-meson width as a function of the value $M$ is assumed to have to
the form 
\begin{eqnarray}
\Gamma_{\rho}(M)=\Gamma _{\rho }\frac{p^{*3}(M,\sqrt{s_0}/2 ,\sqrt{s_0}/2 )}{%
p^{*3}(m_{\rho },\sqrt{s_0}/2 ,\sqrt{s_0}/2 )}\frac{m_{\rho
}^{2}+p^{*2}(m_{\rho },\sqrt{s_0}/2 ,\sqrt{s_0}/2 )}{m_{\rho }^{2}+p^{*2}(M,%
\sqrt{s_0}/2 ,\sqrt{s_0}/2 )}  \label{VII.2}
\end{eqnarray}
where $\Gamma _{\rho }$ is the total experimental width and $\sqrt{s_0} =
2\mu$ is the two-pion threshold energy. In the narrow-width limit, the
integral from the left side of Eq.(\ref{VII.1}) coincides with $B(\rho
\rightarrow e^{+}e^{-})$.

\subsection{$\omega $-meson decays}

The differential branching for the $\omega $-meson decays are shown in Figs. 
\ref{fig16} (a,b). The $\omega $ decays are dominated through the $\omega
\rightarrow \pi ^{0}e^{+}e^{-}$ decay mode. The other modes give only a
small correction to the background. It is interesting that strength of the
channel $\omega \rightarrow \pi ^{0}e^{+}e^{-}$ is one order of magnitude
greater than strength of the direct channel $\omega \rightarrow e^{+}e^{-}$.
The $e^{+}e^{-}$ pairs from these two decays have quite different invariant
masses. The background appears below $500$ MeV. Like in the case of the $%
\rho $-meson, we plot also the weighted $\omega \rightarrow \ell ^{+}\ell
^{-}$ distributions according to Eq.(\ref{VII.1}) with substitutions $%
m_{\rho }\leftrightarrow m_{\omega }$ and $\Gamma _{\rho }\leftrightarrow
\Gamma _{\omega }$ for the threshold energy $\sqrt{s_0} = 3\mu$.

\subsection{$\phi $-meson decays}

The differential branching ratios for the $\phi $-meson decays are shown in
Figs. \ref{fig17} (a,b). We do not consider two-meson finals states. The
dominant ones could be $\phi \rightarrow \pi \pi \gamma ^{*}$ and $\phi
\rightarrow \pi ^0\eta \gamma ^{*}$. The first decay with a real photon in
the $\pi ^0\pi ^0$ mode was observed experimentally \cite{MNA,VEPP}. It has
branching $B^{\exp }(\phi \rightarrow \pi ^0\pi ^0\gamma )=(1.14\pm
0.22)\times 10^{-4}$. This decay dominates through the $f_0\gamma $
mechanism. One can expect that branching $B(\phi \rightarrow \pi ^0\pi
^0e^{+}e^{-})$ is two orders of magnitude smaller than branching of the
radiative decay, and so smaller than all $\phi $-meson branching ratios
listed in Table 3. The $\phi \rightarrow \pi \eta \gamma $ decay was also
measured in the SND experiment \cite{MNA,VEPP} to give $B^{\exp }(\phi
\rightarrow \pi ^0\eta \gamma )=(1.3\pm 0.5)\times 10^{-4}$. The branching $%
B(\phi \rightarrow \pi \eta e^{+}e^{-})$ is also expected to be small. The
measured $\phi \rightarrow \pi \pi \gamma $ and $\phi \rightarrow \pi ^0\eta
\gamma $ widths are significantly greater than predictions of Ref. \cite
{ABram} where formation of the intermediate $f_0\gamma $ and $a_0\gamma $
states is not considered. At present there is, however, yet no experimental
evidence for an $a_0$-meson structure in the $\phi \rightarrow \pi ^0\eta
\gamma $ decay, and thus it is not clear how to interpret the excess in the $%
\phi \rightarrow \pi ^0\eta \gamma $ mode. Consequently, the calculation of
the branching ratios for the dilepton $\phi \rightarrow \pi ^0\eta \ell
^{+}\ell ^{-}$ modes is still unclear.

We do not consider decay modes with kaons in the final states. The invariant
mass of the $e^{+}e^{-}$-pairs in such decays does not exceed $50$ MeV, so
they occur in the region excluded by the experimental cuts $M>50$ MeV. In
Figs. \ref{fig17} (a,b), the $\phi \rightarrow \ell ^{+}\ell ^{-}$
distributions are plotted according to Eq.(\ref{VII.1}) modified for the $%
\phi$-meson. The dominant two-kaon decay channel with $\sqrt{s_0}=2\mu_K$
where $\mu_K$ is the kaon mass and also the three-pion decay channel with $%
\sqrt{s_0}=3\mu$ are taken into account in the $M$-dependent width $%
\Gamma_{\phi}(M)$.

\subsection{$\eta $- and $\eta ^{\prime }$-mesons decays}

The differential branching ratios for the $\eta $- and $\eta ^{\prime }$%
-mesons are shown in Figs. \ref{fig18} (a,b). The $\eta $-meson dominates
through the $\eta \rightarrow \gamma \ell ^{+}\ell ^{-}$ modes. The dominant 
$e^{+}e^{-}$ modes for the $\eta ^{\prime }$-meson are $\eta ^{\prime
}\rightarrow \gamma e^{+}e^{-}$ at $M>250$ MeV and $\eta ^{\prime
}\rightarrow \pi ^{+}\pi ^{-}e^{+}e^{-}$ at $M\lesssim 250$ MeV. The $\eta
^{\prime }\rightarrow \gamma \ell ^{+}\ell ^{-}$ modes display clear
structures connected to the $\omega $- and $\rho $-meson contributions to
the $\eta ^{\prime }\gamma \gamma $ transition form factors. The $\eta
\rightarrow \pi ^{0}\pi ^{0}\ell ^{+}\ell ^{-}$ decay is forbidden by $C$%
-parity conservation, similarly for the $\eta ^{\prime }$-meson.

\subsection{$\pi ^{0}$-, $f_{0}$-, and $a_{0}$-mesons decays}

The differential branching ratios for the $\pi ^{0}$-, $f_{0}$-, and $a_{0}$%
-mesons are shown in Figs. \ref{fig19} (a,b). At $M\lesssim 500$ MeV, the $%
f_{0}$-meson decays dominate through the $f_{0}\rightarrow \pi ^{+}\pi
^{-}\ell ^{+}\ell ^{-}$ mode. The four-body $a_{0}$-meson decay also becomes
dominant with decreasing the values of $M$. The dilepton spectra of the $%
f_{0}$- and $a_{0}$-mesons display clear structures connected to the $\omega 
$- and $\rho $-meson contributions to the transition form factors. 

\section{Conclusion}


The present work is an attempt to approach the DLS puzzle through including
new meson decay channels. We studied decay modes of the light mesons below
the $\phi (1020)$-meson into $e^{+}e^{-}$ and $\mu ^{+}\mu ^{-}$ pairs.
Besides the direct decays and some Dalitz decays, which are well known in
the literature, we presented a systematic investigation of the decay modes
which contribute to the dilepton background and have not yet been taken into
account in previous studies. These are special Dalitz decays as, {\it e.g.} $%
\eta ^{\prime }\rightarrow \gamma \ell ^{+}\ell ^{-}$, $f_0\rightarrow
\gamma \ell ^{+}\ell ^{-}$ and most of the decays to four-body final states.
Although many of these processes are found to give small contributions to
the total dilepton branching ratios, this is not the case for all of them:

We found that in the $\rho ^0$-meson decays, the dominant contribution to
the background below $350$ MeV comes from the $\rho ^0\rightarrow \pi
^{+}\pi ^{-}e^{+}e^{-}$ decay. The decay modes $\rho ^0\rightarrow \pi
^{+}\pi ^{-}e^{+}e^{-}$ and $\rho ^{\pm }\rightarrow \pi ^{\pm }\pi
^0e^{+}e^{-}$ increase the number of the $e^{+}e^{-}$ yield with invariant
mass $M>100$ MeV by about $30\%$ compared to the direct mode $\rho
^0\rightarrow e^{+}e^{-}$ at $M<1$ GeV, if one assumes that production cross
sections for the $\rho ^{\pm }$- and $\rho ^0$-mesons are equal. In case of
the $f_0$-meson, the $f_0\rightarrow \pi ^{+}\pi ^{-}e^{+}e^{-}$ decay is in
the interval $100-400$ MeV one to three orders of magnitude more important
than the next to the dominant $f_0\rightarrow \gamma e^{+}e^{-}$ mode. In
most other cases, the four-body decays give only small contributions and
thus there neglection appears to be justified.

The calculated branching ratios can further be used for experimental
searches for the dilepton meson decays.

The production rates of mesons in $p+p$, $p+A$, and in particular in
heavy-ion ($A+A$) collisions differ by orders of magnitude and depend
sensitively on the available energy of the system. The relative strength of
specific channels for different mesons can hardly be estimated without
complete transport calculations. However, for every meson we can decide
which channels are more important. The relative strength of e.g. $\rho
^{0}\rightarrow \pi ^{0}\ell ^{+}\ell ^{-}$ and $\rho ^{0}\rightarrow \pi
^{+}\pi ^{-}\ell ^{+}\ell ^{-}$ decay modes can be found without knowing the
cross section for the $\rho ^{0}$-meson production.

Since theoretical models for the dileption production have to be adjusted to 
$p+p$ and possibly $\pi +p$ reactions before applied to $A+A$ reactions, a
detailed knowledge of all relevant decay channels and background processes
is indispensable. In particular, the lack in the understanding of the excess
of dilepton pairs below the $\rho $-meson peak in the BEVALAC as well as in
the CERN data shows that scenarios with a reduced $\rho $-meson mass can be
treated at present as a plausible hypothesis only. To draw definite physical
conclusions, one needs to eliminate possible trivial explanations like those
connected to existence of the nondirect dilepton decays of light unflavored
mesons.

After submitting this paper, we got to know on papers by Koch \cite{koch93}
and Lichard \cite{lichard95} where a quite complete set of the Dalitz decays
(without scalar mesons) and also a few four-body channels were investigated.
In Ref. \cite{koch93}, four-body channels are discussed within an
approximation suggested by Jarlskog and Pilkuhn \cite{JP67}, which consists
in neglecting the $M$-dependence of the matrix elements for emission of the
virtual photons with mass $M$. We performed straightforward calculations
without approximations and included additional Dalitz and four-body
channels. In Ref. \cite{lichard95}, the decay mode $\rho ^{0}\rightarrow \pi
^{+}\pi ^{-}e^{+}e^{-}$ is analyzed in details. Our expressions (VI.13) and
(VI.18) coincide with the corresponding formulae from the Lichard paper
where, however, the pion form factor was neglected. The pion form factor has
a small impact on the total $e^{+}e^{-}$ decay rate. The dimuon mode, where
higher invariant masses are involved, is enhanced by about 40\%.

\begin{acknowledgments}
The authors are indebted to A. Bramon for useful remarks.  
M.I.K. is grateful to the Institute for Theoretical 
Physics of University of T\"ubingen for kind hospitality. 
The work is supported by GSI (Darmstadt) under the contract T\"UF\"AST.
\end{acknowledgments}
\newpage



%
%
\newpage

\begin{table}[tbp]
\begin{center}
\begin{tabular}{llllllllll}
$f_{\rho \pi \pi }$ & $g_{f\pi \pi }$ & $g_{a\eta \pi }$ & $f_{\rho \omega
\pi }$ & $f_{\omega \gamma \pi }$ & $f_{\rho \gamma \pi }$ & $f_{\omega
\gamma \eta }$ & $f_{\rho \gamma \eta }$ & $f_{\omega \gamma \eta ^{\prime
}} $ & $f_{\rho \gamma \eta ^{\prime }}$ \\ \hline
6.03 & 1.6 & 2.4 & 16 & 2.33 & 0.74 & 0.60 & 1.86 & 1.29 & 0.45
\end{tabular}
\end{center}
\caption{Meson-meson and meson-photon coupling constants. The coupling
constant $f_{\rho \pi \pi }$ is dimensionless, $g_{f\pi \pi }$ and $g_{a\eta
\pi }$ are in units of GeV and $f_{\rho \omega \pi }$ ... $f_{\rho \gamma
\eta ^{\prime }}$ are in units of 1/GeV. The meson-photon coupling constants
are determined from the radiative meson decays. The extracted values are in
good agreement with predictions from the $SU(3)$ symmetry. The first three
coupling constants are determined from the dominant meson decays.}
\label{table1}
\end{table}

\newpage

\begin{table}[tbp]
\begin{center}
\begin{tabular}{lll}
\ Decay mode & $\;\;\;B^{th}$ & $\;\;\;\;\;\;\;B^{\exp }$ \\ \hline
$\rho ^{\pm }\rightarrow \pi ^{\pm }\pi ^{0}\gamma $ & $4.0\times 10^{-3}$ & 
\\ 
$\rho ^{0}\rightarrow \pi ^{+}\pi ^{-}\gamma $ & $1.2\times 10^{-2}$ & $%
(0.99\pm 0.16)\times 10^{-2}$ \\ 
$\rho ^{0}\rightarrow \pi ^{0}\pi ^{0}\gamma $ & $1.2\times 10^{-5}$ &  \\ 
$\rho ^{0}\rightarrow \pi ^{0}\eta \gamma $ & $3.8\times 10^{-10}$ &  \\ 
$\omega \rightarrow \pi ^{+}\pi ^{-} \gamma $ & $3.2\times 10^{-4}$ & $%
<3.6\times 10^{-3}$ \\ 
$\omega \rightarrow \pi ^{0}\pi ^{0} \gamma $ & $3.1\times 10^{-5}$ & $%
(7.2\pm 2.5)\times 10^{-5}$ \\ 
$\omega \rightarrow \pi ^{0}\eta \gamma $ & $2.1\times 10^{-7}$ &  \\ 
$\eta \rightarrow \pi ^{+}\pi^{-} \gamma $ & $6.9\times 10^{-2}$ & $(4.78
\pm 0.12)\times 10^{-2}$ \\ 
$\eta^{\prime} \rightarrow \pi ^{+}\pi^{-} \gamma $ & $2.5\times 10^{-1}$ & $%
(2.8 \pm 0.4)\times 10^{-1}$ \\ 
$f_{0}\rightarrow \pi ^{+}\pi ^{-}\gamma $ & $1.1\times 10^{-2}$ &  \\ 
$a_{0}^{\pm }\rightarrow \pi ^{\pm }\eta \gamma $ & $2.4\times 10^{-3}$ & 
\end{tabular}
\end{center}
\caption{Branching ratios of radiative meson decays. For channels with the
photon bremsstrahlung, the branching ratios are calculated for photon
energies above 50 MeV. The experimental branching of the $\eta^{\prime}
\rightarrow \pi ^{+}\pi^{-} \gamma $ decay is taken from Ref. [52], other
data are from Ref. [35].}
\label{table2}
\end{table}

\newpage

\begin{table}[tbp]
\begin{center}
\begin{tabular}{lllll}
\ Decay mode & $\;\;\;B_{e^{+}e^{-}}^{th}$ & $\;\;\;\;\;\;\;B_{e^{+}e^{-}}^{%
\exp }$ & $\;\;\;B_{\mu ^{+}\mu ^{-}}^{th}$ & $\;\;\;\;\;\;\;B_{\mu ^{+}\mu
^{-}}^{\exp }$ \\ \hline
$\rho ^{0}\rightarrow \ell ^{+}\ell ^{-}$ & input & $(4.48\pm 0.22)\times
10^{-5}$ & $4.\,5\times 10^{-5}$ & $(4.60\pm 0.28)\times 10^{-5}$ \\ 
$\rho \rightarrow \pi \ell ^{+}\ell ^{-}$ & $4.1\times 10^{-6}$ &  & $%
4.6\times 10^{-7}$ &  \\ 
$\rho ^{0}\rightarrow \eta \ell ^{+}\ell ^{-}$ & $2.7\times 10^{-6}$ &  & $%
7.0\times 10^{-11}$ &  \\ 
$\rho ^{\pm }\rightarrow \pi ^{\pm }\pi ^{0}\ell ^{+}\ell ^{-}$ & $5.4\times
10^{-5}$ &  & $1.8\times 10^{-7}$ &  \\ 
$\rho ^{0}\rightarrow \pi ^{+}\pi ^{-}\ell ^{+}\ell ^{-}$ & $1.7\times
10^{-4}$ &  & $6.7\times 10^{-7}$ &  \\ 
$\rho ^{0}\rightarrow \pi ^{0}\pi ^{0}\ell ^{+}\ell ^{-}$ & $7.5\times
10^{-8}$ &  & $2.4\times 10^{-9}$ &  \\ 
$\rho \rightarrow \pi \eta \ell ^{+}\ell ^{-}$ & $1.9\times 10^{-12}$ &  & 
&  \\ \hline
$\omega \rightarrow \ell ^{+}\ell ^{-}$ & input & $(7.15\pm 0.19)\times
10^{-5}$ & $7.1\times 10^{-5}$ & $<1.8\times 10^{-4}$ \\ 
$\omega \rightarrow \pi ^{0}\ell ^{+}\ell ^{-}$ & $7.9\times 10^{-4}$ & $%
(5.9\pm 1.9)\times 10^{-4}$ & $9.2\times 10^{-5}$ & $(9.6\pm 2.3)\times
10^{-5}$ \\ 
$\omega \rightarrow \eta \ell ^{+}\ell ^{-}$ & $6.0\times 10^{-6}$ &  & $%
1.8\times 10^{-9}$ &  \\ 
$\omega \rightarrow \pi ^{+}\pi ^{-}\ell ^{+}\ell ^{-}$ & $3.9\times 10^{-6}$
&  & $2.9\times 10^{-8}$ &  \\ 
$\omega \rightarrow \pi ^{0}\pi ^{0}\ell ^{+}\ell ^{-}$ & $2.0\times 10^{-7}$
&  & $7.4\times 10^{-9}$ &  \\ 
$\omega \rightarrow \pi ^{0}\eta \ell ^{+}\ell ^{-}$ & $8.7\times 10^{-10}$
&  &  &  \\ \hline
$\phi \rightarrow \ell ^{+}\ell ^{-}$ & input & $(3.00\pm 0.06)\times
10^{-4} $ & $3.0\times 10^{-4}$ & $(2.48\pm 0.34)\times 10^{-4}$ \\ 
$\phi \rightarrow \pi ^{0}\ell ^{+}\ell ^{-}$ & $1.6\times 10^{-5}$ & $%
<1.2\times 10^{-4}$ & $4.8\times 10^{-6}$ &  \\ 
$\phi \rightarrow \eta \ell ^{+}\ell ^{-}$ & $1.1\times 10^{-4}$ & $(1.3
\;^{+\;0.8}_{-\;0.6})\times 10^{-4}$ & $6.8\times 10^{-6}$ &  \\ \hline
$\eta \rightarrow \gamma \ell ^{+}\ell ^{-}$ & $6.5\times 10^{-3}$ & $%
(4.9\pm 1.1)\times 10^{-3}$ & $3.0\times 10^{-4}$ & $(3.1\pm 0.4)\times
10^{-4}$ \\ 
$\eta \rightarrow \pi ^{+}\pi ^{-}\ell ^{+}\ell ^{-}$ & $3.6\times 10^{-4}$
& $(1.3\;^{+\;1.2}_{-\;0.8})\times 10^{-3}$ & $1.2\times 10^{-8}$ &  \\ 
\hline
$\eta ^{\prime }\rightarrow \gamma \ell ^{+}\ell ^{-}$ & $4.2\times 10^{-4}$
&  & $8.1\times 10^{-5}$ & $(1.04\pm 0.26)\times 10^{-4}$ \\ 
$\eta ^{\prime }\rightarrow \omega \ell ^{+}\ell ^{-}$ & $2.0\times 10^{-4}$
&  &  &  \\ 
$\eta ^{\prime }\rightarrow \pi ^{+}\pi ^{-}\ell ^{+}\ell ^{-}$ & $1.8\times
10^{-3}$ &  & $2.0\times 10^{-5}$ &  \\ \hline
$f_{0}\rightarrow \gamma \ell ^{+}\ell ^{-}$ & $2.2\times 10^{-7}$ &  & $%
2.8\times 10^{-8}$ &  \\ 
$f_{0}\rightarrow \pi ^{+}\pi ^{-}\ell ^{+}\ell ^{-}$ & $1.4\times 10^{-4}$
&  & $4.1\times 10^{-7}$ &  \\ \hline
$a_{0}^{0}\rightarrow \gamma \ell ^{+}\ell ^{-}$ & $6.0\times 10^{-8}$ &  & $%
7.4\times 10^{-9}$ &  \\ 
$a_{0}\rightarrow \pi \eta \ell ^{+}\ell ^{-}$ & $4.0\times 10^{-5}$ &  & $%
1.4\times 10^{-9}$ &  \\ \hline
$\pi ^{0}\rightarrow \gamma \ell ^{+}\ell ^{-}$ & $1.18\times 10^{-2}$ & $%
(1.198\pm 0.032)\times 10^{-2}$ &  & 
\end{tabular}
\end{center}
\caption{The integral branchings ratios of the unflavored meson decays to
electron-positron and muon-antimuon pairs. The experimental data are from
Ref. [35].}
\label{table3}
\end{table}
%
%
\newpage

\begin{center}
{\bf FIGURE CAPTIONS}
\end{center}

Fig. 1: Direct decays of vector mesons into electron-positron and
muon-antimuon pairs.

Fig. 2: Decays of scalar mesons into a photon and a virtual photon.

Fig. 3: Decays of scalar mesons into a photon and a virtual photon.

Fig. 4: Magnetic dipole transitions $V\rightarrow P\gamma ^{*}$.

Fig. 5: $\eta $-meson decays through a two-step mechanism.

Fig. 6: $\rho ^{0}$-meson decays into a virtual photon and two charged pions.

Fig. 7: $\rho ^{0}$-meson decays into a virtual photon and two neutral pions.

Fig. 8: $\omega $-meson decays into a virtual photon and two neutral pions.

Fig. 9: $\omega $-meson decays into a virtual photon and two charged pions.
The first two diagrams (a) and (b) are gauge invariant. They correspond to
the two-step mechanism of the $\omega $-meson decays. The next two diagrams
(c) and (d) occurring due to the $\omega \rho $ mixing correspond to the
photon bremsstrahlung.. The last diagram (e) restores the gauge invariance
of the previous two diagrams.

Fig.10: $\rho $ -meson decays into a virtual photon and $\eta $- and $\pi $%
-mesons.

Fig.11: $\omega $ -meson decays into a virtual photon and $\eta $- and $\pi
^{0}$-mesons.

Fig.12: $\rho ^{\pm }$-meson decays into a virtual photon and $\pi ^{\pm }$-
and $\pi ^{0}$-mesons. The sum of the diagrams (a), (b), and (c) is gauge
invariant. The diagram (d) is gauge invariant.

Fig.13: $f_{0}(980)$ -meson decays into a virtual photon and two charged
pions.

Fig.14: $a_{0}(980)$ -meson decays into a virtual photon and $\eta $- and $%
\pi $-mesons.


Fig.15: The differential branching ratios for the $\rho $-meson decays into $%
e^{+}e^{-}$ (a) and $\mu ^{+}\mu ^{-}$ (b) channels as functions of the
invariant mass, $M$, of the dilepton pairs. The solid curves (8 and 9) are
the total ratios for the $\rho ^{0}$- and $\rho ^{\pm }$ -mesons.

Fig.16: The differential branching ratios for the $\omega $-meson decays
into $e^{+}e^{-}$ (a) and $\mu ^{+}\mu ^{-}$ (b) channels as functions of
the invariant mass, $M$, of the dilepton pairs. The solid curve is the total
ratio. The background is dominated through the $\pi e^{+}e^{-}$ and $\pi \mu
^{+}\mu ^{-}$ Dalitz decays.

Fig.17: The differential branching ratios for the $\phi $-meson decays into $%
e^{+}e^{-}$ (a) and $\mu ^{+}\mu ^{-}$ (b) channels versus the invariant
mass, $M$, of the dilepton pairs.

Fig.18: The differential branching ratios for the $\eta $- and $\eta
^{\prime }$-mesons decaying into $e^{+}e^{-}$ (a) and $\mu ^{+}\mu ^{-}$ (b)
channels versus the invariant mass, $M$, of the dilepton pairs. The solid
curves give the total ratios. The narrow structure in the $\eta ^{\prime
}\rightarrow \gamma e^{+}e^{-}$ decay is connected to the $\omega $-meson
contribution to the $\eta ^{\prime }\rightarrow \gamma \gamma ^{*}$
transition form factor (see Eq.(IV.5)). The same holds for the $\eta
^{\prime }\rightarrow \gamma \mu ^{+}\mu ^{-}$ decay.

Fig.19: The differential branching ratios for the $\pi ^{0}$-, $f_{0}(980)$%
-, and $a_{0}(980)$-mesons into $e^{+}e^{-}$ (a) and $\mu ^{+}\mu ^{-}$ (b)
channels versus the invariant mass, $M$, of the dilepton pairs. The solid
curves No. 6 (a) and No. 5 (b) give the total branching ratios,
respectively, of the $e^{+}e^{-}$ and $\mu ^{+}\mu ^{-}$ modes of the $f_{0}$%
-meson. The structures in the $f_{0}$- and $a_{0}$-mesons Dalitz decays $%
\gamma \ell ^{+}\ell ^{-}$ are connected to the $\omega $- and $\rho $%
-mesons contributions to the $f_{0}$ and $a_{0}$ transition form factors.

%
%

\newpage

\newpage
\begin{figure}[h]
\begin{center}
\leavevmode
\epsfxsize = 11cm
\epsffile[85 400 460 690]{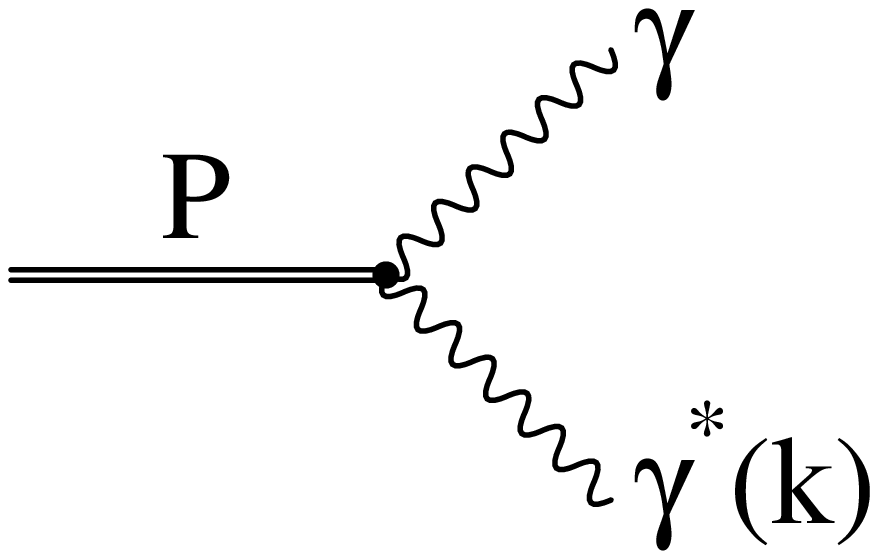} 
\end{center}
\caption{}
\label{fig2}
\end{figure}

\newpage
\begin{figure}[h]
\begin{center}
\leavevmode
\epsfxsize = 11cm
\epsffile[85 400 460 690]{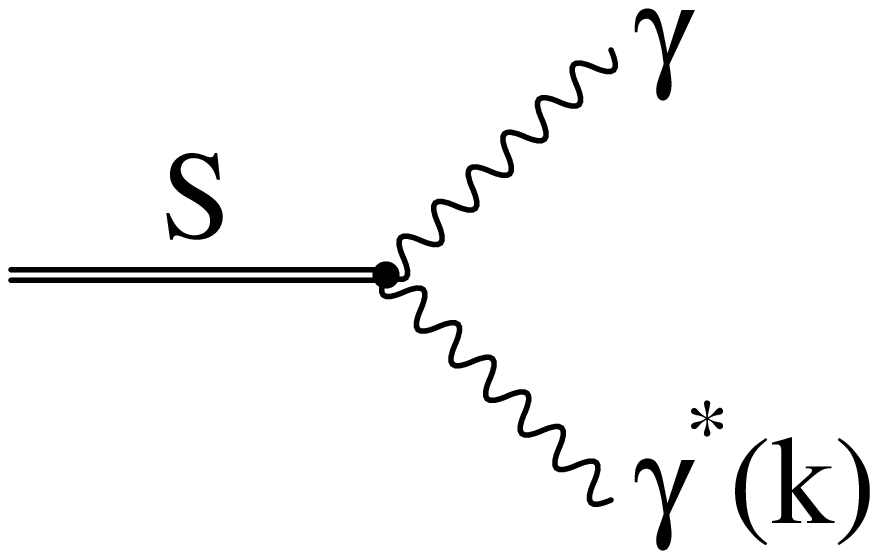} 
\end{center}
\caption{}
\label{fig3}
\end{figure}

\newpage
\begin{figure}[h]
\begin{center}
\leavevmode
\epsfxsize = 11cm
\epsffile[85 400 460 690]{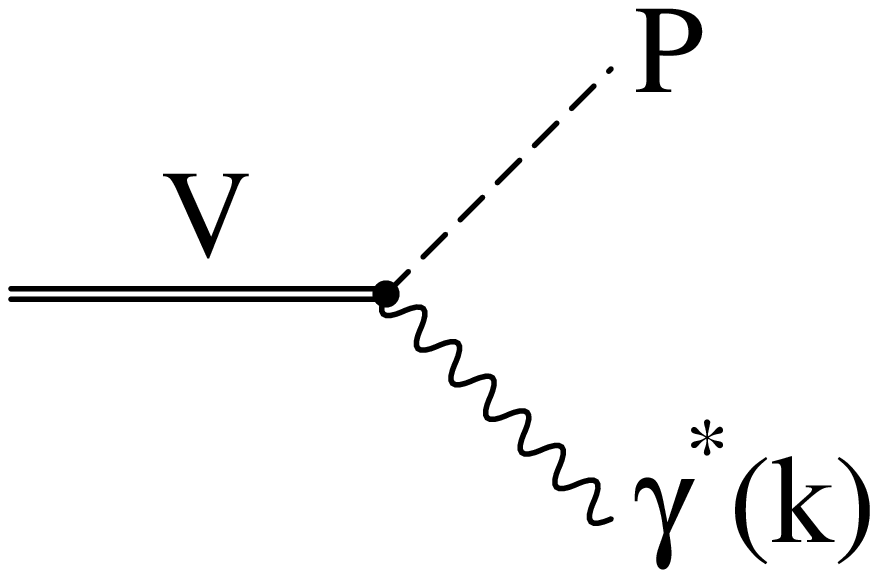} 
\end{center}
\caption{}
\label{fig4}
\end{figure}

\newpage
\begin{figure}[h]
\begin{center}
\leavevmode
\epsfxsize = 11cm
\epsffile[85 400 460 690]{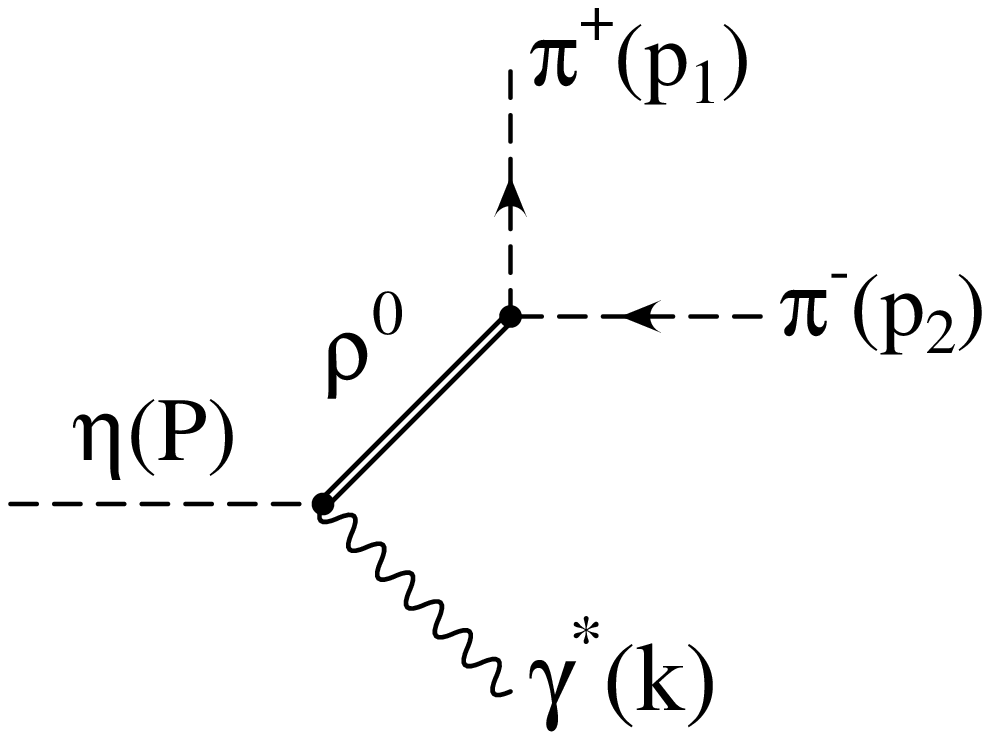} 
\end{center}
\caption{}
\label{fig5}
\end{figure}

\newpage
\begin{figure}[h]
\begin{center}
\leavevmode
\epsfxsize = 15cm
\epsffile[30 210 560 600]{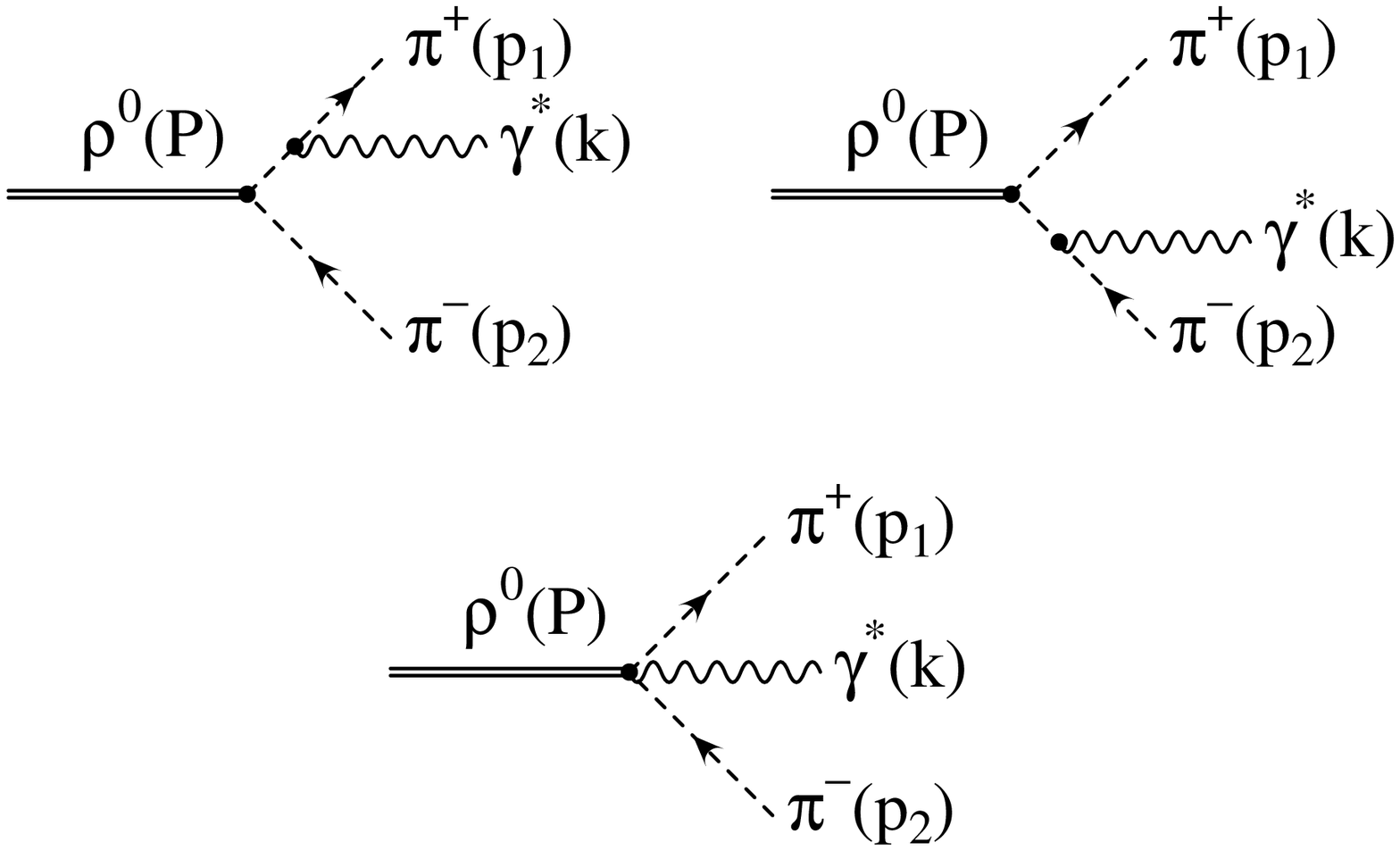}
\end{center}
\caption{}
\label{fig6}
\end{figure}


\newpage
\begin{figure}[h]
\begin{center}
\leavevmode
\epsfxsize = 11cm
\epsffile[130 340 460 670]{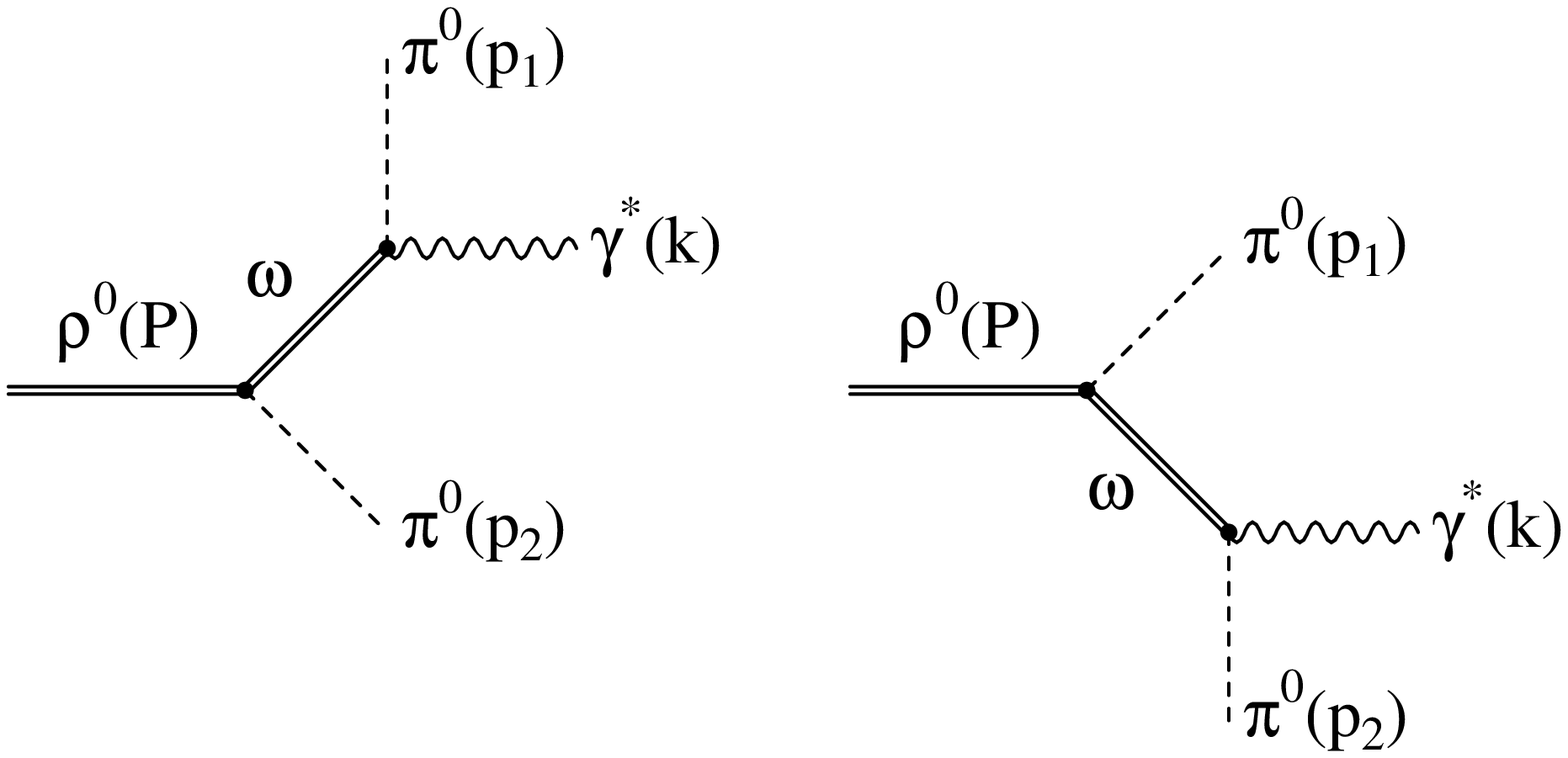}
\end{center}
\caption{}
\label{fig7}
\end{figure}

\newpage
\begin{figure}[h]
\begin{center}
\leavevmode
\epsfxsize = 11cm
\epsffile[130 340 460 670]{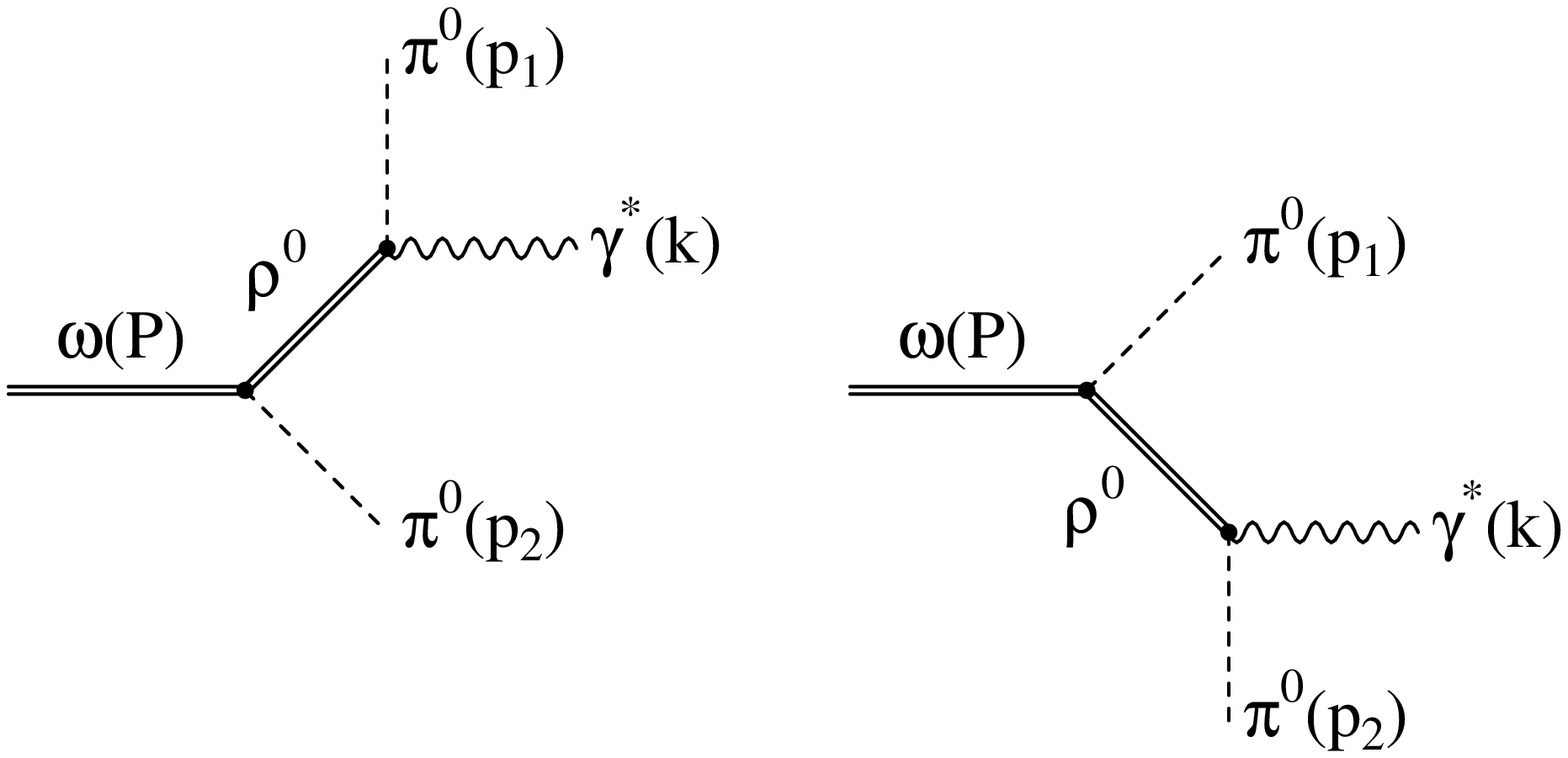}
\end{center}
\caption{}
\label{fig8}
\end{figure}

\newpage
\begin{figure}[h]
\begin{center}
\leavevmode
\epsfxsize = 15cm
\epsffile[10 10 580 740]{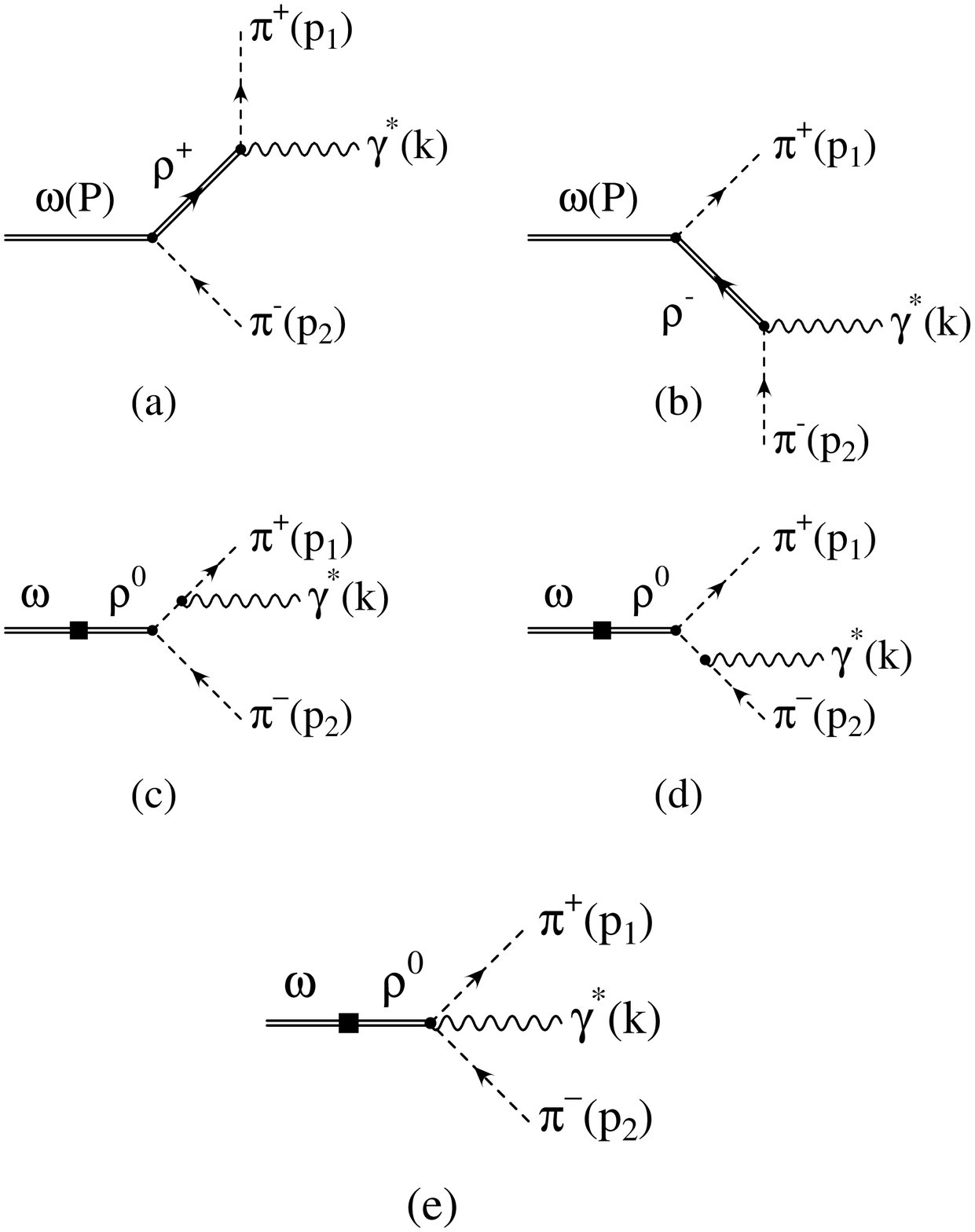} 
\end{center}
\caption{}
\label{fig9}
\end{figure}

\newpage
\begin{figure}[h]
\begin{center}
\leavevmode
\epsfxsize = 11cm
\epsffile[85 400 460 690]{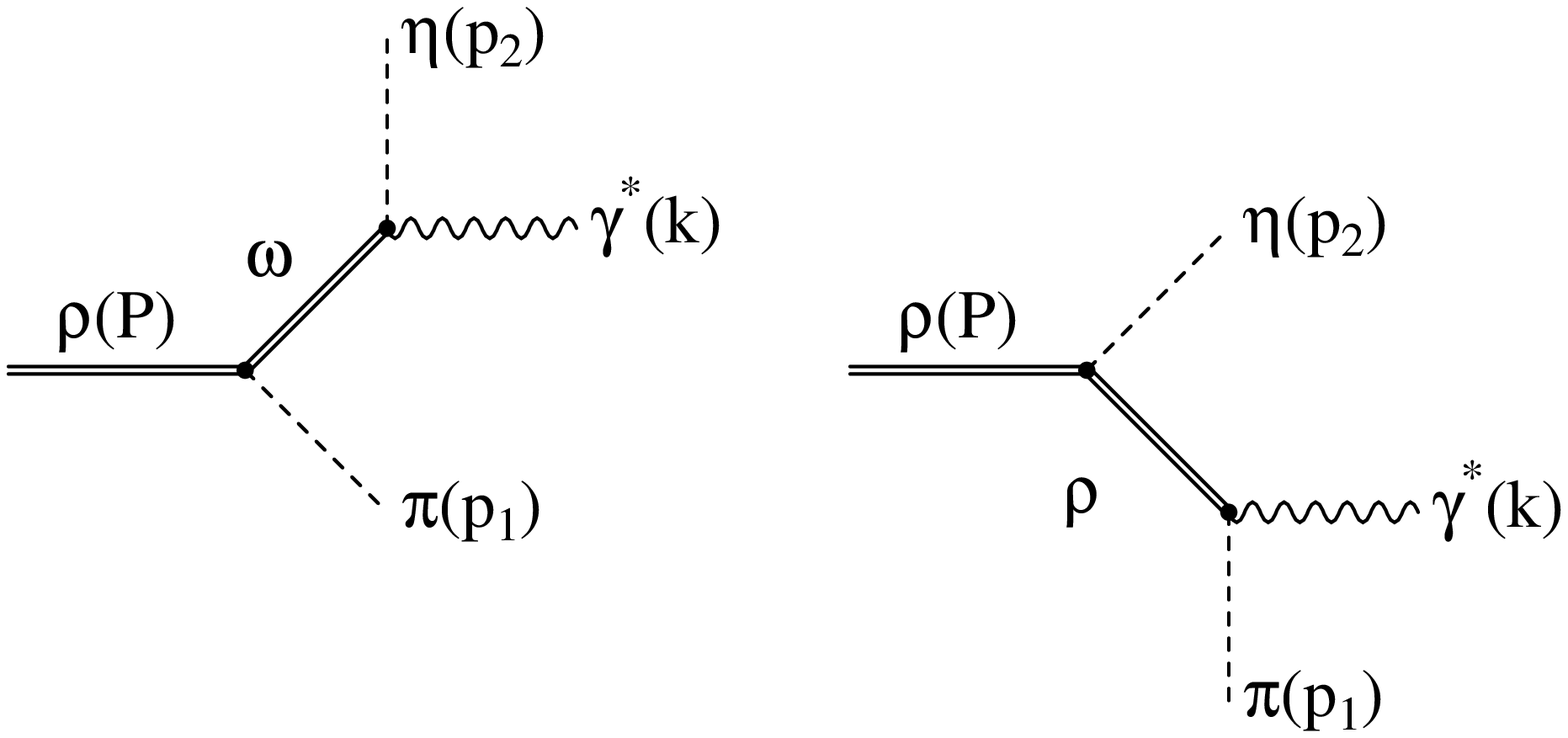} 
\end{center}
\caption{}
\label{fig10}
\end{figure}

\newpage
\begin{figure}[h]
\begin{center}
\leavevmode
\epsfxsize = 11cm
\epsffile[85 400 460 690]{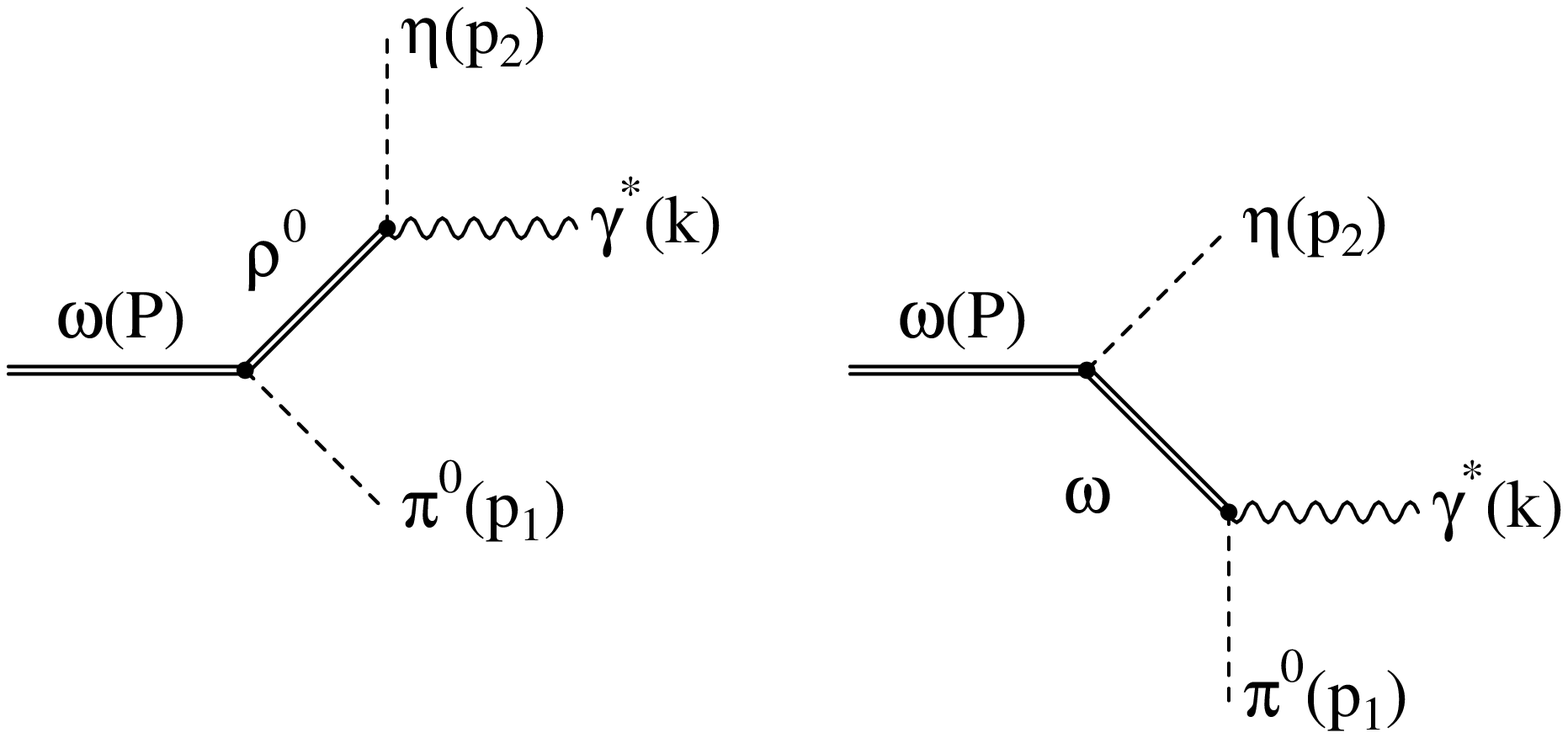} 
\end{center}
\caption{}
\label{fig11}
\end{figure}

\newpage
\begin{figure}[h]
\begin{center}
\leavevmode
\epsfxsize = 15cm
\epsffile[20 150 580 690]{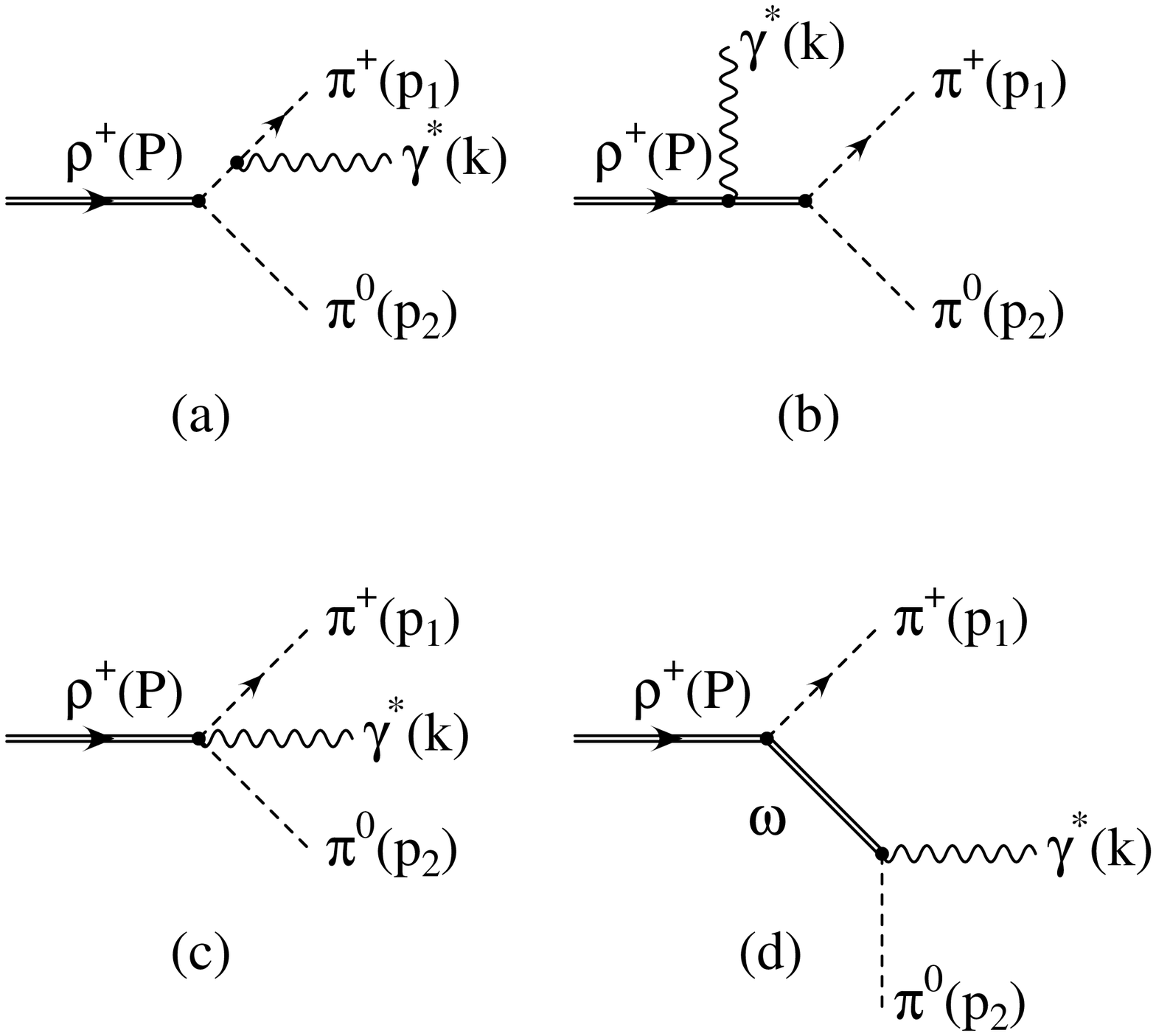} 
\end{center}
\caption{}
\label{fig12}
\end{figure}

\newpage
\begin{figure}[h]
\begin{center}
\leavevmode
\epsfxsize = 11cm
\epsffile[130 340 460 670]{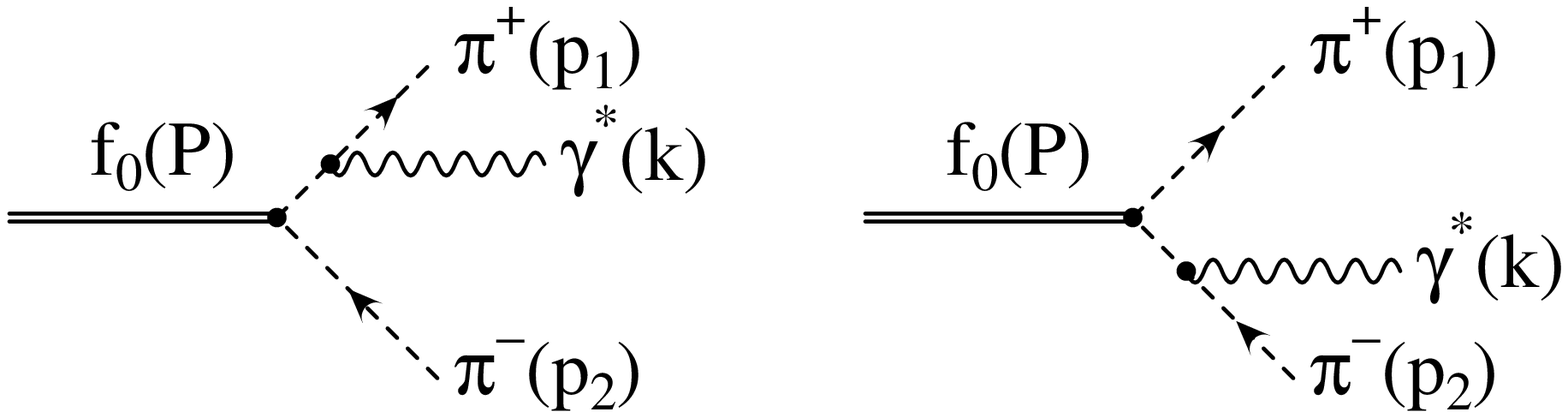}
\end{center}
\caption{}
\label{fig13}
\end{figure}

\newpage
\begin{figure}[h]
\begin{center}
\leavevmode
\epsfxsize = 15cm
\epsffile[20 460 560 690]{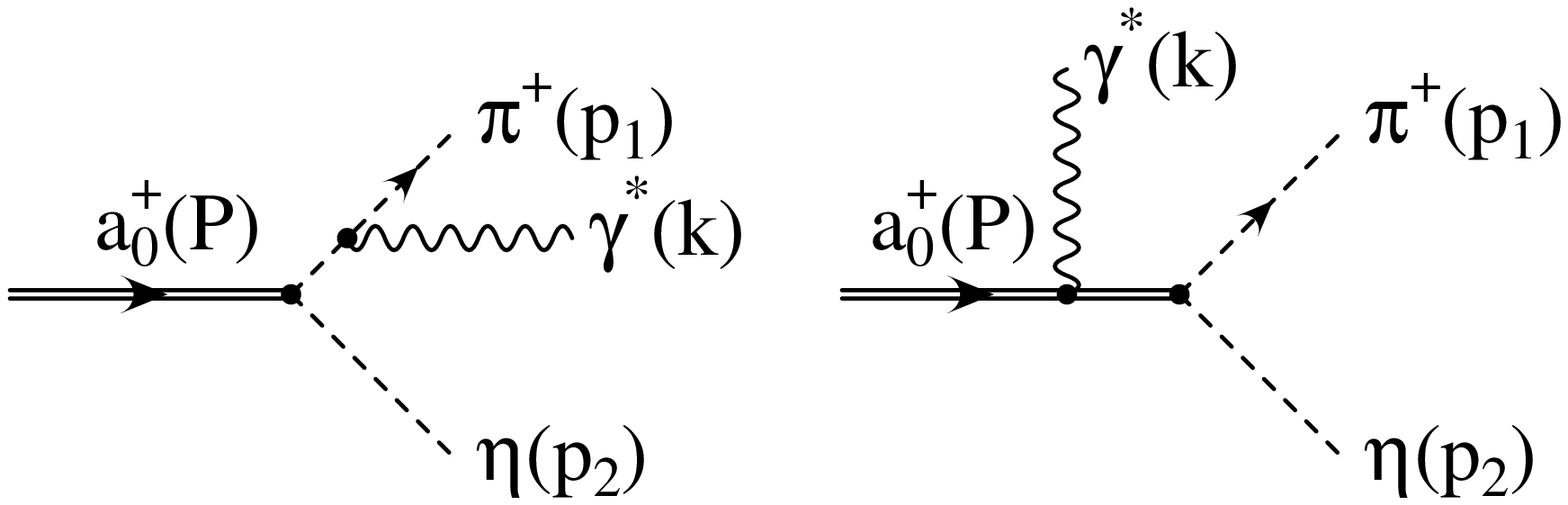}
\end{center}
\caption{}
\label{fig14}
\end{figure}


\newpage
\begin{figure}[h]
\begin{center}
\leavevmode
\epsfxsize = 13cm
\epsffile[40 40 530 410]{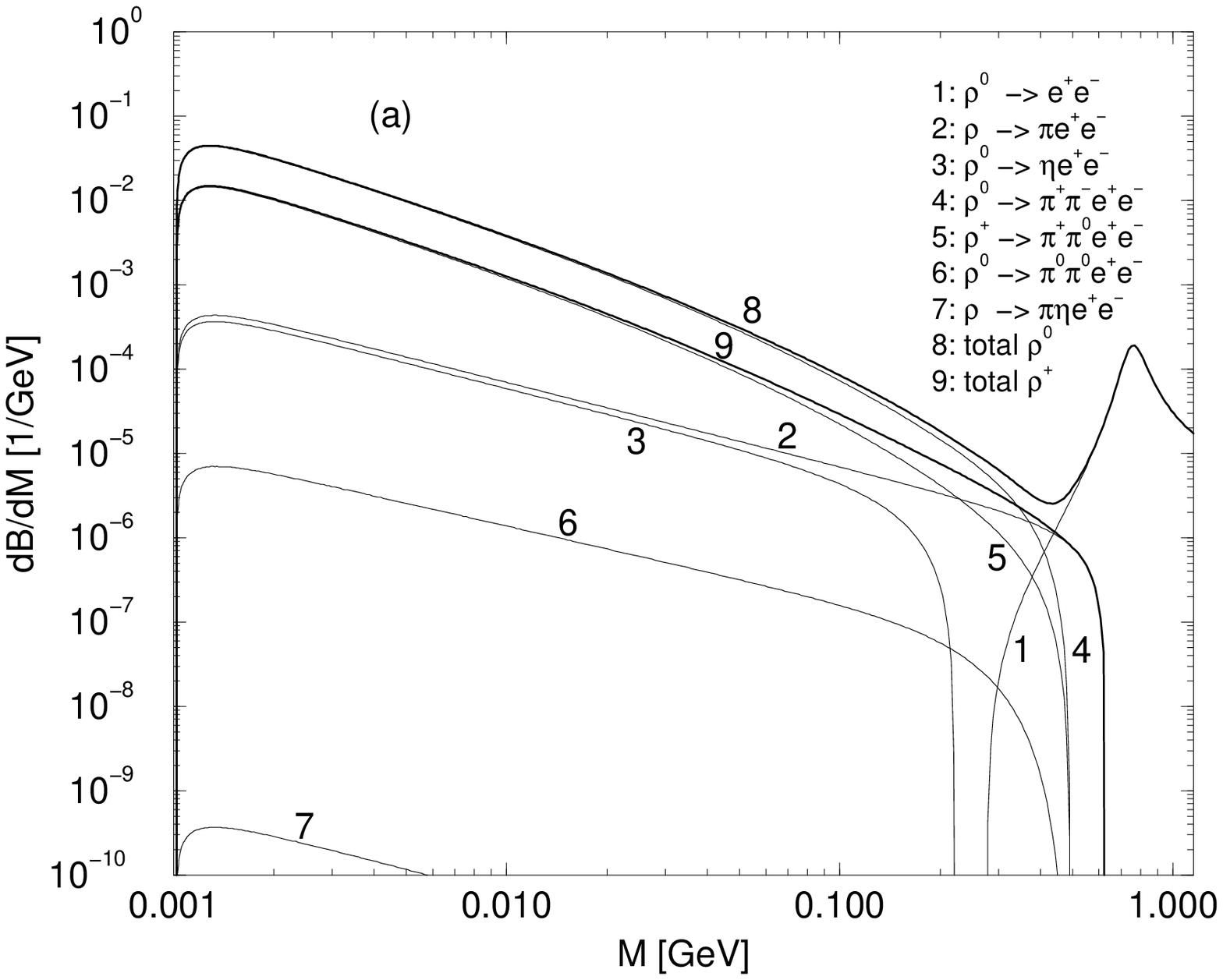} 
\vskip 1cm
\leavevmode
\epsfxsize = 13cm
\epsffile[40 40 530 410]{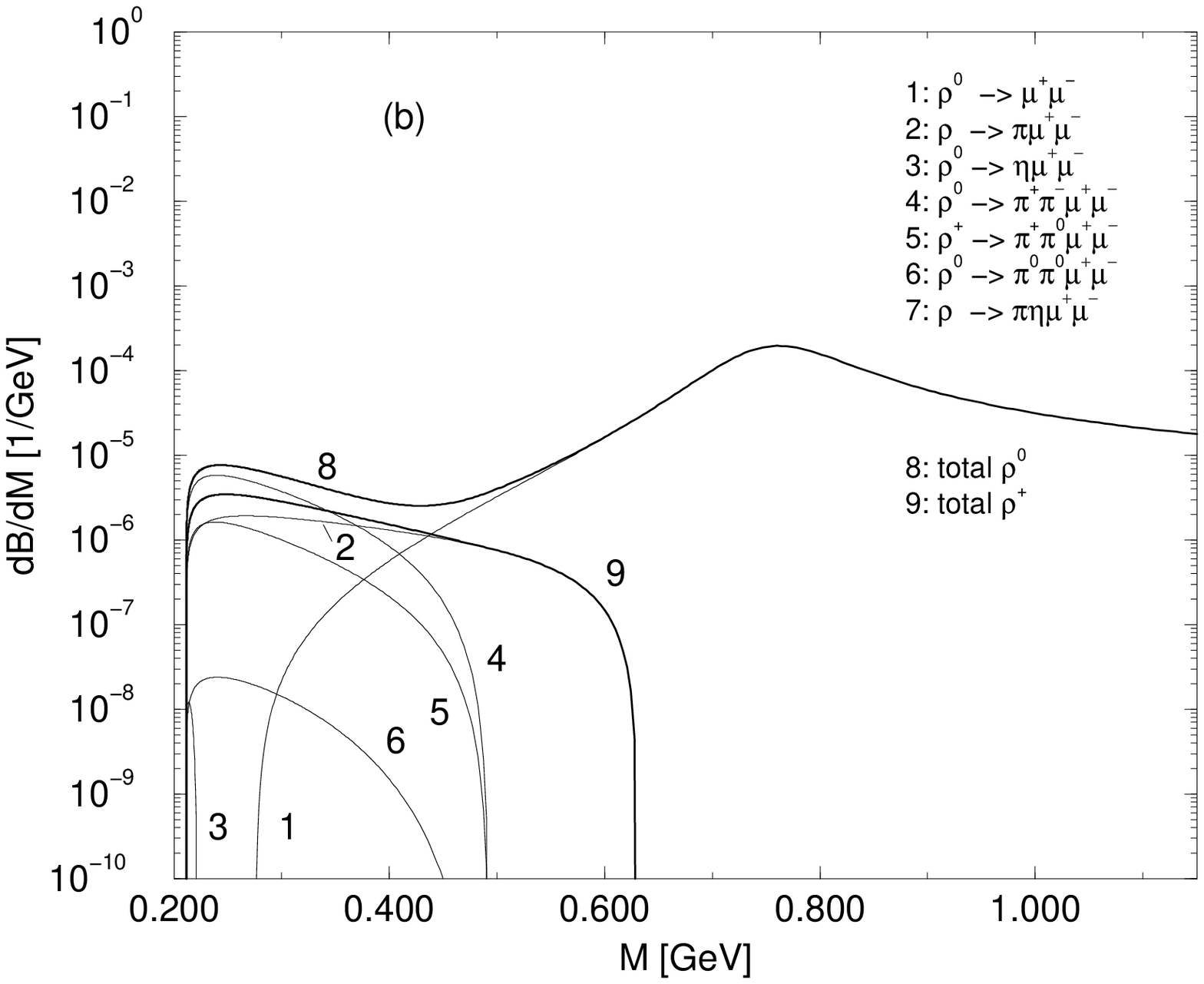} 
\end{center}
\caption{}
\label{fig15}
\end{figure}

\newpage
\begin{figure}[h]
\begin{center}
\leavevmode
\epsfxsize = 13cm
\epsffile[40 40 530 410]{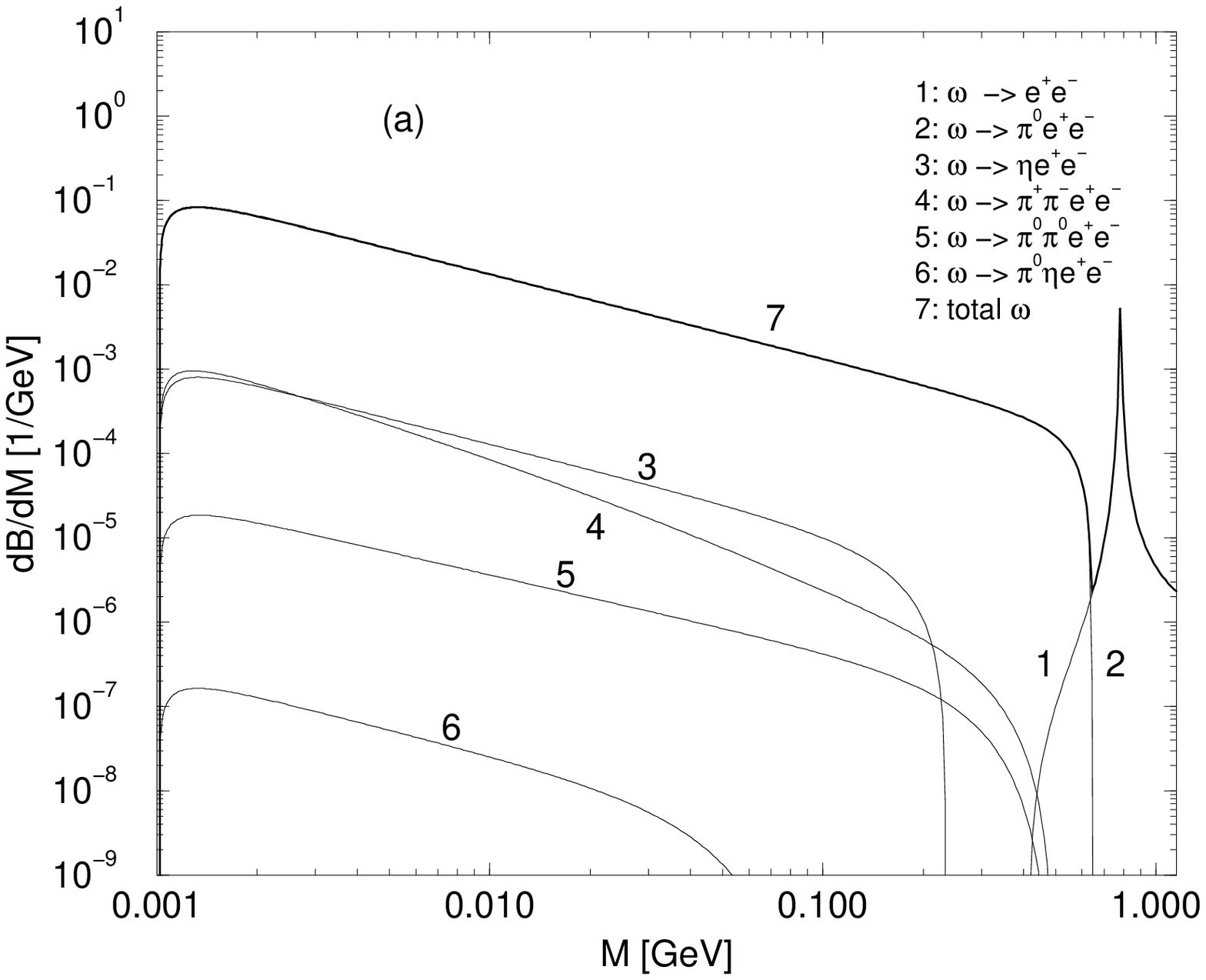} 
\vskip 1cm
\leavevmode
\epsfxsize = 13cm
\epsffile[40 40 530 410]{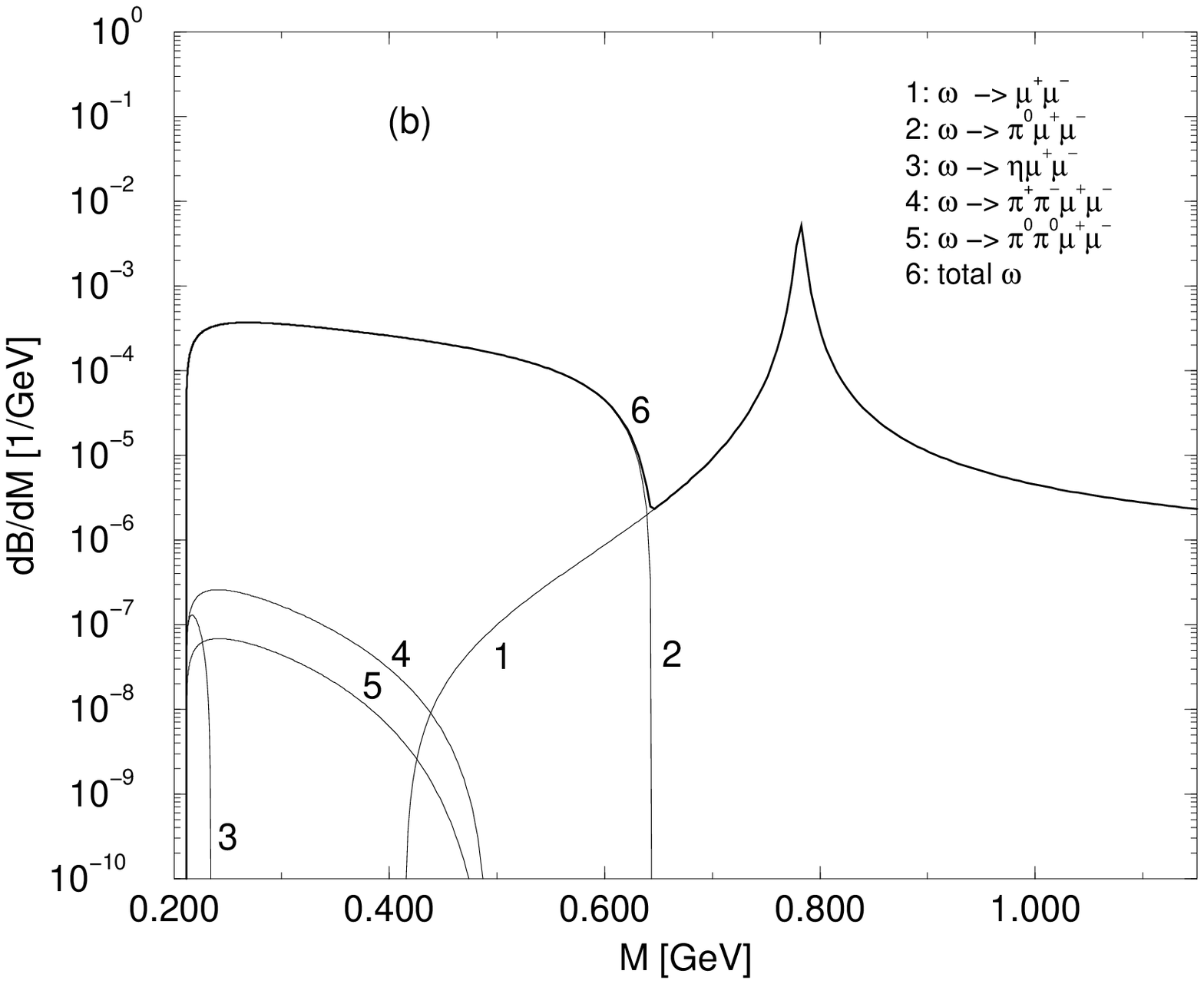} 
\end{center}
\caption{}\label{fig16}
\end{figure}

\newpage
\begin{figure}[h]
\begin{center}
\leavevmode
\epsfxsize = 13cm
\epsffile[40 40 530 410]{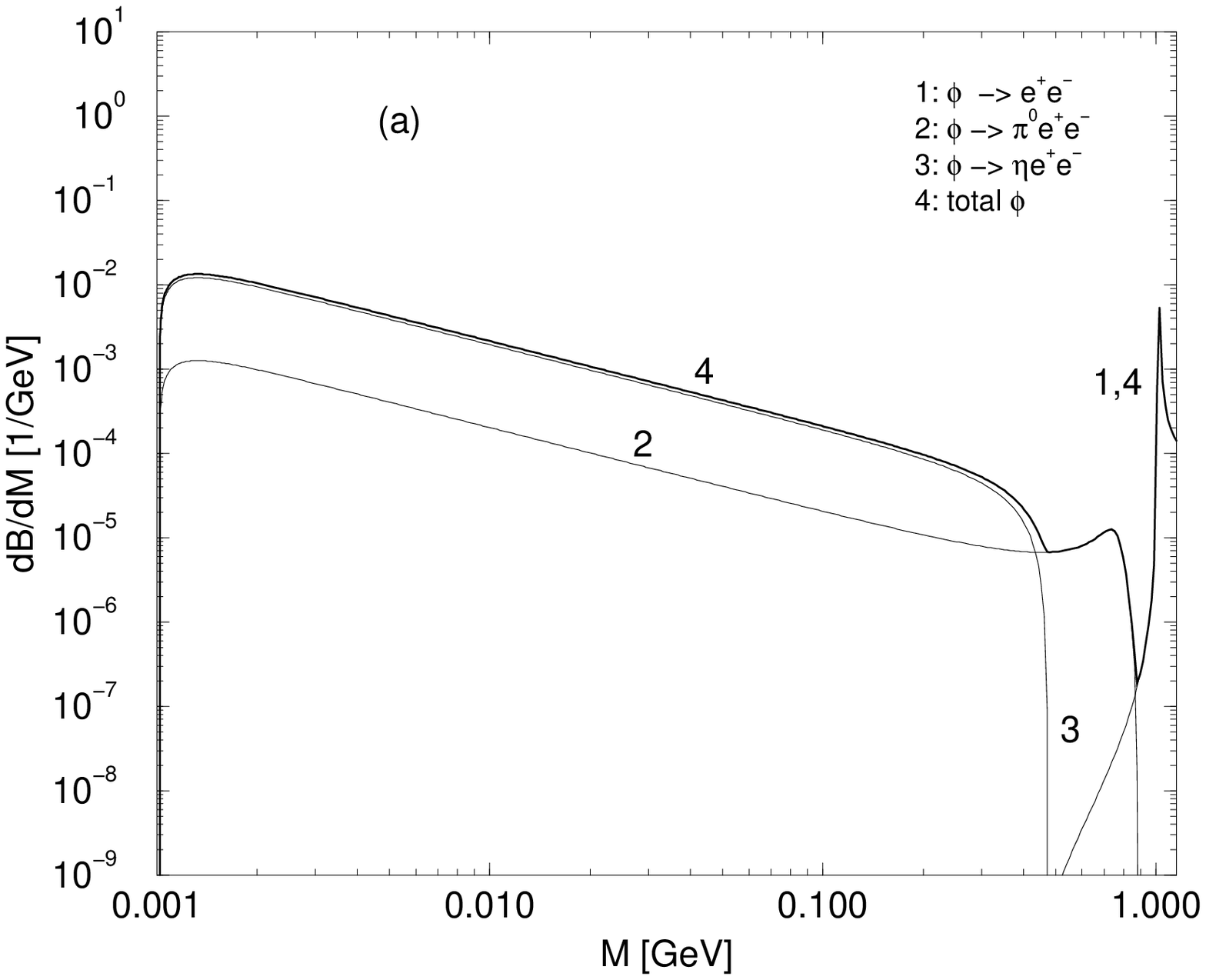} 
\vskip 1cm
\leavevmode
\epsfxsize = 13cm
\epsffile[40 40 530 410]{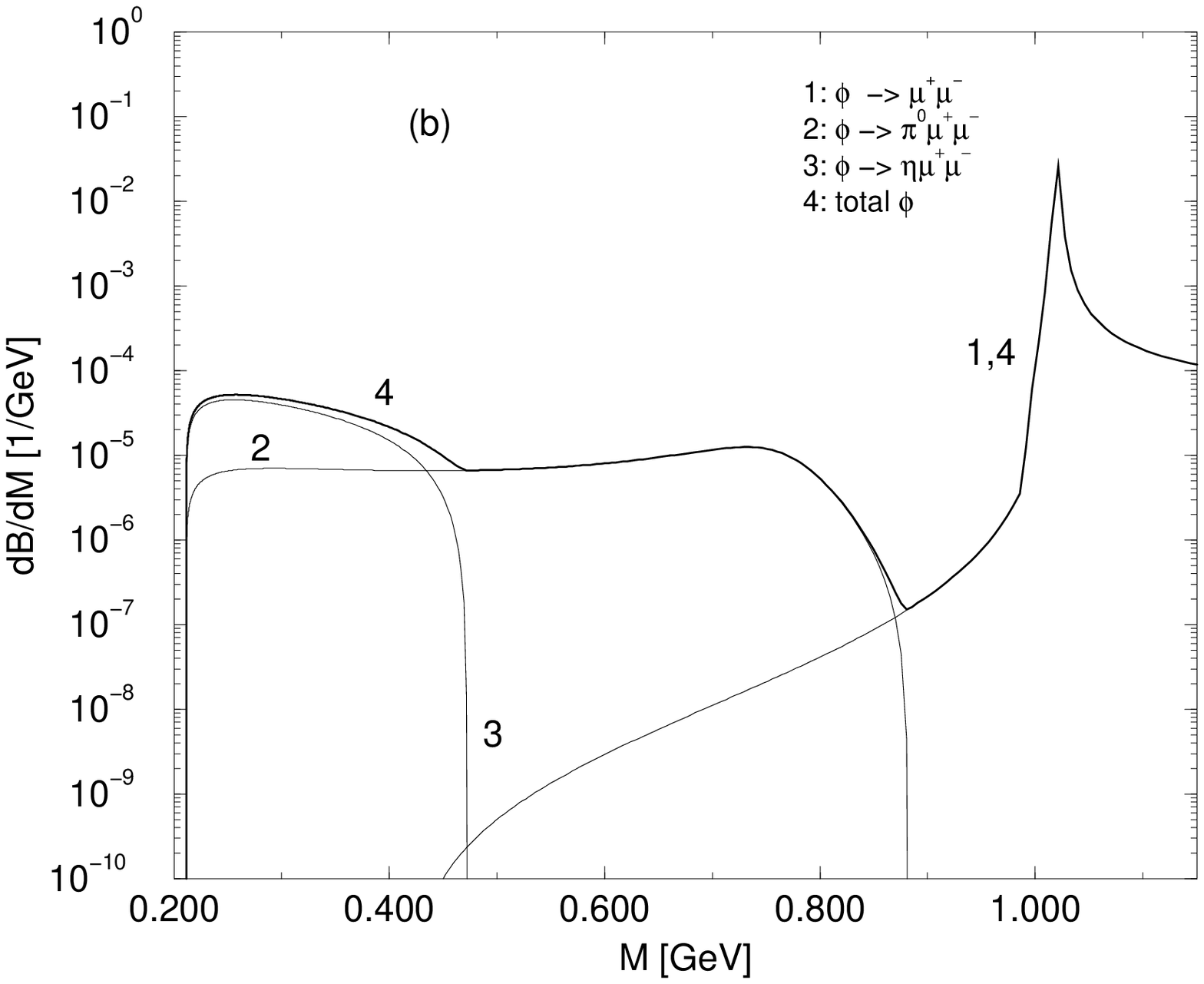} 
\end{center}
\caption{}\label{fig17}
\end{figure}

\newpage
\begin{figure}[h]
\begin{center}
\leavevmode
\epsfxsize = 13cm
\epsffile[40 40 530 410]{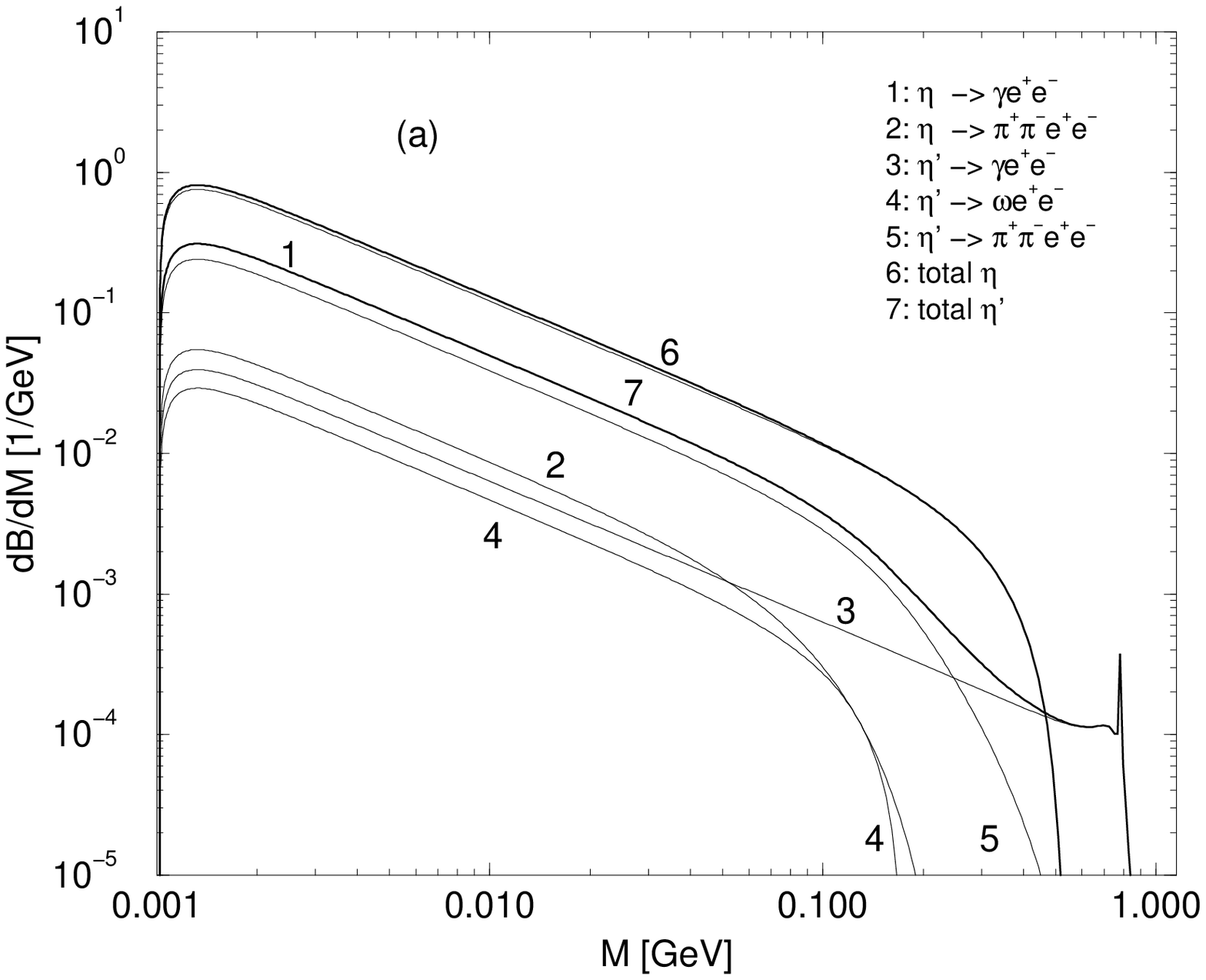} 
\vskip 1cm
\leavevmode
\epsfxsize = 13cm
\epsffile[40 40 530 410]{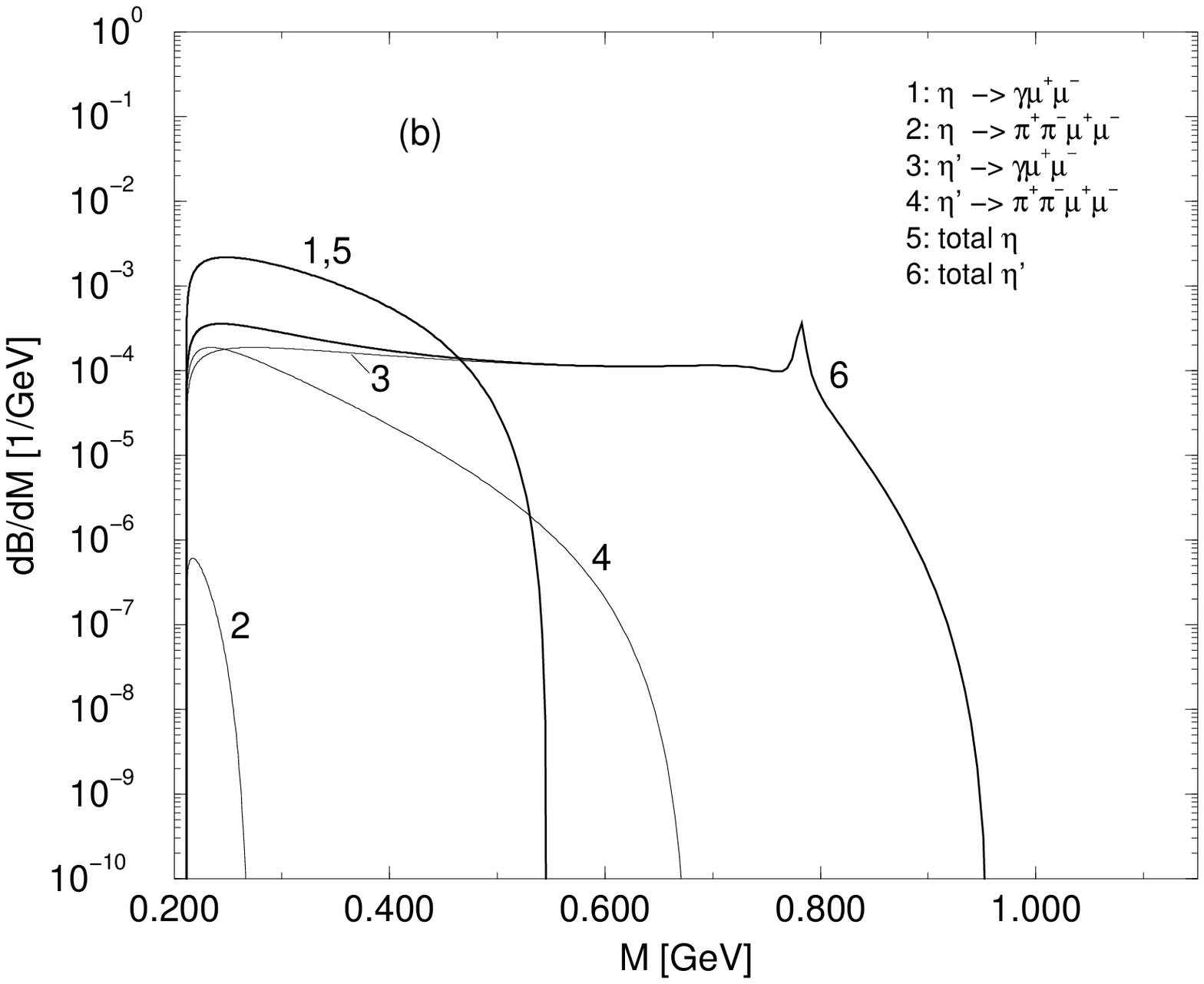} 
\end{center}
\caption{}\label{fig18}
\end{figure}

\newpage
\begin{figure}[h]
\begin{center}
\leavevmode
\epsfxsize = 13cm
\epsffile[40 40 530 410]{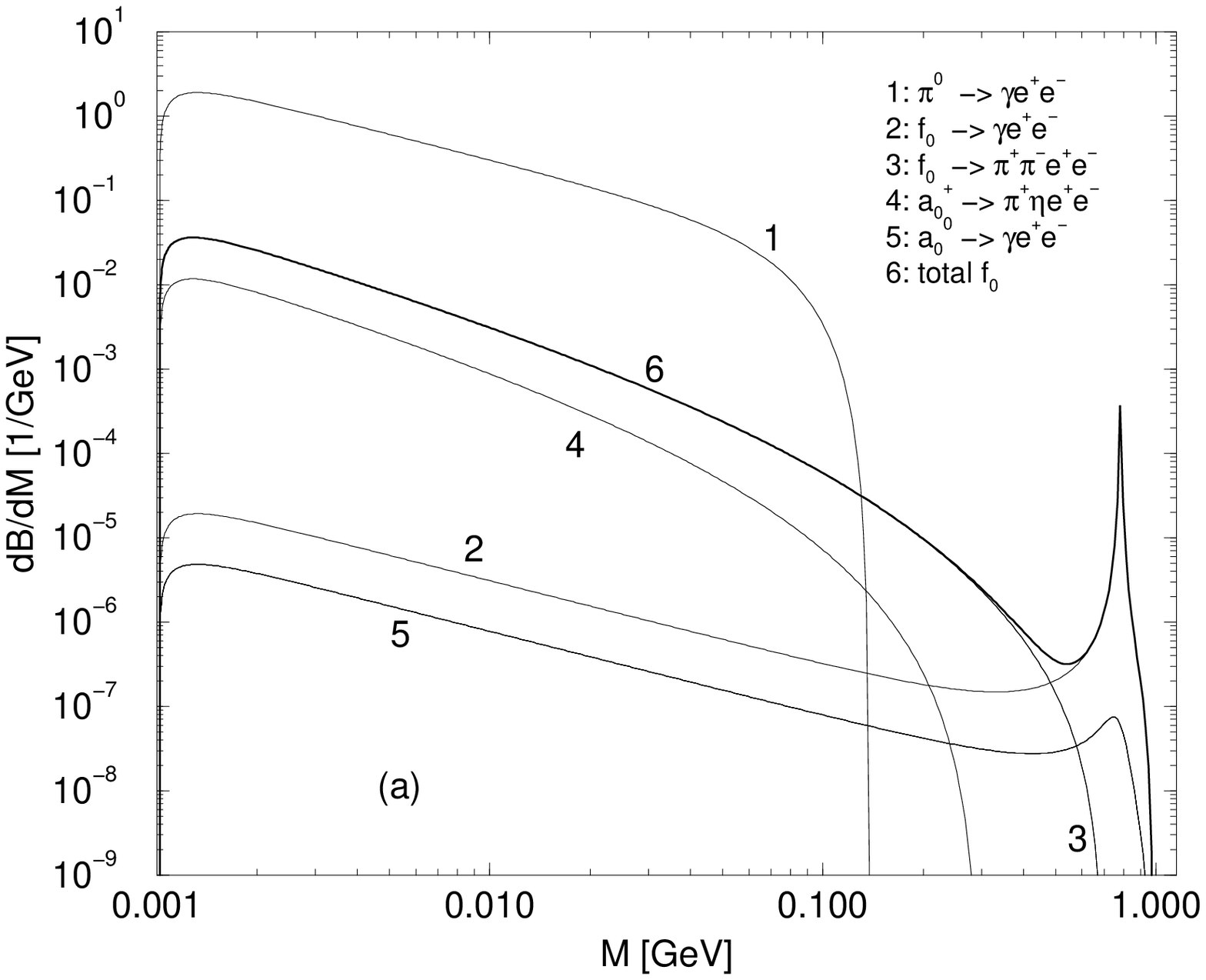} 
\vskip 1cm
\leavevmode
\epsfxsize = 13cm
\epsffile[40 40 530 410]{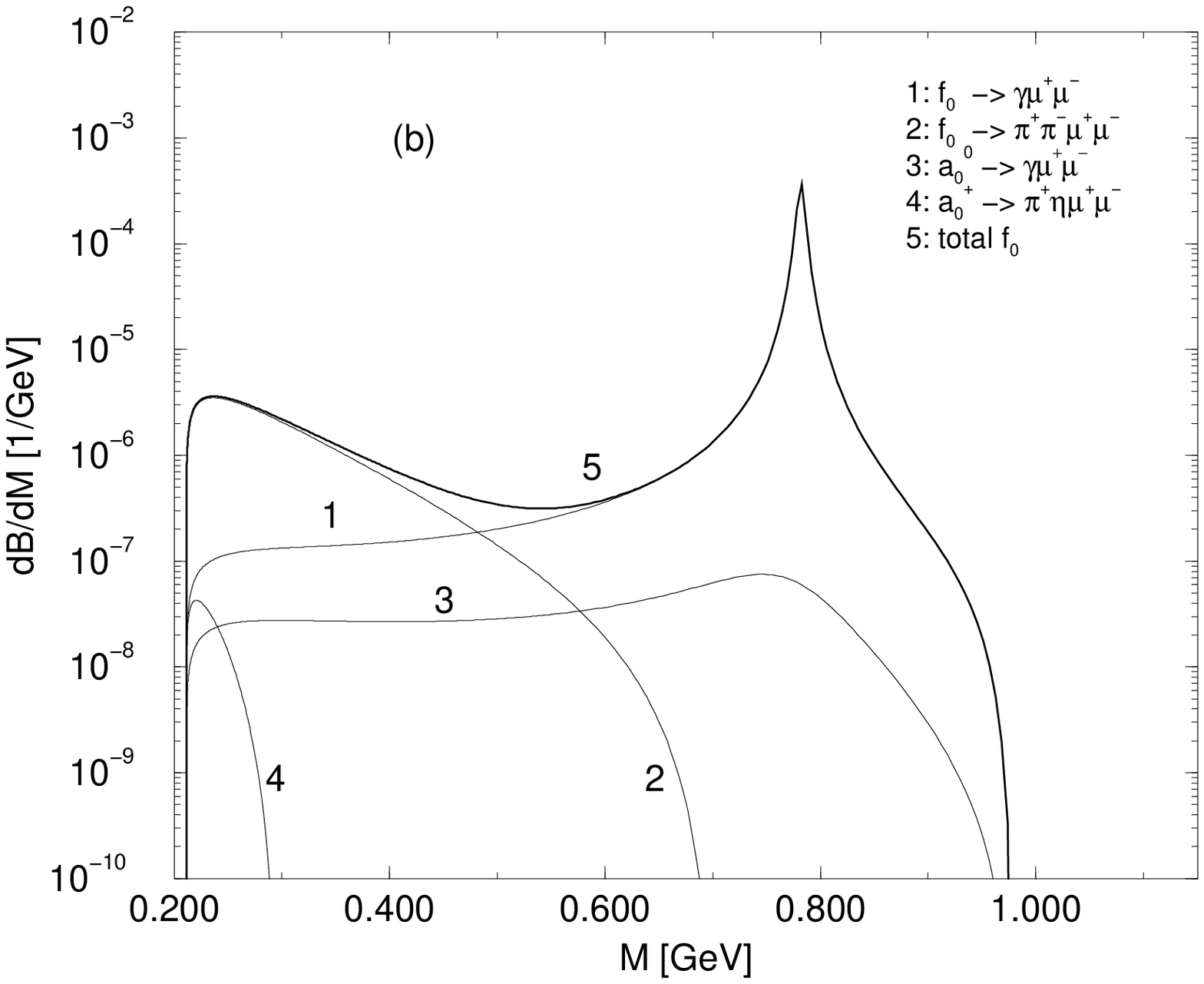} 
\end{center}
\caption{}\label{fig19}
\end{figure}


\begin{references}
\bibitem{Wal}  J.D. Walecka, Ann. Phys. (N.Y.) {\bf 83}, 491 (1974).

\bibitem{Chin}  S.A. Chin, Ann. Phys. {\bf 108}, 301 (1977).

\bibitem{QCD}  E.G. Drukarev and E.M. Levin, JETP Lett. {\bf 48}, 338
(1988); Nucl. Phys. {\bf A511}, 679 (1988).

\bibitem{MES}  C. Amadi and G.E. Brown, Phys. Rep. {\bf 234}, 1 (1993); 
\newline
T. Hatsuda H. Suomi and H. Kuwabara, Progr. Theor. Phys. {\bf 95}, 1009
(1996).

\bibitem{Fae}  T. Maruyama, K. Tsushima and A. Faessler, Nucl. Phys. {\bf %
A535}, 497 (1991); \newline
K. Tsushima, T. Maruyama and A. Faessler, Nucl. Phys. {\bf A537}, 303 (1992).

\bibitem{herrmann93}  M. Herrmann, B. Friman and W. N\"{o}renberg, Nucl.
Phys. {\bf A560}, 411 (1993).

\bibitem{rapp97}  R. Rapp, G. Chanfray and J. Wambach, Nucl. Phys. {\bf A617}%
, 472 (1997).

\bibitem{peters98}  W. Peters, M. Post, H. Lenske, S. Leupold, and U. Mosel,
Nucl. Phys. {\bf A632}, 109 (1998).

\bibitem{Bia}  N. Bianchi {\it et al.}, Phys. Lett. {\bf B299}, 219 (1993);
Phys. Lett. {\bf B325}, 333 (1994).

\bibitem{Kon}  L.A. Kondratyuk, M.I. Krivoruchenko, N. Bianchi, E. De
Sanctis and V. Muccifora, Nucl. Phys. {\bf A579}, 453 (1994).

\bibitem{Wei}  V. Weisskopf, Physikalische Zeitschrift {\bf 34}, 1 (1933).

\bibitem{SOB}  I.I. Sobelman, {\it Introduction to the Theory of Atomic
Spectra}, Pergamon Press, Oxford {\it e.a.}, 1972.

\bibitem{BR95}  G.E. Brown and M. Rho, Phys. Rev. Lett. {\bf 66}, 2720
(1991).

\bibitem{ceres}  G. Agakichiev {\it et al.}, Phys. Rev. Lett. {\bf 75}, 1272
(1995); nucl-ex/9712008;\newline
A. Drees, Nucl. Phys. {\bf A610}, 536c (1996).

\bibitem{helios}  M. Masera, Nucl. Phys. {\bf A590}, 93c (1995).

\bibitem{koch92}  P. Koch, Phys. Lett. {\bf B288}, 187 (1992).

\bibitem{li95}  G.Q. Li, C.M. Ko, G.E. Brown, Phys. Rev. Lett. {\bf 75},
4007 (1995).

\bibitem{Cassing95}  W. Cassing, W. Ehehalt and C.M. Ko, Phys. Lett. {\bf %
B363}, 35 (1995);\newline
W. Cassing, W. Ehehalt I. Kralik, Phys. Lett. {\bf B377}, 5 (1996).

\bibitem{Cassing98}  W. Cassing, E.L. Bratkovskaya, R. Rapp and J. Wambach,
Phys. Rev. {\bf C57}, 916 (1998).

\bibitem{DLS}  R.J. Porter et al., Phys. Rev. Lett. {\bf 79}, 1229 (1997).

\bibitem{ernst}  C. Ernst, S.A. Bass, M. Belkacem, H. St\"{o}cker and W.
Greiner, Phys. Rev. {\bf C58}, 447 (1998).

\bibitem{BK}  E.L. Bratkovskaya and C.M. Ko, Phys. Lett. {\bf B445}, 265
(1999).

\bibitem{brat98}  W. Cassing, E.L. Bratkovskaya, R. Rapp and J. Wambach,
Nucl. Phys. {\bf A634}, 168 (1998).

\bibitem{BC}  W. Cassing and E. Bratkovskaya, Phys. Rep. {\bf 308}, 65
(1999).

\bibitem{Wil}  W.K. Wilson {\it et al}., Phys. Rev. {\bf C57}, 1865 (1998).

\bibitem{pppd}  E.L. Bratkovskaya, W. Cassing, M. Effenberg and U. Mosel,
preprint nucl-th/9903009.

\bibitem{hades}  J. Friese for the HADES Collaboration, to be published in
Prog. Part. Nucl. Phys., Vol. {\bf 42}.

\bibitem{Dav}  R.M. Davidson, in: Proceedings of the workshop on the
structure of the $\eta ^{\prime }$-meson, New Mexico State University and
CEBAF, Las Cruces, New Mexico, March 8-9, 1996.

\bibitem{SAK}  J.J. Sakurai, {\it Currents and Mesons}, University of
Chicago Press, Chicago (1969).

\bibitem{LGL}  L.G. Landsberg, Phys. Rep. {\bf 128}, 301 (1985).

\bibitem{Bayu}  Yu.D. Bayukov {\it et al.}, Yad. Fiz. {\bf 57}, 421 (1994).

\bibitem{QCR}  V.A. Matveev, R.M. Muradyan and A.N. Tavkhelidze, Lett. Nuovo
Cim. {\bf 7}, 719 (1973); \newline
S.J. Brodsky and G.R. Farrar, Phys. Rev. Lett. {\bf 31}, 1153 (1973); Phys.
Rev. {\bf D11}, 1309 (1975).

\bibitem{VZ}  A.I. Vainstein and V.I. Zakharov, Phys. Lett. {\bf 72B}, 368
(1978).

\bibitem{MNA}  M.N. Achasov {\it et al}. (SND Collaboration), Phys. Lett. 
{\bf 440B}, 442 (1998).

\bibitem{VEPP}  M.N. Achasov {\it et al.}, {\it Experiments at VEPP-2M with
SND detector}, Preprint hep-ex/9809013.

\bibitem{HJB}  H.-J. Behrend {\it et al}., Z. Phys. {\bf C49}, 401 (1991).

\bibitem{BL}  S.J. Brodsky and G.P. Lepage, Phys. Rev. {\bf D24}, 1808
(1981).

\bibitem{BES}  A. Bramon, R. Escribano and M.D. Scadron, Eur. Phys. J. {\bf %
C7}, 271 (1999).

\bibitem{PDG}  Particle Data Group, Eur. Phys. J. {\bf C31} (1998).

\bibitem{NAT}  N.A. Tornvist, Phys. Rev. Lett. {\bf 49}, 624 (1982).

\bibitem{RLJ}  R.L. Jaffe, Phys. Rev. {\bf D15}, 267 (1977); {\bf D15}, 281
(1977).

\bibitem{CLO}  F.E. Close, N. Isgur and S. Kumano, Nucl. Phys. {\bf B389},
513 (1993).

\bibitem{AIva}  N.N. Achasov and V.N. Ivanchenko, Nucl.Phys. {\bf B315}, 465
(1989).

\bibitem{PJain}  P. Jain {\it et al}., Phys. Rev. {\bf D37}, 3252 (1988).

\bibitem{Ulf}  Ulf-G. Meissner, Phys. Rep. {\bf 161}, 213 (1988); \newline
F. Klingl, N. Kaiser and W. Weise, Z. Phys. {\bf A356}, 193 (1996).

\bibitem{CHL}  C.H. Lai and G. Quigg, Preprint FN-296, Fermilab USA (1976).

\bibitem{ARGUS}  N. Albrecht {\it et al.}, (ARGUS Collaboration), Phys.
Lett. {\bf B185}, 223 (1987).

\bibitem{DOL}  S.I. Dolinsky {\it et al.}, Phys. Rep. {\bf 202}, 99 (1991).

\bibitem{HPM}  G. Hoehler et al., Nucl. Phys. {\bf 111}, 505 (1976).

\bibitem{NFF}  S. Dubnicka, Nuovo Cim. {\bf 100A}, 1 (1990).

\bibitem{Dzh}  R.I. Dzhelyadin et al., Phys. Lett. {\bf B102}, 296 (1981).

\bibitem{LiGale}  G.Q. Li, C. Gale, Phys. Rev. {\bf C58}, 2914 (1998). 

\bibitem{GMSW}  M. Gell-Mann, D. Sharp and W.E. Wagner, Phys. Rev. Lett. 
{\bf 8}, 261 (1952).

\bibitem{RAG66}  R.A. Grossman, LeRoy R. Price and F.S. Crawford, Jr., Phys.
Rev. {\bf 146}, 993 (1966).

\bibitem{FF}  W. R. Frazer and J. Fulco, Phys. Rev. Lett. {\bf 2}, 365
(1959); Phys. Rev. {\bf 117}, 1603 (1960); 1609 (1960); \newline
G. Gounaris and J. J. Sakurai, Phys. Rev. Lett. {\bf 21}, 244 (1968).

\bibitem{JGL}  J.G. Layter {\it et al.}, Phys. Rev. {\bf D7}, 2565 (1973).

\bibitem{HA87}  H. Albrecht {\it et al.}, Phys. Lett. {\bf B199}, 457 (1987).

\bibitem{FBU}  F. Butler {\it et al.}, Phys. Rev. {\bf D42}, 1368 (1990).

\bibitem{SIB}  S.I. Bityukov {\it et al.}, Z. Phys. {\bf C50}, 451 (1991).

\bibitem{CPIC}  C. Picotto, Phys. Rev. {\bf D45}, 1569 (1992).

\bibitem{FO}  S. Fajfer and R.J. Oakes, Phys. Rev. {\bf D44}, 1599 (1991).

\bibitem{PS}  P. Singer, Phys. Rev. {\bf 130}, 2441 (1963); {\bf 161}, 1694
(1967).

\bibitem{Lubl}  M. Lublinsky, Phys. Rev. {\bf D55}, 249 (1997).

\bibitem{ABram}  A. Bramon, A. Grau and G. Pancheri, Phys. Lett. {\bf B283},
416 (1992).

\bibitem{ABK}  N.N. Achasov and G.N. Shestakov, Sov. J. Nucl. Phys. {\bf 9},
19 (1978).

\bibitem{MIK}  M.I. Krivoruchenko, Yad. Fiz. {\bf 45}, 169 (1987).

\bibitem{Bre}  Lee Brekke, Ann. Phys. (NY) {\bf 240}, 400 (1995).

\bibitem{koch93}  P. Koch, Z. Phys. {\bf C57}, 283 (1993).

\bibitem{lichard95}  P. Lichard, Phys. Rev. {\bf D51}, 6017 (1995).

\bibitem{JP67}  C. Jarlskog and H. Pilkuhn, Nucl. Phys. {\bf B1} (1967) 264.
\end{references}
\end{document}